\documentclass[11pt]{article}
\usepackage{dsfont}
\usepackage{pifont}
\usepackage{amsfonts,amstext}
\usepackage{amsmath,amssymb,graphicx}
\usepackage{hyperref}
\usepackage[numbers,sort&compress]{natbib}
\usepackage{slashed}
\usepackage{multicol,color,longtable}
\usepackage[T1]{fontenc}
\usepackage{hyphenat}
\usepackage{mathrsfs,mathdots,mathtools}
\usepackage{array}
\usepackage{lscape,pdflscape}
\usepackage{multirow}
\usepackage{xtab,caption}
\usepackage{float}
\usepackage{empheq,comment}
\usepackage{stackengine}
\usepackage[makeroom]{cancel}
\usepackage{stmaryrd}
\usepackage{bbm,bm}
\usepackage{xspace}
\usepackage{afterpage}
\usepackage{url}
\usepackage{helvet}         
\usepackage{courier}        
\usepackage{type1cm}        
\usepackage{multicol}        
\usepackage[bottom]{footmisc}
\usepackage{soul}

\newcolumntype{P}[1]{>{\centering\arraybackslash}p{#1}}
\definecolor{darkred}{rgb}{0.65,0.15,0}
\definecolor{newgreen}{rgb}{0.2,0.62,0.14}
\hypersetup{pdfborder={0 0 0},colorlinks=true,urlcolor=blue,citecolor=blue,linkcolor=darkred,linktocpage=true}

\newcommand{\be}{\begin{equation}}
\newcommand{\ee}{\end{equation}}
\newcommand{\bea}{\setlength\arraycolsep{2pt} \begin{eqnarray}}
\newcommand{\eea}{\end{eqnarray}}

\newcommand{\nn}{\nonumber}
\newcommand{\cM}{\mathcal{M}}
\newcommand{\cV}{\mathcal{V}}

\newcommand{\cF}{\mathcal{F}}

\newcommand{\w}[1]{\\[0.#1cm]}

\makeatletter

\@addtoreset{equation}{section}
\makeatother

\hypersetup{colorlinks=true,urlcolor=blue}

\def\ft#1#2{{\textstyle{\frac{\scriptstyle #1}{\scriptstyle #2} } }}

\def\sg{{\mathfrak g}}
\def\cA{{\cal A}}
\def\cO{{\cal O}}
\def\cP{{\cal P}}
\def\cD{{\cal D}}
\def\cF{{\cal F}}
\def\cG{{\cal G}}
\def\cH{{\cal H}}
\def\cK{{\cal K}}
\def\cL{{\cal L}}
\def\cM{{\cal M}}
\def\cN{{\cal N}}
\def\cV{{\cal V}}
\def\cO{{\cal O}}
\def\cP{{\cal P}}
\def\cQ{{\cal Q}}

\def\ba{\begin{array}}
\def\ea{\end{array}}

\def\a{\alpha}\def\ad{{\dot\alpha}}
\def\ab{{\bar\alpha}}\def\ba{{\bar\alpha}}
\def\b{\beta}\def\bd{{\dot\beta}}\def\bb{{\bar\beta}}
\def\c{\gamma}
\def\g{\gamma}
\def\d{\delta}
\def\bd{{\bar\delta}}
\def\e{\epsilon}\def\eps{\epsilon}\def\ve{\varepsilon}
\def\vp{\varphi}

\def\L{\Lambda}
\def\m{\mu}
\def\n{\nu}
\def\r{\rho}\def\rh{\rho}
\def\s{\sigma}
\def\t{\tau}

\def\O{\Omega}
\def\x{\xi}
\def\p{\psi}\def\bpsi{{\bar\psi}}
\def\hI{{\hat I}}
\def\hJ{{\hat J}}
\def\R1{{\rm dim}\,R_1}
\def\adot{{\dot a}}

\def\dA{{\dot A}}

\def\hR{{\widehat R}}
\def\del{\partial}
\def\tr{{\rm tr}}
\def\mR{{\mathbb{R}}}
\def\tI{{\tilde I}}
\def\tJ{{\tilde J}}
\def\tK{{\tilde K}}
\def\vb{{\overline v}}
\def\rmi{{\rm i}}
\def\dg{\dagger}
\def\bi{{\bar \imath}}
\def\bj{{\bar \jmath\,}}

\def\bb{{\bar\beta}}
\def\zb{{\bar z}}

\newcommand{\eq}[1]{(\ref{#1})}

\begin{document}

 {\flushright  MI-TH-192\\[15mm]}

\begin{center}

{\LARGE \bf Survey of Supergravities}\\[5mm]

\vspace{6mm}
\normalsize
{\large  Ergin Sezgin }

\vspace{10mm}

{\it Mitchell Institute for Fundamental Physics and Astronomy\\ Texas A\&M University
College Station, TX 77843, USA}

\vspace{10mm}

\hrule

\vspace{5mm}

 \begin{tabular}{p{14cm}}

A large class of supergravities in diverse dimensions are surveyed. 
This includes maximal supergravities, their general  gaugings in the framework 
of embedding tensor formalism, supergravities with less than maximal 
supersymmetry, their matter couplings and general gaugings. The emphasis is 
on summarizing their most general form to date, and primarily in their 
component formulation. A class of exceptional field theories in an extended 
geometric framework are summarized briefly. For most of the  supergravities surveyed, the bosonic 
part of the Lagrangians and supertransformations up to leading terms in fermions 
are given.

\end{tabular}

\vspace{6mm}
\hrule
\end{center}

\newpage

\setcounter{tocdepth}{2}
\tableofcontents



\newcommand{\TableA}{
\begin{table}[H]
\footnotesize{ \begin{tabular}{|c|c|c|c|c|c|}
\hline
$D$ & $N$ & Scalar Manifold $G/H$    & $n_V^{\rm tot}$ & $n_T$ & Comment \\
\hline
10 & (1,1) & $\mR $ & ---  & --- & \\
\hline
 & (2,0) & $SU(1,1)/U(1)$ & ---  & --- & \\
\hline
9 & 2 & $\mathbb{R} \times SL(2,\mathbb R)/SO(2)$ & 3 & --- & \\
&&&&&\\
  &  1  &  $\mathbb{R} \times SO(n_V,1)/SO(n_V)$  &  $n_V+1$ & --- &  \\
&&&&&\\
\hline
&&&&&\\
8 & 2 &  $\Big(SL(3,\mathbb R)/SO(3)\Big)\times \Big(SL(2,\mR)/SO(2)\Big)$  & $3+3$ & ---  &\\
&&&&&\\
  &1& $\mathbb{R} \times SO(n_V,2)/(SO(n_V)\times SO(2))$ & $n_V+2$ & ---  &\\
  &&&&&\\
  \hline
  &&&&&\\
7 & 4 & $SL(5,\mathbb R)/SO(5) $ & 10 & --- &  \\
  &&&&&\\
  & 2 & $\mathbb{R} \times SO(n_V,3)/(SO(n_V)\times SO(3))$ &  $n_V+3$ & ---  &\\
  &&&&&\\
  \hline
  &&&&&\\
6 & (2,2) & $SO(5,5)/(SO(5)\times SO(5))$ & $16_s$ & --- &  \\
  &&&&&\\
 & (3,1) & $F_{4(4)}/(USp(6)\times USp(2))$ & (14,1)  & (6,2) &  {\rm no\, graviton}\\
  &&&&&\\
 & (4,0) & $E_{6(6)}/USp(8)$ & --   & 27 & {\rm no\, graviton} \\
&&&&&\\
 & (3,0) & $SU^\star(6)/USp(6)$ & --   & 14 &  {\rm no\, graviton}  \\
 &&&&&\\ 
  & (2,1) & $SU^\star(4)/USp(4)$ & (4,2)  & 5+1 & $N=(1,0)$ twin  \\
  &&&&& \\ 
    & (2,0) &  $SO(n_T,5)/(SO(n_T)\times SO(5))$  & ---    & $n_T$  &  \\
  &&&&&\\
  & (1,1) & $\mathbb{R} \times SO(n_V,4)/(SO(n_V)\times SO(4))$ & $n+4$ & ---  &\\
 &&&&&\\
      & (1,0)  &  $SO(n_T,1)/SO(n_T)$  & $n_V$ &  $n_T$   & \\
  &&&&&\\
      & &   $SO(9,1)/SO(9)$  &  $16_s$ &  9   & magical \\
  &&&&&\\
      & &   $SO(5,1)/SO(5)$  & $8_s$   &   5  & magical, $N=(2,1)$ twin  \\
  &&&&&\\
      & &   $SO(3,1)/SO(3)$   & $4_s$  &   3  & magical  \\
  &&&&&\\
      & &   $SO(2,1)/SO(2)$  & $2_s$   &   2  & magical  \\
  &&&&&\\
    \hline
\end{tabular}
  \caption{ \footnotesize Scalar manifolds from $n_V$ vector and $n_T$ tensor multiplet couplings for supergravities in $D \ge 6$ dimensions with $N$ supersymmetry. $n_V^{\rm tot}$ is the total number of vector fields, and  $n_V$ is the total number of vector multiplets. The last four entries in $6D$ arise in magical supergravities discussed section 7.5, and their consecutive circle reductions yield the last four entries for $D=5,N=2$ and $D=4, N=2$ in table 5, and for $D=3,N=4$ in table 6. For ``twin'' supergravities see sections 7.6 and 8.2.} 
}
\end{table}}


\newcommand{\TableB}{
\begin{table}[H]
{\footnotesize
\begin{tabular}{|c|c|c|c|c|}
\hline
&&&&\\
D & N  & Scalar Manifold $G/H$   & $n_V^{\rm tot}$  & Comment\\
\hline
5 & 8 & $E_6/USp(8)$ & 27  &  \\
  &&&&\\
   & 6 &  $SU^{*}(6)/USp(6)$ &   14+1    &  N=2 twin \\
&&&&\\
     & 4 & $\mathbb{R} \times SO(n_V,5)/(SO(n_V)\times SO(5))$ & $n_V+5$   &  \\
&&&&\\
     & 2 &  $SO(n_V,1)/SO(n_V)$ & $n_V+1$   &\, VSR, $n_V>1$, nonsym. r-map image\, \\
&&&&\\
&& $SO(1,1)$ & 1+1  & VSR, sym. r-map image \\
&&&&\\
&& $\mathbb{R} \times SO(n_V-1,1)/SO(n_V-1)$ & $n_V+1$   &  VSR \\

&&&&\\
  &&  $E_{6(-26)}/F_4$  & 26+1  & VSR\\
&&&&\\
  &&  $SU^*(6)/Sp(3)$ & 14+1  & VSR,  N=6 twin \\
&&&&\\
  &&  $SL(3,C)/SU(3)$ & 8+1  &   VSR\\
&&&&\\
 &&  $SL(3,R)/SO(3)$  & 5+1   &   VSR\\
\hline
4  & 8 & $E_{7(7)}/SU(8)$ & 28  &\\
   &&&&\\
   & 6 & $SO^\star(12)/U(6)$ & 16 & N=2 twin \\
   &&&&\\
   & 5 &  $SU(5,1)/U(5)$  &  10  &  \\
  &&&&\\
   & 4 & $(SL(2,\mR/SO(2))$ &&\\
  && $\times SO(n_V, 6)/(SO(n_V)\times SO(6))$ & $n_V+6$   & $n\ge 1$ \\
  &&&&\\
   & 3 & $SU(n_V,3)/S[U(n_V)\times U(3)] $ & $n_V+3$  &   $n\ge 3$ \\
  &&&&\\
& 2 & $SU(n_V,1)/(SU(n_V)\times U(1))$ & $n_V+1$  & SK, not an r-map image\\
  &&&&\\
  && $SU(1,1)/U(1)$ & 1+1  & SK, r-map of pure 5D sugra\\
  &&&&\\
  &&  $(SU(1,1)/U(1))$ &&\\
  &&  $\times SO(n_V-1,2)/(SO(n_V-1)\times SO(2))$ &  $n_V+1$  & VSK\\
   &&&&\\
   && $\big[SU(1,1)/U(1)\big]^2$ &  2   & VSK \\
  &&&&\\
   && $E_{7(-25)}/(E_6 \times U(1))$ &  28   &  VSK \\
  &&&&\\
  &&  $SO^\star(12)/U(6)$  & 16 &VSK, N=6 twin   \\
  &&&&\\
  & &  $SU(3,3)/S[U(3)\times U(3)]$ & 10 & VSK \\
  &&&&\\
  &&   $Sp(6,\mR)/U(3)$   &  7 &  VSK \\
  \hline
  \end{tabular}}
  \caption{ \small Symmetric scalar manifolds for supergravities in $D=4,5$ with $N$ supersymmetry, and vector/tensor multiplet couplings. VSR and (V)SK refer to very special real and (very) special K\"ahler manifolds discussed in section 8.4 and 9.6, respectively. In $5D, N=2$ supergravity, if a subgroup $K\subset G$ can be gauged, then $n_V={\rm dim}\, K$ and $n_T=n_V^{\rm tot} - {\rm dim}\, K$ vectors need to be dualized to two-forms that lie in symplectic representation(s) of $K$, as discussed in section 8.4. The ``twins" are discussed in sections 8.2 and 9.2. For the r-map, see \cite{deWit:1992cr}.}
  \end{table}
  }


\newcommand{\TableCC}{
\begin{table}[H]
{\footnotesize
\begin{tabular}{|c|c|c|c|c|c|}
\hline
   &&&\\
   D & $N$  & Scalar Manifold $G/H$      & Comment\\
   &&&\\
   \hline
   &&&\\
3 & 16 & $E_{8(8)}/SO(16)$  &  \\
   &&&\\
   & 12 & $E_{7(-5)}/(SO(12)\times Sp(1))$ &  $N=4$ twin \\
   &&&\\
   & 10 &  $E_{6(2)}/(SO(10)\times U(1))$   &  \\
  &&&\\
  & 9 & $F_{4(-20)}/SO(9)$  & \\
  &&&\\
  & 8 & $SO(n,8)/(SO(n)\times SO(8))$  &  $N=4$ twin for $n=4$ \\
  &&&\\
  & 6 & $SU(n,4)/S[U(n)\times U(4)]$  & $N=4$ twin for $n=2$ \\
  &&&\\
  & 5 & $Sp(n,2)/(Sp(n)\times Sp(2))$ &   $N=4$ twin for $n=1$\\
  &&&\\
  & 4 & $Sp(n+1,1)/(Sp(n+1)\times Sp(1))$ &\ QK, not a $c$-map image, $N=5$ twin \\
  &&&\\
  && $U(2,1)/(U(2)\times U(1))$ &  special QK, $c$-map of pure $4D$ sugra\\
  &&&\\
  && $G_{2(2)}/SO(4)$   &  special QK \\
  &&&\\
  && $SU(n+2,2)/(SU(n)\times SU(2)\times U(1))$  & special QK, $N=6$ twin  \\
  &&&\\
  && $SO(n+4,4)/(SO(n+4)\times SO(4))$  & very special QK, $N=8$ twin \\
  &&&\\
  && $SO(3,4)/[SU(2)]^3$  & very special QK \\
  &&&\\
   & & $E_{8(-24)}/(E_7\times Sp(1))$ & very special QK   \\
  &&&\\
  &&  $E_{7(-5)}/(SO(12)\times Sp(1))$  &  very special QK, $N=12$ twin    \\
  &&&\\
  && $E_{6(2)}/(SU(6)\times Sp(1))$ &  very special QK \\
  &&&\\
  &&  $F_{4(4)}/(Sp(3)\times Sp(1))$  &   very special QK \\
  &&&\\
  \hline
  \end{tabular}}
  \caption{ \small Symmetric scalar manifolds in $3D$ supergravities for $4\le N\le 16$. The (very) special QK manifolds are discussed in section 7.4. For ``twin" supergravities, see \cite{Roest:2009sn}, and for the $c$-map, see \cite{Cecotti:1988qn}. The scalar manifold in $3D$ is an arbitrary Riemannian manifold for $N=1$, K\"ahler for $N=2$, quaternionic for $N=3$, in general a product of two quaternionic manifolds for $N=4$ and symmetric homogeneous space for $N>4$.}
  \end{table}
  }

\newcommand{\TableD}{
\begin{table}
\begin{center}
\footnotesize{\begin{tabular}{|c|c|c|c|c|}
\hline
$G_T$ & vectors & vectors reality  & $\Gamma^I_{AB}$ & two-forms \\[1ex] \hline
&&&&\\[-2ex]
${SO}(9,1)$ &
${\bf 16}_c$
& MW & $\Gamma^I_{AB}$
& ${\bf 10}$
\\[1ex]\hline&&&&\\[-2ex]
${SO}(5,1)\times {USp}(2)$ &
${\bf (4}_c,{\bf 2)}$
& SMW,\, $A=(\alpha r)$ &
$\Gamma^I_{\alpha r,\beta s} = \Gamma^I_{\alpha\beta} \epsilon_{rs}$
& ${\bf (6,1)}$
\\[1ex]\hline&&&&\\[-2ex]
${SO}(3,1)\times {U}(1)$ &
${\bf (2,1)}_+ + {\bf (1,2)}_-$
&W,\, $A=\{\alpha,\dot{\beta}\}$
&
$\left(
\begin{array}{cc}
0 & \Gamma^I_{\alpha\dot{\beta}}\\
\bar\Gamma^I_{\dot\alpha{\beta}} & 0
\end{array}
\right)$
& ${\bf (2,2)_0}$
\\[3ex]\hline&&&&\\[-2ex]
${SO}(2,1)$  &
${\bf 2}$&
M & $\Gamma^I_{AB}$
& ${\bf 3}$
\\[1ex]\hline
\end{tabular}}
\caption{
{\small
The first column shows the full global symmetry groups
of magical supergravities, the second column gives the representation content of the vector fields under these groups, whose reality properties
are listed in the third column: Majorana (M), Weyl (W), Majorana-Weyl (MW), symplectic Majorana-Weyl (SMW). The last column gives the two-form representation content.
 }}
\label{tab:reality}
\end{center}
\end{table}}


\newcommand{\TableE}{
\begin{table}[htb]
\begin{center}
\begin{tabular}{ccccccc}
$D=6$ &&$D=5$ && $D=4$ && $D=3$
\\[.6ex]\hline\\[0ex]
$\frac{SO(9,1)}{SO(9)}$ &
$\longrightarrow$ &
$\frac{E_{6(-26)}}{F_4}$ &
$\longrightarrow$ &
$\frac{E_{7(-25)}}{E_6 \times SO(2)}$ &
$\longrightarrow$ &
$\frac{E_{8(-24)}}{E_7 \times SU(2)}$
\\[2ex]
$\frac{SO(5,1)}{SO(5)}$ &
$\longrightarrow$ &
$\frac{SU^*(6)}{USp(6)}$ &
$\longrightarrow$ &
$\frac{SO^*(12)}{U(6)}$ &
$\longrightarrow$ &
$\frac{E_{7(-5)}}{SO(12) \times SU(2)}$
\\[2ex]
$\frac{SO(3,1)}{SO(3)}$ &
$\longrightarrow$ &
$\frac{SL(3,\mathbb{C})}{SO(3)}$ &
$\longrightarrow$ &
$\frac{SU(3,3)}{SU(3)\times SU(3)\times U(1)}$ &
$\longrightarrow$ &
$\frac{E_{6(+2)}}{SU(6) \times SU(2)}$
\\[2ex]
$\frac{SO(2,1)}{SO(2)}$ &
$\longrightarrow$ &
$\frac{SL(3,\mathbb{R})}{SO(3)}$ &
$\longrightarrow$ &
$\frac{Sp(6,\mathbb{R})}{U(3)}$ &
$\longrightarrow$ &
$\frac{F_{4(+4)}}{USp(6)\times USp(2)}$
\\[2ex]
\hline
\end{tabular}
\end{center}
\label{tab:cosets}
\caption{\small Scalar target spaces of magical supergravities in $6,5,4$ and $3$  dimensions. }
\end{table}}


\section{Introduction}

After simple supergravity in four dimension was discovered in 1976 \cite{Freedman:1976xh,Deser:1976eh}, supergravities in diverse dimensions were constructed at a rapid pace in subsequent years. The importance of the fact that they appear as low energy limits of string theory has been widely recognized. In a  bottom to top approach, supergravities continue to be of great relevance also in studying the effective theories of quantum gravity. Select papers were put together with brief commentaries on supergravities in diverse dimensions (including conformal supergravities), their compactifications, and anomalies in 1989 \cite{Salam:1989fm}, where an extensive list of references was provided. Since then, the literature on supergravities has expanded greatly, driven considerably by advances made in string theory. Some of the noteworthy developments are as follows: (a) duality symmetries have been explored with remarkable results, (b) exceptional field theories have emerged as manifestly duality invariant  formulations of supergravities in diverse dimensions, (c) generalized geometries have been a fertile area of study, (d) the embedding tensor formalism  has made it possible to study systematically all possible gaugings of supergravities, (e) progress has been made in the construction of 
higher derivative extensions of supergravities, (f) anomalies have been probed in more depth, and (g) consistency of higher derivative extended  supergravities at different levels has been pursued vigorously in the frame of the swampland program, thereby making progress in addressing the question of uniqueness of  the string theory as a UV complete theory of quantum gravity.

Here we shall provide a brief survey of a large class of supergravity theories that have been constructed so far. Similarities between them will become apparent as we span different dimensions and amount of supersymmetry. One might hope to relate them all to string/M  theory by several mechanisms that have been discovered. However, as the amount of supersymmetry decreases, more general couplings arise. These may ultimately be related to string/M theory as well by means of yet to be discovered mechanisms. In any event, it is worthwhile to pursue the goal of establishing as many connections as possible among them.

Surveying a large class of supergravities (pure, matter coupled and gauged) in a limited space is a challenging task. Several omitted topics go beyond the scope of this relatively brief survey. Among the omitted topics are: higher derivative extensions\footnote{Higher derivative supergravities in diverse dimensions will be reviewed elsewhere \cite{OPS}.}, supergravities on manifold with boundaries, compactifications, consistent truncations,  Killing spinors and exact solutions, AdS/CFT correspondence, quantum supergravity and  amplitudes. We touch very little the subjects of superspace and anomalies, and we survey briefly the exceptional field theories.

In this chapter, starting from $11D$ and ending up with $1D$, we survey the supergravities that have been constructed so far at the level of two derivatives. In sections 2-10, we give the field contents, the bosonic parts of the Lagrangians and the supertransformations of the fermionic fields in leading order in fermions. The focus on the supertransformations of the fermions is due to their relevance to the definition of Killing spinors. The fact that they are given in leading order in fermions will not be repeated henceforth. Only in $11D$ the full action and supertransformations are given. In section 12, we summarize the $E_{n(n)}$ exceptional field theories for $n=6,7,8$, and $N=1$ supersymmetric double field theory. The properties of spinors in arbitrary dimensions and the issue of conventions are discussed in appendix A, the embedding tensor formalism is summarized in appendix B, and a comprehensive table of symmetric coset spaces arising in pure and matter coupled supergravities is provided in tables 4,5 and 6 in appendix C.   

\section{D=11}

\subsubsection*{Standard $11D$ supergravity}

As is well known, eleven is  highest  dimension in which a supergravity can exist, if fields with spin $s>2$ are to be avoided\footnote{We assume $(10,1)$ signature. A locally supersymmetric action in $(10,2)$ dimensions is discussed in \cite{Castellani:2017vbi}.}
The supergravity multiplet in  eleven dimensions is the unique supermultiplet with spin $s\le 2$ fields, and it consists of the vielbein $e_\mu{}^a$, the real 3-form potential $A_{\m\n\rh}$ and the gravitino $\psi_\m$, which is a Majorana spinor. The 11D supergravity was constructed in \cite{Cremmer:1978km}. In the rest of this survey, we give the bosonic parts of the supergravity Lagrangians and the supersymmetry transformations of the fermions only.  However, given the special place $11D$ supergravity occupies in the landscape of supergravities, we shall make an exception and recall the full 2-derivative action, and its full supersymmetry. The full action given is given by \cite{Cremmer:1978km}\footnote{ In the conventions used here $\{\Gamma_a,\Gamma_b\} =2\eta_{ab}$ with $\eta_{ab} ={\rm diag}(-,++...+),\  \gamma^{a_1...a_{11}} = -\epsilon^{a_1...a_{11}},\hfill\break \bar\psi = \psi^\dagger i\gamma_0, \  D_\mu(\omega) \epsilon  = 
\partial_\mu\epsilon +\frac14 \omega_\mu{}^{ab} \gamma_{ab}\epsilon$ and $R= e^\mu_a e^\nu_b R_{\mu\nu}{}^{ab}$.} 
\bea
{\cal L}= e R(\omega)-\frac{1}{48} e F_{\mu\nu\rho\sigma} F^{\mu\nu\rho\sigma} - \frac{1}{144^2} \varepsilon^{\mu_1...\mu_{11}} F_{\mu_1...\mu_4} F_{\mu_5...\mu_8} A_{\mu_9...\mu_{11}}
\nn\\
+ e\bar\psi_\mu \Gamma^{\mu\nu\rho} D_\nu \left(\frac{\omega+\widehat\omega}{2}\right) \psi_\rho  + \frac{1}{192} e\, \bar\psi_{[\lambda} \Gamma^\lambda\Gamma^{\mu\nu\rho\sigma} \Gamma^\tau \psi_{\tau]}
 \left(F_{\mu\nu\rho\sigma} + \widehat F_{\mu\nu\rho\sigma} \right)\ ,
\label{CSJ}
\eea  
where
\bea
\widehat\omega_{\mu ab} &=& \omega^{(0)}_{\mu ab} -\frac14 (\bar\psi_\mu\gamma_a\psi_b -\bar\psi_\mu\gamma_b\psi_a 
+\bar\psi_a\gamma_\mu\psi_b )\ ,
\nn\\
\omega_{\mu ab} &=& \widehat\omega_{\mu ab} -\frac{1}{8}\bar\psi_c \gamma_{\mu ab}{}^{cd} \psi_d\ ,
\nn\\
F_{\mu\nu\rho\sigma} &=& 4 \partial_{[\mu} A_{\nu\rho\sigma]}\ ,\qquad \widehat F_{\mu\nu\rho\sigma} = F_{\mu\nu\rho\sigma} -3 {\bar\psi}_{[\mu} \gamma_{\nu\rho} \psi_{\sigma]}\ ,
\eea
and $ \omega^{(0)}_{\mu ab} $ is the spin connection without torsion. The local supersymmetry transformations are
\bea
\delta e_\mu{}^a &=& -\frac12 \bar\epsilon \gamma^a \psi_\mu\ ,\qquad \delta A_{\mu\nu\rho} = \frac32 \bar\epsilon\gamma_{[\mu\nu} \psi_{\rho]}\ ,
\nn\w2
\delta\psi_\mu &=& D_\mu (\widehat\omega)\epsilon +\frac{1}{288} \left(\gamma_\mu{}^{\nu\rho\sigma\tau} -8\delta_\mu^\nu \gamma^{\rho\sigma\tau} \right) \widehat F_{\nu\rho\sigma\tau}\epsilon\ .
\eea
The equations of motion also have a rigid scaling symmetry, also known as the trombone symmetry\footnote{This amusing  terminology first appeared in \cite{Cremmer:1997xj}, where it is motivated by arguing that ``it allows one to scale magnitudes in and out". As discussed in detail in \cite{LeDiffon:2008sh}, in a generic two-derivative supergravity theory in $D$ dimensions, the global trombone symmetry is present with $g_{\m\n} \to \lambda^2 g_{\m\n}$ and $A_{\m_1...\m_p} \to \lambda^p A_{\m_1...\m_p}$, 
scalar fields remaining invariant, and $\psi_\m \to \lambda^{1/2} \psi_\m, \chi \to \lambda^{-1/2}\chi$, thus the Lagrangian  scaling homogeneously as $\cL \to \lambda^{D-2}\cL$.}, under which $g_{\m\n} \to \lambda^2 g_{\m\n}$ and  $A_{\m\n\r} \to \lambda^3 A_{\m\n\r}$.
The equations of motion admit $11D$ Minkowski spacetime as a vacuum solution, but not (A)dS spacetime. Furthermore, the action does not admit the introduction of a cosmological constant \cite{Bautier:1997yp}. A vast literature has accumulated on $11D$ supergravity and especially on its compactifications. For a more detailed review of the $11D$ supergravity itself, and the duality symmetries that emerge in its toroidal compactifications, see the chapter by Samtleben in this volume.

\subsubsection*{Modified $11D$ supergravity EOMs}

In \cite{Howe:1997he} it was observed that the EOMs of $11D$ supergravity admit a slight modification in which the 
standard spin connection $D$ can be replaced by a conformal one taking the form ${\hat D}= D+ 2k$, provided that the conformal part of the curvature vanishes, namely $dk=0$. In simply connected spaces this implies that the one form $k$ 
is exact and this modification amounts to a field redefinition. However, in non-simply connected spacetimes this modification  is non-trivial, as was shown in \cite{Howe:1997qt}. This slightly modified $11D$ supergravity is formulated at the level of EOMs only, since it arises in an on-shell Weyl superspace formulation. It was shown in \cite{Howe:1997he} that  its  dimensional reduction on a circle with a nontrivial Wilson line gives rise to gauged type IIA supergravity in $10D$. This 
$11D$ theory was called MM theory in \cite{Chamblin:2001dx} , where its properties were further discussed. In particular, it was shown that it admits $dS_D \times S^{10-D}\times S^1$ solution for any $D$, and that while this solution  admits a 
Killing spinor, it is not globally well defined. It has later been shown that the $10D$ EOM's that are obtained from this theory  can also be obtained by a Scherk-Schwarz reduction of $11D$ supergravity on a circle, using 
the on-shell rigid scale invariance of 11D supergravity equations of motion \cite{Lavrinenko:1997qa}\footnote{Nonetheless, see \cite{Chamblin:2001dx} for  cautionary remarks about the comparison of the mechanism described in \cite{Howe:1997qt}, with the Scherk-Schwarz reduction described in \cite{Lavrinenko:1997qa}.} 

\subsubsection*{Massive $11D$ supergravity}

Motivated by the problem of embedding of type IIA supergravity with Romans mass term into $11D$, which is often referred to as the massive type IIA supergravity, one can construct an action in $11D$ by introducing an auxiliary non-dynamic vector field $k^\mu$ with respect to which the Lie derivative of the metric and the 3-form potential, namely $\cL_k g_{\m\n}=\cL_k C_{\m\n\r}=0$ \cite{Bergshoeff:1998vv}. The bosonic part of the resulting action is given in \cite[eq. (1.15)]{Bergshoeff:1998vv}, where it has been argued to provide a target space background for massive branes. Indeed, its dimensional reduction on a circle produces the type IIA supergravity with the Romans mass deformation \cite{Bergshoeff:1998vv}.

\subsubsection*{\bf $11D$ supergravity with both three-form and six-form} 

As in any supergravity theory, one my ask whether a formulation exists in which a set of potential fields may be Hodge-dualized. For early treatment of the role of the six-form potential in $11D$ supergravity, see \cite{Nicolai:1980kb,Fre:1984pc}. The story for dualization of  graviton is very complicated and beyond the scope of this survey. Focusing on the three-form potential, one cannot simply add a Lagrange multiplier term $F_4\wedge dA_6$ and treat $F_4$ as an independent variable, since it does not appear only through its field strength in the action. Rather, one may look for an action in which both $A_3$ and $A_6$ appear. Putting the gravitini aside, the key to this formulation is the duality equation 
\be
{\cal O} :=  F_7 - \star F_4=0\ ,
\ee
where $F_7 = dA_6 - A_3 \wedge dA_3$. There have been several proposals for writing down an action for a $p$-form potential and its dual $(D-p-2)$-form in $D$-dimensions, from which the desired equations motion can be derived. For an extensive list of references on this subject, see for example, see \cite{Sen:2015nph,Bansal:2021bis}. Here we focus on $11D$, and briefly mention few of the proposals. In one of them, manifest Lorentz covariance is maintained at the cost of introducing a timelike unit vector built out of a real scalar $a(x)$ and a non-polynomial action \cite{Bandos:1997gd}.  In another approach based on \cite{Siegel:1983es}, an action with a term of the form $\lambda\,\cO^2$ added to $\cL_{CJS}$, where $\lambda$ is a Lagrange multiplier field with suitably symmetries, was considered in \cite{Nishino:1998rr}. It has been argued in \cite{Henneaux:1988gg}, however, while such actions are permissible classically, they have difficulties quantum mechanically if not properly treated. Nonetheless, it may be worth mentioning that, curiously enough a pseudo-Lagrangian of the form  
\begin{align}
\cL =& \cL_{CJS} + \cO_{\m\n\r\s} \cO^{\m\n\r\s}\ .
\label{LBKS}
\end{align}
follows from a consistent truncation of a ``master exceptional field theory" based on $E_{11}$ duality symmetry \cite{Bossard:2021ebg}. The ``pseudo" means that the duality equation is to be imposed by hand after obtaining the equations of motion. 
Next we mention the approach of \cite{Bunster:2011qp} where an action is written but at the expense of sacrificing manifest Lorentz covariance. See \cite{Kleinschmidt:2022qwl} where this formalism has been applied to $11D$ supergravity. Finally, there exists a proposal \cite{Sen:2015nph}, which is inspired by string field theory, and has been applied to type IIB supergravity so far, which is manifestly Lorentz covariant, and has finitely many auxiliary fields, and has been asserted that it is amenable to BV quantization.

\subsubsection*{M-Branes and Superspace} 

$11D$ supergravity has $M2$-brane~\cite{Duff:1990xz} and $M5$-brane solutions~\cite{Gueven:1992hh}. This is consistent with  fact that there exist actions for $M2$-brane and $M5$-branes which describe their propagation in $11D$ spacetime. In fact their description is best achieved in $(11|32)$ dimensional superspace, and relies critically on the presence of a worldvolume local fermionic symmetry known as  $\kappa$-symmetry. See~\cite{Bergshoeff:1987cm} for the construction of the $M2$-brane action, and~\cite{Howe:1997fb} for  $M5$-brane field equations, and~\cite{Bandos:1997ui} for an $M5$-brane action. It is a remarkable fact that the $\kappa$-symmetry of these actions imposes constraints on  the target superspace torsion and the super four-form (these are spelled out in~\cite{Bergshoeff:1987cm}) which imply uniquely~\cite{Howe:1997he} the standard $11D$ supergravity field equations \cite{Cremmer:1978km} up to field redefinitions~\cite{Cremmer:1980ru,Brink:1980az}.  

It is worth noting that $11D$ supergravity also admits the pp-wave solution \cite{Hull:1984vh} which may be viewed in as a 1-brane, as well as KK monopole solution which may be considered as a 6-brane \cite{Townsend:1995kk}. Finally, $M9$ brane solutions have also been studied \cite{Bergshoeff:1998bs,Bergshoeff:1998re}, including their relationship to the boundaries of the $M_{10}\times S^1/\mathbb{Z}_2$ that arise in the Horava-Witten construction of $11D$ supergravity on a manifold with boundary \cite{Horava:1995qa,Horava:1996ma}. 

\subsubsection*{Signature of spacetime}

It was shown in~\cite{Hull:1998ym} that  M-theory in $(1,10)$ dimensions is linked via dualities to a theory in $(2,9)$ and $(5,6)$ dimensions, referred to as $M^\star$ and $M^\prime$ theories, respectively. Various limits of these were shown to give rise to type IIA-like string theories in $(0,10), (1,9), (2,8), (4,6)$ and $(5,5)$ dimensions, and to type IIB-like string theories in $(1,9), (1,9), (3,7)$ and $(5,5)$ dimensions. 

\section{D=10}

\subsection{ Type IIA supergravity and its massive deformations}

The $N=(1,1)$ supergravity in $10$D is usually referred to as type IIA or simply IIA supergravity, as it arises in the low energy limit of type IIA string. Its field content is $(e_\mu{}^a, B_{\mu\nu}, \phi, C_\mu, C_{\mu\nu\rho}, \psi_\mu, \chi )$.  The full action and supersymmetry transformations were constructed in \cite{Giani:1984wc,Campbell:1984zc}. Later,  the action was extended by Romans \cite{Romans:1985tz} by introduction of a mass parameter $m$.  The bosonic part of this extended action in string frame, and in the notations and conventions of ~\cite{Bergshoeff:2001pv}, is given by
\bea
S &=& -\frac{1}{{\kappa}^2} \int d^{10}x \sqrt{-g} \Bigg\{ e^{-2\phi} \Big[ R-4 (\partial\phi)^2 +\frac12 H^2\Big]  + \frac12 m^2 + \frac12 G_2 \cdot G_2
\nn\w2
&& + \frac12 G_4 \cdot G_4 
-\star \Big[ \frac12 dC_3 \wedge dC_3 \wedge B +\frac16 m\, dC_3 \wedge B \wedge B \wedge B 
\nn\w2
&& +\frac{1}{40}m^2\, B\wedge B \wedge B\wedge B\wedge B \Big]  \Bigg\}\ ,
\label{2A}
\eea
where
\begin{align}
G_2 =& dC_1 + m\,B\ , \qquad 
G_4 = dC_3 + dB \wedge C_1 + \frac12 m B\wedge B\ .
\end{align}
The fermionic terms and the supertransformations can be found in \cite[eqs. (2.33) and (2.34)]{Bergshoeff:2001pv}. 
This action is invariant under the rigid scale transformations under which $e^\phi \to\alpha e^\phi, \ C_3\to \alpha C_3$ and $B\to \alpha^{-2} B$ provided that one also scales $m \to \alpha^5 m$ \cite{Romans:1985tz}. If $m=0$, which is a smooth limit which gives the usual type IIA supergravity, then one must transform instead $C_1 \to \a^3 C_1$ \cite{Giani:1984wc,Campbell:1984zc}. Note that this is a genuine (off-shell) symmetry, as opposed to the (on-shell) trombone symmetry. In the presence of the Romans mass parameter $m$, the theory no longer admits $10D$ Minkowski spacetime as a vacuum solution but it has a domain-wall solution. Furthermore, the two-form potential has eaten the vector field to become massive. 

The massive IIA supergravity can be obtained from the massive $11D$ supergravity summarized earlier by dimensional reduction on a circle \cite{Bergshoeff:1998vv}. As mentioned earlier, there also exists another massive deformation of type IIA supergravity, which can be obtained from $11D$ supergravity by Scherk-Schwarz reduction on a circle, in which the $11D$ on-shell trombone symmetry is employed, thereby giving rise to a gauged theory in $10D$.  In this gauged type IIA theory, the vector eats a scalar, and the three-form eats the two-form to become massive. The bosonic field equations of the gauged type IIA theory are provided in \cite[Sec. 5]{Lavrinenko:1997qa}, and they clearly exhibit the Stuckelberg symmetries associated with the Higgs mechanism responsible for the 1-form and 3-form fields becoming massive. See also \cite{Kerimo:2004qx} where the generalized reduction of the $11D$ supertransformations to $10D$, as well as vacuum solutions, are given. 
The resulting theory can also be viewed as type IIA supergravity in which a combination of the on-shell trombone symmetry (leaving the $10D$ scalars invariant) and an off-shell  $GL(1)$ symmetry (leaving the of the$10D$ metric invariant) of type IIA supergravity is gauged. This phenomenon is explained in detail in \cite{LeDiffon:2008sh} where a systematic framework for the classification and construction of these theories is set up, using the embedding tensor formalism (see appendix B here). This theory has no $D8$ brane solution but it admits a supersymmetric and non-static cosmological solution\cite{Howe:1997qt}, as well as nonsupersymmetric de Sitter solution \cite{Lavrinenko:1997qa,Chamblin:2001dx}.

\subsubsection*{The democratic formulation} 

 One may consider the dualization of the RR fields $C_3, C_1$ in \eq{2A} to obtain an action in which the dual potentials $A_5, A_7$ appear instead. To do this, the field $C_3$ needs to be redefined so that all RR fields appear only under derivatives in the action. This is possible by defining $A_3= C_3 -C_1 \wedge B$. However, adding the Lagrange multiplier terms $\int \left( dC_1 \wedge dA_7-dA_3\wedge dA_5\right)$, and treating $G_2$ and $G_4$ as integration variables, leads to coupled and complicated nonlinear  equations in terms of the dual field strengths. One can formally solve these equations at the expense of having inverses of $B$ dependent functionals, and it doesn't seem to be illuminating. However, it is possible to write down a simple pseudo-action in which both the RR fields and their duals appear. This is referred to as the democratic formulation, and the bosonic part of the action takes the form \cite{Bergshoeff:2001pv}
\bea
S &=& -\frac{1}{\kappa^2}} \int d^{10}x \sqrt{-g} \Bigg\{ e^{-2\phi} \Big[ R-4 ({\partial\phi)^2 +\frac12 H^2 \Big]  + \frac12 \sum_{n=0}^5 G_{2n}\cdot G_{2n}\Bigg\}\ ,
\label{A1}
\eea
where 
\be
G=dC-dB \wedge C +G_0 e^B\ ,
\label{A2}
\ee
in which  formal sums $C=C_1+C_3+...+C_9$ and $G=G_0+ G_2+ ...+ G_{10}$ are understood. Note that there are no explicit Chern-Simons term in the action. This is a pseudo-action because the duality equations must be put in by hand, {\it after} varying  action. These duality equations take  the form $G_{2n} + \Psi_{2n}= (-1)^{[n]} \star G_{(10-2n)}$, where $\Psi_{(2n)}$ are certain fermionic bilinear terms that can be found in \cite{Bergshoeff:2001pv}, where the fermionic terms in the action and the supertransformations that leave the action \eq{A1} invariant are also given. 

\subsubsection*{Superspace}

Type IIA supergravities provided consistent target spaces for super $D$-branes, as well as Type IIA string. The requirement of local $\kappa$-symmetry of their worldvolume description imposes constraints on target superspace torsion and appropriate superforms. Some of these constraints are conventional and may differ as they amount to field redefinitions. Up to such field redefinitions, superspace constraints that describe Type IIA supergravity are given in ~\cite{Carr:1986tk,Cederwall:1996ri,Bergshoeff:1997cf,Howe:2004ib}\footnote{The relation between those of~\cite{Bergshoeff:1997cf} to others remains to be shown.}.

\subsection{ Type IIB supergravity }

 The $N=(2,0)$ supergravity in $10D$ is usually referred to as type IIB supergravity as it arises in the low energy limit of type IIB string theory. The multiplet is $(e_\mu{}^a, B_{\mu\nu}, \phi, C, C_{\mu\nu}, C_{\mu\nu\rho\sigma}, \psi_\mu, \chi)$ where the $C$-fields are real $0,2,4$ form fields and  the fermions are Weyl. On-shell the field strength of $C_4$ is self-dual, and this is an obstacle for writing a (standard) manifestly Lorentz invariant action.  On the other hand, manifestly Lorentz covariant field equations of motion have have been constructed \cite{Schwarz:1983qr,Howe:1983sra}. A pseudo-action can also be constructed in the sense that the correct equations of motion are obtained by using  self-duality equation by hand {\it after}  the variation of the action. The bosonic part of such an action, in the string frame is given by \cite{Bergshoeff:1995as,Bergshoeff:1995sq}
\bea
S &=& \frac12 \int d^{10} x\, \sqrt{-g} \Bigg\{ e^{-2\phi} \Big[-R+4\partial_\mu \phi \partial^\mu \phi -\frac34 H_{\mu\nu\rho} H^{\mu\nu\rho}\Big]
\nn\w2
&& -\frac12 \partial_\mu C \partial^\mu C -\frac34\left(F_{\mu\nu\rho}-CH_{\mu\nu\rho}\right)\left(F^{\mu\nu\rho}-CH^{\mu\nu\rho}\right)
\nn\w2
&& -\frac56 F_{\mu_1...\mu_5} F^{\mu_1...\mu_5} -\frac{1}{48} \varepsilon^{\mu_1...\mu_{10}} H_{\mu_1...\mu_3} F_{\mu_4...\mu_6} C_{\mu_7...\mu_{10}}\Bigg\}\ ,
\eea
where 
\be
H=dB\ ,\qquad F_3=dC_2\ ,\qquad F_5 = dA_4 +\frac34 \left( B\wedge F_3 -C_2 \wedge H \right)\ .
\ee
The fermionic terms to be added to this pseudo-action and the attendant supersymmetry transformations remain to be spelled out. See, however, \cite{Tanii:2014gaa,Becker:2006dvp} for more information. The $SL(2,\mR)$ symmetry of the action above is not manifest but it is in a convenient form for passing to the democratic formulation. Indeed, a democratic formulation of this pseudo-action in which duals of RR fields are introduced is also available and it takes  same form as in eqs. \eq{A1} and \eq{A2}, now with $n=1/2,...,9/2$, and $C=C_1+C_2+...+C_8$ and $G=G_1+...+ G_9$. In this case too the fermion terms and supertransformations can be found in \cite{Bergshoeff:2001pv}. 

It is often very useful to express the type IIB supergravity equations in manifestly in $SL(2,\mR)$ invariant form. The bosonic part of the type IIB (pseudo)action in Einstein frame is given by \cite{Polchinski:1998rr}
\bea
S_{IIB}^{(0)} &=& \frac{1}{2\kappa_{10}^2} \int \left( R-2 P_\m {\bar P}^\m- \frac{|G_3|^2}{2\cdot 3!} -\frac{|F_5|^2}{4\cdot 5!}\right) \star \mathbbm{1}
\nn\w2
&& +\frac{1}{8i\kappa_{10}^2} \int C_4\wedge G_3 \wedge {\bar G}_3\ ,
\eea
where
\begin{align}
\tau &=C_0+ie^{-\phi}\ ,\quad P_\m= \frac{i}{2} (\tau_2)^{-1} \nabla_\m\tau \ ,\quad G_3= (\tau_2)^{-1/2} \left(F_3-\tau H_3\right)\ ,
\nn\w2
H_3&=dB_2\ ,\quad  F_3=dC_2\ ,\quad F_5=dC_4-\frac12 H_3\wedge C_2 +\frac12 F_3\wedge B_2\ ,
\end{align}
and $\tau= \tau_1 + i\tau_2$.  The supertransformations of the fermionic fields are \cite{Green:1998by}
\bea
\delta \psi_\m &=& D_\m \e +\frac{1}{480} i\gamma^{\m_1...\m_5} \gamma_\m F_{\m_1...\m_5} +\frac{1}{96} \Big(\gamma_\m{}^{\n\r\s} -9\delta_\m^\n \gamma^{\r\s} \Big) G_{\n\r\s}\ ,
\nn\w2
\delta \chi &=& i\gamma^\m \e^\star P_\m\ .
\eea
As is well known, the gravitational anomalies cancel exactly in this theory \cite{Alvarez-Gaume:1983ihn}. As for the composite local $U(1)$ symmetry, it was noted in \cite{Marcus:1985yy} that it is anomalous. The anomaly vanishes upon the use of Maurer-Cartan equation though, and therefore there are no anomalies in the fundamental theory but the composite $U(1)$ vector field can not become dynamical \cite{Moore:1984dc}. 

\subsubsection*{Superspace and generalized type IIB supergravity}

While  the standard on-shell superspace constraints for Type IIB supergravity~\cite{Howe:1983sra}  are sufficient for the  type IIB superstring action to be $\kappa$-symmetric~\cite{Grisaru:1985fv}, it was shown in \cite{Wulff:2016tju} that they are not necessary. Instead, it was found that they yield a generalized version of type IIB supergravity equations of motion, bosonic part of which were found in \cite{Arutyunov:2015mqj}, if the target space admits a Killing vector. The  equations of motion involve  Killing vector $K$ and additional vector field $Z$ with $K^\mu Z_\mu=0$. For their description in exceptional field theory framework, see~\cite{Baguet:2016prz}. 

\subsection{ Type I, I$^\prime$ and heterotic supergravities coupled to Yang-Mills}

 The fields of $N=(1,0)$ supergravity are $(e_\mu{}^a, B_{\mu\nu}, \phi, \psi_\mu, \chi)$, and those of the Yang-Mills multiplet are $(A_\mu, \lambda)$, where $(\psi_\mu, \lambda)$ are left-handed and $\chi$ is right-handed Majorana-Weyl.  The coupling of the two multiplets was achieved long ago~\cite{Bergshoeff:1981um,Chapline:1982ww}, and the bosonic part of  Lagrangian in string frame, and setting $\kappa=1$, is given by
\begin{align}
\cL=&  ee^{2\phi} \Big[ 
\frac{1}{4} R(\omega) + g^{\m\n} \partial_\m \phi \partial_\n \phi
- \frac{1}{12} H_{\m\n\rh} H^{\m\n\rh} -\frac14 \beta\,F^I_{\m\n} F^{I \m\n} \Big]\ ,
\label{2f}
\end{align}
where $\beta=1/g_{YM}^2$, and letting $F= F^I T^I$ with $T^I$ in the adjoint representation, 
\begin{align}
H = dB -\beta  {\rm Tr}\left( A\wedge dA +\frac23 A\wedge A\wedge A\right)\ ,
\label{d2}
\end{align}
where ${\rm Tr} (T^I T^J)=-\delta^{IJ}$.  The supertransformations of the fermions are
\be
\delta\psi_\m
= D_\m(\omega_+) \epsilon \ , \qquad 
\delta \chi = \frac12 \gamma^\m \epsilon \partial_\m \phi -\frac{1}{12} H_{\m\n\rh} \gamma^{\m\n\rh} \epsilon\ ,
\ee
where $\omega_{+\m ab} = \omega_{\m ab}+ H_{\m ab}$.
In a celebrated paper~\cite{Green:1984sg} it was shown that the one loop gravitational, gauge and mixed anomalies cancel only for gauge groups $SO(32), E_8 \times E_8\ , E_8 \times U(1)^{248}$ and $U(1)^{496}$. Several years later, it was shown that the latter two theories do not have consistent coupling to quantum gravity, and as such they are in swampland~\cite{Adams:2010zy,Kim:2019vuc}. The case of $SO(32)$, together with the identification of $\beta$ with the string tension $\alpha'$, corresponds to the low energy effective action of Type I string. Performing the following field redefinitions in this action,  with the primed field denoting the  Type I fields (see, for example~\cite{Tseytlin:1995bi}) 

\be
g'_{\m\n}= e^{-\phi} g_{\m\n}\ ,\qquad \phi'=-\phi\ ,\qquad B_{\m\n}'=B_{\m\n}\ ,\qquad A_\m'=A_\m\ ,
\ee
gives the low energy effective action of heterotic string with $SO(32)$ gauge symmetry~\cite{Tseytlin:1995bi}. These field redefinitions signify  strong-weak coupling duality originally proposed in~\cite{Witten:1995ex}. Taking the gauge group to be $E_8\times E_8$, it is related to $11D$ supergravity on a particular limit of $11D$ supergravity reduced on $S^1/Z_2$. It has also been shown that the case of $E_8\times E_8$ in $10D$ is related to Type I$^\prime$ (also known as Type IA) case, in which a suitable boundary term needs to be added to the effective action of  the Type I case~\cite{Polchinski:1995df}. 

\subsubsection*{\bf Dualization }

$N=(1,0)$ supergravity plus Yang-Mills system in which a six-form potential, instead of the two-form potential is employed was constructed directly long ago in \cite{Chamseddine:1981ez}. The result may also be obtained by dualization. Note that both potentials describe the same number of degrees of freedom on-shell. Furthermore, the six-form potential couples to a five-brane, and as such it is reasonable to expect string-five-brane duality at work~\cite{Strominger:1990et,Duff:1990ya}. Focusing on the supersymmetry aspects, the dualization of the two-form formulation proceeds by adding the following total derivative Lagrange multiplier term to  Lagrangian \eq{2f}
\be
\Delta \cL = \frac{1}{2! \times 5!} \e^{\m_1...\m_{10}}  \big(\partial_{\m_1} C_{\m_2...\m_7} \big)\big(\partial_{\m_8} B_{\m_9\m_{10}}\big) = \frac16 e\, {\widetilde G}^{\m\n\rh}\Big( H_{\m\n\rh}+ \beta X_{\m\n\rh}(A) \Big)\ ,
\ee
where $C^{(6)}$ is the dual potential and $G^{(7)}=dC^{(6)}$. Treating $H$ as an independent variable and integrating it out in the Lagrangian $\cL_{\rm dual} = \cL + \Delta \cL$ yields  the theory in its dual formulation. Going to  the fivebrane frame by letting $g_{\m\n} \to g'_{\m\n}= e^{-2\phi/3} g_{\m\n}$ yields
\bea
\cL_{\rm dual} &=& e e^{-2\phi/3} \Big( \frac14 R -\frac{1}{2\times 7!} G_{\mu_1...\m_7} G^{\mu_1...\mu_7}\Big) -\frac14 \beta e {\rm Tr}\left(F_{\m\n} F^{\m\n}\right) 
\nn\w2
&& -\frac{4!\times 6!} \beta \epsilon^{\m_1...\m_{10}} C_{\m_1...\m_6} {\rm Tr} \big(F_{\m_7\m_8} F_{\m_9\m_{10}} \big)\ .
\eea
In these conventions, the fermionic terms and supertransformations can be easily obtained from the expressions provided in~\cite{Chang:2022urm}. Note  the absence of $(\partial\phi)^2$ term in the Lagrangian.  Diagonalizing  the coupled field equations, the correct number of degrees of freedom follows nonetheless.

\subsubsection*{\bf Superspace}

The superspace formulation of $N=(1,0)$ supergravity in $(10|16)$ dimensional target superspace is useful in tackling variety of problems more conveniently then in component formulation, such as the worldsheet description of type I and heterotic string propagating in curved background. The pure $N=(1,0)$ supergravity in superspace was formulated long ago in \cite{Nilsson:1981bn}, and with different set of conventional constraints in \cite{Witten:1985nt}. Coupling Yang-Mills fields as well, the superspace Bianchi identity for the super three-form field strength gets modified as $DH= \alpha' \tr F\wedge F$, and together with the super torsion constraints, it has been analyzed in \cite{Atick:1985de}, thereby providing a superspace formulation of heterotic supergravity in $10D$.

\section{D=9}

\subsection{Maximal $9D$ gauged supergravity}

Maximal supergravity in $9D$ has the field content
\be 
\{ e_\m{}^a, \varphi, \tau, A_\m^I, B_{\m\n}^i, C_{\m\n\r};\, \psi_\m, \lambda, {\tilde\lambda}  \}\ ,
\label{9e}
\ee
where $\tau \equiv \chi+ ie^{-\varphi}$, and $I=(0,r)$, with $r,i=1,2$, the scalars $(\chi, \varphi)$ parametrize the coset $SL(2,\mR)/SO(2)$, the fermions are Dirac, and $\lambda$ and $\tilde\lambda$ independent of each other. There is also a global scaling symmetry, $\mR$, parametrized by the  scalar $\phi$, and the on-shell global trombone symmetry $\mR^+$. The $SO(2), SO(1,1)^+$ and $\mR$ subgroups of $SL(2,\mR)$, $\mR^+$, and a two dimensional non-abelian Lie group $A(1)$, were gauged, one at a time, in \cite{Bergshoeff:2002nv} by means of  Scherk-Schwarz reductions of $11D, IIA$ and $IIB$ supergravities. These can be viewed as one-parameter deformations of the unique ungauged $N=2$ supergravity. Considering a combination of these deformations, five two-parameter deformations were found in \cite{Bergshoeff:2002nv}. 

A direct analysis of gauging was later carried out in \cite{Fernandez-Melgarejo:2011nso} where the embedding tensor formalism was employed, confirming these results. Denoting the  generators of the algebra $SL(2,\mR)\times \mR\times \mR^+$ by $T^A$ and using the notation $A_\m^I = (A_\m, A_\m^r)$, one can introduce  the embedding tensor $\theta_I{}^A$, so that the gauge group generator can be written as

\be
X_I = \theta_I{}^A T_A\ ,\quad I=1,2,3\ ,\quad A=1,...,5\ .
\ee
Note that the magnetic fields (the duals of the electric ones listed in \eq{9e}) are not present in this gauging, though their representation content has been discussed in \cite{Fernandez-Melgarejo:2011nso}. The closure of the supersymmetry algebra puts linear constraints on $\theta_I{}^A$ which are solved such that its independent components are \cite{Fernandez-Melgarejo:2011nso}
\be
\theta_0{}^m\ , \quad \theta_0{}^5\ ,\quad \theta_r{}^4\ ,\quad \theta_r{}^5\ , \qquad  m=1,2,3\ ,\ r=1,2\ ,
\ee
forming a triplet, two doublets and a singlets of $SL(2,\mR)$. Gauge invariance furthermore imposes the quadratic constraints\cite{Fernandez-Melgarejo:2011nso}
\bea
&& \theta_0{}^m \left(12\theta_r{}^4 + \theta_r{}^5 \right) =0\ ,\quad \theta_r{}^4 \theta_0{}^5=0\ ,\quad \theta_r{}^5 \theta_0{}^5=0\ ,
\nn\w2
&& \theta_s{}^4 \theta_0{}^m T_{m r}{}^s =0\ ,\quad \epsilon^{rs} \theta_r{}^4  \theta_s{}^5=0\ ,
\eea
where $T_{mr}{}^s$ are  $2\times 2$ the representation matrices of the gauge generators $T_A$ for  $A=m$. The undeformed bosonic Lagrangian in these conventions, the covariant field strengths and the supertransformations, including the shift functions that arise in the supertransformations of the fermions, are given in \cite{Fernandez-Melgarejo:2011nso}. See also \cite{Bergshoeff:2002nv}, and both of these papers for the earlier references that cover specific gaugings of maximal $9D$ supergravity. The action\footnote{In the case of gauged trombone symmetry, the theory can be formulated only at the level of equations of motion, since the undeformed Lagrangian scales under this global symmetry. For a detailed treatment of gauging the trombone symmetries, see \cite{LeDiffon:2008sh}.} remains to be spelled out for the most general gauged maximal supergravity in $9D$. For a subset of these theories, the superpotential from which the potential can be deduced has been given in \cite{Bergshoeff:2002nv} where solutions including 1/2 BPS domain-wall, and maximally symmetric one with constant scalars, namely (A)dS and Minkowski spacetimes, are given. 

\subsection{ Half-maximal $9D$ supergravity coupled to vector multiplets}

This theory was constructed by Noether procedure in \cite{Gates:1984kr}, the result of which we summarize below. Combining $n$-copies of the Maxwell multiplet consisting of the fields $(A_\mu, \phi, \lambda)$ with the supergavity multiplet gives the field content
\be
\{ e_\m{}^a, B_{\m\n}, \phi, \cV_M{}^A, A_\m^M ;\,\psi_\m, \chi, \lambda^a \}\ ,
\ee
where $M, A=0,1,..,n,\ a=1,..,n$, $\varphi$ is the dilaton, $\cV_M^A= \left(\cV_M{}^0, \cV_M{}^a \right)$ is the $SO(n,1)/SO(n)$ coset representative, and the fermions are pseudo-Majorana. The vector $A_\m^0$ belongs to supergravity multiplet. The $SO(n,1)$ invariant tensor $\eta={\rm diag} (-1,+1,...,+1)$ and the positive definite scalar matrix $\cM$ are given by
\be
\eta_{MN} =- \cV_M{}^0 \cV_N{}^0 + \cV_M{}^a \cV_N{}^a\ ,\qquad \cM_{MN} = \cV_M{}^0 \cV_N{}^0 + \cV_M{}^a \cV_N{}^a\ ,
\ee
and the field strengths are defined as 
\be
P_\m^a = \cV_0{}^I \partial_\m \cV_I{}^a\ ,\qquad F_{\m\n}^M=2\partial_{[\m} A_{\n]}^M\ ,\qquad H_{\m\n\r}= 3\partial_{[\m} B_{\n\r]} +3\eta_{MN} F_{[\m\n}^M A_{\r]}^N\ . 
\ee
The bosonic part of the ungauged Lagrangian is given by \cite{Gates:1984kr}
\bea
e^{-1} \cL &=& -\frac14 R +\frac74 \partial_\m\varphi \partial^\m \varphi +\frac{1}{12} e^{-4\varphi} H_{\m\n\r} H^{\m\n\r} -\frac14 e^{-2\varphi}\cM_{MN} F_{\m\n}^M F^{\m\n N} 
\nn\w2
&& + \frac12 P_\m^a P^\m_a\ ,
\eea
and the supertransformations of the fermions are
\bea
\delta \psi_\m &=& D_\m\e-\frac{1}{14\sqrt2} e^{-\phi} \cV_M{}^0 F_{\r\s}^M \left(\gamma_\m{}^{\r\s}-12\delta_\m^\r\gamma^\s\right) \e +\frac{1}{42} e^{-2\phi} H_{\r\s\tau}\left(\gamma_\m{}^{\r\s\tau} -\frac{15}{2} \delta_\m^\r\gamma^{\s\tau} \right)\e\ ,
\nn\w2
\delta\chi &=& -\frac{\sqrt7}{2} \gamma^\m \e \partial_\m \phi +\frac{1}{6\sqrt7} e^{-2\phi} H_{\m\n\r} \gamma^{\m\n\r} \e -\frac{1}{2\sqrt{14}} e^{-\phi} \cV_M{}^0 F_{\m\n}^M \gamma^{\m\n} \e\ ,
\nn\w2
\delta\lambda^a &=& -\frac{1}{\sqrt2} \gamma^\m P_\m^a \e -\frac{1}{2\sqrt2} e^{-\phi} \cV_M{}^a F_{\m\n}^M \gamma^{\m\n}\e\ .
\eea
The theory also has the (off-shell) global $SO(1,1)$ scaling symmetry under which $\phi\to \phi+\alpha, A_\m^M \to e^{\a} A_\m^M$ and $B_{\m\n} \to e^{2\alpha} B_{\m\n}$, not to be confused with the (on-shell) trombone symmetry mentioned in footnote 2. The gauging of a subgroup of $SO(1,1)\times SO(n,1)$ by employing the $(n+1)$ gauge field present in the theory has not been carried out so far, but it is expected to be very similar to the gauging of a subgroup of $SO(1,1) \times SO(n,5)$ in $N=4, 5D$ supergravity \cite{Schon:2006kz} summarized in section 8.3.

\section{D=8}

\subsection{Maximal $8D$ gauged supergravity}

This theory was constructed in \cite{Salam:1984ft} by Scherk-Schwarz reduction of $11D$ supergravity on $SU(2)$ group manifold, and it gives a gauged maximal supergravity in $9D$ with scalars parametrizing the coset $SL(3,\mR)/SO(3)\,\times\, SL(2,\mR)/SO(2)$, and local gauge symmetry $SU(2) \rtimes \mathbb R^3$. The Lagrangian, the supertransformations, the definition of field strengths and the gauge transformations are given in \cite{Salam:1984ft}. In this result the gauge group $SO(3)$ is the maximal compact subgroup of $SL(3,\mR)$ which employs three of the six vector fields present in the  maximal supergravity multiplet. The six gauge fields can be assembled into $(3,2)$ representation of $SL(3,\mR)\times SL(2,\mR)$, and the general gaugings can thus be studied systematically by means of the embedding tensor formalism, and this was done in \cite{Puigdomenech:2008kia,LassoAndino:2016vmh}, as we summarize below.  The supergravity multiplet is supplemented by dual three-form $S^\a$ and four-form $W^M$ required by the tensor hierarchy associated with the embedding tensor formalism, without upsetting the total count of physical degrees of freedom such that the total set of fields are
\be
\{ e_\m{}^a,\ \phi,\  B, \ \cV_M{}^i, \ A_\m^{\a M},\ B_{\m\n\,M},\ C_{\m\n\r}{}^\a,\ S_{\m\n\r}{}^\a,\ W_{\m\n\r\s}{}^M ;\, \psi_\m,\ \chi_i^A \}\ , 
\label{fcm9}
\ee
where the $3\times 3$  matrix $\cV_M{}^i$ is the representative of the $SL(3,R)/SO(3)$ coset, the scalars $(B,\phi)$ parametrize the coset $SL(2,\mR)/SO(2)$, the index  $A=1,2$, and the fermions are pseudo-Majorana. The index $\alpha=1,2$ labels the $SL(2,\mR)$ doublet. The fields $S^\alpha$ and $W^M$ are dually related to the ordinary supergravity multiplet fields $C_\a$ and $B_M$, respectively. Duality equations are to be imposed by hand, and they ensure that the physical degrees of freedom remain as $128_B+128_F$. The most general gauging is governed by the embedding tensors $\theta_{\alpha M,K}{}^L$ and $\theta_{\alpha M,\beta}{}^\gamma$. Thus the gauge group generators $G\subset SL(3,\mR)\times SL(2,\mR)$ can be written as
\be
X_{\alpha M} = \theta_{\alpha M, K}{}^L\,t_L{}^K + \theta_{\alpha M, \beta}{}^\gamma  t_\gamma{}^\beta\ .
\ee
As a consequence of the linear and quadratic constraints on the embedding tensors, it has been shown that they can be parametrized by $\xi_{\alpha M}$ and $f_\alpha{}^{MN}= f_\alpha{}^{(MN)}$ as follows \cite{Puigdomenech:2008kia} 
\bea
\theta_{\alpha M},{}^\beta{}_\gamma &=& \xi_{\gamma M}\delta_\alpha^\beta -\frac12 \delta^\beta_\gamma \xi_{\alpha M}\ ,
\nn\w2
\theta_{\alpha M},{}^ K{}_L &=& f_\alpha{}^{KR} \epsilon_{RML} -\frac34 \left( \xi_{\alpha L}\delta_M^K -\frac13 \xi_{\alpha M} \delta^K_L \right)\ ,
\eea
with the conditions that 
\be
\xi_{\alpha[M} \xi_{\beta|N]}=0\ ,\quad f_\alpha{}^{MN} \xi_{\beta N}=0\ ,\quad \epsilon^{\alpha\beta} \left( f_\alpha{}^{MK} f_\beta{}^{NL} \epsilon_{PKL} -f_\alpha{}^{MN}  \xi_{\beta P}\right)  =0\ .
\ee
For a discussion of the solutions to the constrains which determine the possible gauge groups, see \cite{Puigdomenech:2008kia}. They include the 1-dimensional subgroups of $SL(2,\mR)$ (rescalings, shifts and the Borel subgroup), and the Borel subgroup of $SL(3,\mR)$. The field strengths for the fields $(A,B,C)$, see \eq{fcm9}, are given in \cite{Puigdomenech:2008kia}, but not the action and supersymmetry transformations. In \cite{LassoAndino:2016lwl}, where the embedding tensor is also employed, the  focus is on the $SO(3)$ gauging, and it is shown that this gauging, which makes use of $A_\m^{1M}$ obtained by the Scherk-Schwarz reduction of $11D$ supergravity \cite{Salam:1984ft}, is related by an $SL(2,\mR)$ transformation to the gauging obtained by $A_\m^{2M}$, obtained from the  dimensional reduction of massive 11D supergravity \cite{Meessen:1998qm}. 

\subsection{Half-maximal $8D$ gauged supergravity coupled to vector multiplets}

This theory was constructed by Noether procedure in \cite{Salam:1985ns} where the $(n+2)$ parameter subgroup of the global $SO(n,2)$ symmetry was gauged. We summarize these results next. Combining $n$-copies of the vector multiplet consisting of the fields $(A_\mu, 2\phi, \lambda) $ with the supergavity multiplet gives the field content
\be
\{ e_\m{}^r, B_{\m\n}, \varphi, \cV_M{}^A , A_\m^M  ;\,\psi_\m, \chi, \lambda^a \}\ ,
\ee
where $M,A=1,...,n+2,\ a=1,..,n$, the scalar $\varphi$ is the dilaton, $\cV_M{}^A$ is the representative of the $SO(n,2)/SO(n)\times SO(2)$ coset, and the fermions are pseudo-Majorana. 
Writing $A_\m^M = \left(A_\m^m, A_\m^\a\right)$ with $m=1,2$ and $\a=1,...,n$, the vectors $A_\m^m$ belong to the supergravity multiplet. The $SO(n,2)$ invariant tensor $\eta={\rm diag} (-1,-1,+1,...,+1)$ and the positive definite scalar matrix $\cM$ are given by
\be
\eta_{MN} =-\cV_M{}^i\cV_N^i +\cV_M{}^a \cV_N{}^a\ ,\qquad \cM_{MN}= \cV_M{}^i\cV_N^i +\cV_M{}^a \cV_N{}^a\ ,
\ee
where $i=1,2$. Using all the vector fields to gauge a $(n+2)$ parameter semisimple subgroup $G\subset SO(n,2)$, we introduce the field strengths
\bea
\cF_{\m\n}^I &=& 2\partial_{[\m} A_{\n]}^I +f_{KL}{}^I A_\m^K A_\n^L\ ,
\nn\w2
\cH_{\m\n\r} &=& 3 \del_{[\m} B_{\n\r]}+ 3\cF_{[\m\n}^I A_{\r]}^J \eta_{IJ} - f_{IJ}{}^L \eta_{LK} A_{[\m}^I A_\n^J A_{\r]}^K\ .
\eea
with $f_{IJ}{}^K$ representing  structure constants of the gauge group $G$. The gauged scalar current is given by
\be
\cV^M{}_A \left( \partial_\m \delta_M^N +f_{MP}{}^N A_\m^P\right) \cV_N{}^B = \begin{pmatrix} \cQ_{\m a}{}^b & \cP_{a}{}^i\\ \cP_{\m i}{}^a & \cQ_{\m i}{}^j \end{pmatrix}\ .
\ee
Using the above ingredients, the Lagrangian is given by\footnote{The ungauged theory also has the (off-shell) global $SO(1,1)$ scaling symmetry under which $\phi \to \phi-\alpha, A_\m^M \to e^{\a} A_\m^M$ and $B_{\m\n} \to e^{2\alpha} B_{\m\n}$.}\cite{Salam:1985ns}
\bea
e^{-1} \cL &=& \frac14 R -\frac14 e^\phi \cM_{MN} \cF_{\m\n}^M \cF^{\m\n N}-\frac{1}{12} e^{2\phi} \cH_{\m\n\r} H^{\m\n\r} +\frac38 \del_\m \phi \del^\m\phi
\nn\w2
&& +\frac14 \cP_\m^{ai} \cP^\m_{ai} -\frac18 e^{-\phi} C^a C_a\ ,
\label{Lag8D}
\eea
where 
\be
C^a = f_{IJ}{}^K \cV^I{}_1 \cV^J{}_2 \cV_K{}^a\ ,
\ee
and the supertransformations of the fermions are
\bea
\delta\psi_\m &=& D_\m \e +\frac{i}{12\sqrt2} \cF_{\r\s}^I \left(\cV_M{}^1+i\gamma_9 \cV_M{}^2\right) \left( \gamma_\m{}^{\r\s}-10\delta_\m^\r \gamma^\s\right) \e 
\w2
&& -\frac{1}{36} e^\phi \cH_{\r\s\tau} \left( \gamma_\m{}^{\r\s\tau} -6\delta_\m^\r \gamma^{\s\tau} \right) \e\ ,
\nn\w2
\delta \chi &=& -\frac12 i \partial_\mu \phi \gamma^\mu \epsilon + \frac{1}{6\sqrt2} e^{\phi/2} \cF_{\r\s}^I \left(\cV_M{}^1+i\gamma_9 \cV_M{}^2\right) \gamma^{\mu\nu} \e +\frac{1}{18} i e^{\phi}\cH_{\mu\nu\rho} \gamma^{\mu\nu\rho}\e\ ,
\nn\w2
\delta\lambda^a &=& -\frac{i}{2} \gamma^\m  \left(\cP_\mu^{a1} +i\gamma^9 \cP_\mu^{a 2} \right) \e +\frac{1}{2\sqrt2} e^{\phi/2} \cF_{\m\n}^I L_I{}^a \gamma^{\m\n} \e +\frac{i}{2\sqrt2} e^{-\phi/2} C^a \gamma_9 \e\ ,
\nn
\eea
where $D_\m \e^i= D_\m(\omega,\cQ)\e^i$. The terms arising due to the gauging are remarkably similar to those we shall see in $6D, N=(1,0)$ supergravity. In particular a positive potential, which plays a key role in Minkowski compactification, arises in both cases. 

The most general gaugings of the half-maximal supergravity coupled to arbitrary number of vector multiplets remains to be worked out. The embedding tensor formalism may be employed directly in $7D$ to this end. In a search for higher dimensional origin, on the other hand,  it is natural to consider the framework of the so-called double field theory; see section 12. For example, the Lagrangian \eq{Lag8D} agrees with the result in \cite[eq. (3.43)]{Baron:2017dvb} obtained in a particular reduction of the (gauged) double field theory. 

\section{D=7}

\subsection{ Maximal $7D$ gauged supergravity}

The ungauged maximal supergravity in $7D$ was constructed by Noether procedure in \cite{Sezgin:1982gi}, for the supergravity multiplet
\be
\big( e_\m{}^m, \cV_M{}^{ab}, A_\m^{MN}, B_{\m\n\,M} ; \psi_\m^a, \chi^{abc},  \big)\ ,
\label{7Dmax}
\ee
where $\cV_M{}^{ab}=\cV_M{}^{[ab]}$ is  representative of  coset $SL(5)/SO(5)$, with $M=1,...,5$ and $a=1,...,4$ labeling  4-plet of  $USp(4) \approx SO(5)$. The gauge fields $A_\m^{MN}$ are in the $10$-plet of $SL(5)$. The spinors are symplectic-Majorana, and $\chi^{abc}=\chi^{[ab]c}, \chi^{[abc]}=0, \Omega_{ab}\chi^{abc}=0$ is in the $16$-plet of $USp(4)$. 

It was noted in \cite{Sezgin:1982gi} that the 10 gauge fields could not be used to gauge $SO(5) \subset SL(5)$ because the two-form potentials carry a non-trivial representation of $SL(5)$. Later, using instead the multiplet
\be
\big( e_\m{}^m, \cV_M{}^{ab}, A_\m^{MN}, C_{\m\n\r}{}^M ; \psi_\m^a, \chi^{abc} \big)\ ,
\ee
with the three-form potential satisfying a ``self-duality'' condition of the form $dC^M = m \star C^M$, where $m$ is the gauge coupling constant, the $SO(5)$ gauging was achieved by Noether procedure \cite{Pernici:1984xx}. This construction was motivated by the compactification of $11D$ supergravity on the sphere $S^4$ \cite{Pilch:1984xy}. Subsequently, $SO(4,1)$ and $SO(3,2)$ gaugings were obtained \cite{Pernici:1984zw}. A unified treatment which yields more general gaugings was achieved in \cite{Samtleben:2005bp}, using the embedding tensor formalism for the following multiplet of fields
\be
\big( e_\m{}^m, \cV_M{}^{ab}, A_\m^{MN}, B_{\m\n\,M}, C_{\m\n\r}{}^M ; \psi_\m^a, \chi^{abc}  \big)\ .
\ee
Both the two-form $B_{\mu\nu M}$ and the 3-form $C_{\mu\nu\rho}{}^M$ are now present but, as we shall see below, the $C$-field equation  will relate their field strengths to each other, and consequently the on-shell degrees of freedom remain as $128_B +128_F$. In the rest of this subsection, we follow \cite{Samtleben:2005bp} to summarize their key results. The most general gauging is encoded in a real embedding tensor $\theta_{MN,P}{}^Q= \theta_{[MN],P}{}^Q$, which determines the  gauge group $G \subset SL(5)$ with generators
\be
X_{MN} = \theta_{MN,P}{}^Q t^P{}_Q\ ,
\ee
where $t^P{}_Q$ are the $SL(5)$ generators. Thus, the covariant derivatives are given by $D_\m=\nabla_\m -g A_\m^{MN} X_{MN}$. Supersymmetry requirement imposes a linear constraint on the embedding tensor such that in  the product $\bf{10}\otimes {\bf 24}$, only the representations ${\bf 15}+\bf{\overline 40}$ survive. Therefore, it can be parametrized as
\be
\theta_{MN,P}{}^Q =\delta^Q_{[M} Y_{N]P} -2\e_{MNPRS} Z^{RS,Q}\ ,
\label{et7}
\ee
where $Y_{MN}=Y_{(MN)}$, and $Z^{MN,P}= Z^{[MN],P}$ with $Z^{[MN,P]}=0$. In addition,  a quadratic constraint needs to be imposed on the embedding tensor to ensure  the closure of the gauge algebra. It has been shown that the full content of this quadratic constraint is encoded in the equation \cite{Samtleben:2005bp}
\be
Z^{MN,P} X_{MN}=0\ .
\ee
The building blocks for the action are the covariantly transforming field strengths
\bea
\cF_{\m\n}{}^{MN} &=& 2\partial_{[\m} A_{\n]}^{MN} +g (X_{PQ})_{RS}{}^{MN} A_{[\m}^{PQ} A_{\n]}^{RS} +g Z^{MN,P} B_{\m\n P}\ ,
\nn\w2
\cH_{\m\n\r\, M} &=& 3D_{[\m} B_{\n\rh]M} + 6\e_{MNPQR} A_{[\m}^{NP} \Big( \partial_\n A_{\rh]}^{QR} + \frac23 g X_{ST,U}{}^Q A_\n^{RU} A_{\r]}^{ST} \Big) 
\nn\w2
&& +g Y_{MN} S^N_{\m\n\r}\ ,
\nn\w2
Y_{MN}\, \cG_{\m\n\r\sigma}{}^N &=& Y_{MN} \Big( 4D_{[\m} C_{\n\r\sigma]}{}^N +6 \cF_{[\m\n}^{NP} B_{\r\sigma]\,P} +3 g Z^{NP,Q} B_{[\m\n\,P} B_{\r\s]Q}
\w2
&& +8 \e_{PQRST} A_{[\m}^{NP}A_\n^{QR} \partial_\r A_{\s]}^{ST} + 4g \e_{PQRVW} X_{ST,U}{}^V A_{[\m}^{NP} A_\n^{QR} A_\r^{ST} A_{\s]}^{UW} \Big)\ .
\nn
\eea
where $(X_{PQ})_{RS}{}^{MN}= 2(X_{PQ})_{[R}{}^{[M} \delta_{S]}^{N]}$ and $(X_{MN})_P{}^Q = \theta_{MN,P}{}^Q$. Furthermore, the  4-form field strength arises in the action and transformation rules always under the projection with $Y_{MN}$.
Denoting the gauge transformations associated with the $(A,B,C)$ fields by  $(\Lambda,\Sigma, \Phi)$,the gauge algebra takes the form $[\delta_{\Lambda_1},\delta_{\Lambda_2}]=\delta_\Lambda + \delta_\Xi + \delta_\Phi$ and $[\delta_{\Xi_1},\delta_{\Xi_2}]=\delta_\Phi$, as shown in \cite{Samtleben:2005bp}. 

Few remaining building blocks are constructed as follows. Firstly, the $SL(5)$ Lie algebra valued gauged Maurer-Cartan form decomposes as
\be
\cV_{ab}{}^M \Big( \partial_\m \cV_M{}^{cd} -g A_\m{}^{PQ} X_{PQ,M}{}^N \cV_N{}^{cd} \Big)  = P_{\m ab}{}^{cd} + 2Q_{\m[a}{}^{[c} \delta_{b]}^{d]}\ .
\ee
As usual, $P_{\m ab}{}^{cd}$ facilitates  construction of a kinetic term for the scalars, and $Q_{\m a}{}^b$ is the composite $SO(5)$ connection which is needed in the covariant derivatives of  the fermions.

Another building block for  gauging is the $T$-tensor, which is the embedding tensor suitably dressed up with  coset vielbeins. More specifically, using  the decomposition of its $SL(5)$ content under $USp(4)$ as ${\bf 15} +\overline{\bf 40} \to ( {\bf 1}+ {\bf 14} ) + ({\bf 5}+ {\bf 35} )$, one defines  the corresponding $T$-tensors denoted by $ B, B^{[ab]}{}_{[cd]}, C_{[ab]}$ and $C^{[ab]}{}_{(cd)}$ as
\be 
Y_{MN} = \cV_M{}^{ab} \cV_N{}^{cd} Y_{ab,cd}\ ,\qquad Z^{MN,P}= {\sqrt 2} \cV_{ab}{}^M \cV_{cd}{}^N \cV_{ef}{}^P \Omega^{bd} Z^{(ac)[ef]}\ ,
\ee
where the T-tensor fields are decomposed as 
\bea
Y_{ab,cd} &=& \frac{1}{\sqrt 2} \Big[ (\Omega_{ac}\Omega_{bd} -\frac14 \Omega_{ab}\Omega_{cd} ) B +\Omega_{ae}\Omega_{bf} B^{[ef]}{}_{[cd]}\ ,
\nn\w2
Z^{(ab)[cd]} &=& \frac{1}{16} \Omega^{a[c} C^{d]b} +\frac{1}{16} \Omega^{b[c}C^{d]a} -\frac18 \Omega^{ae}\Omega^{bf} C^{cd}{}_{ef}\ .
\eea
The bosonic part of  gauged supergravity Lagrangian is given by
\bea
e^{-1}\cL &=& -\frac12 R -\cF_{\m\n}^{ab} \cF^{\m\n}_{ab} -\frac16 \cH_{\m\n\r ab} \cH^{\m\n\r}{}_{ab} -\frac12 P_{\m ab}{}^{cd} P^\m{}_{cd}{}^{ab} + e^{-1} \cL_{\rm top}
\nn\w2
&& +\frac{1}{128} g^2 \left( 15 B^2 + 2C^{ab} C_{ab} -2B^{ab}{}_{cd} B^{cd}{}_{ab} -2 C^{[ab]}{}_{(cd)} C_{[ab]}{}^{(cd)} \right) \ ,
\eea
where $\cH_{\mu\nu\rho ab} = \cV^M{}_{ab} \cH_{\mu\nu\rho M}$, and $\cL_{\rm top}$ has a complicated form provided explicitly in~\cite{Samtleben:2005bp}, with the field $C_{\m\n\r}{}^M$ occurring only under the projection with $Y_{MN}$. Even though $\cL_{\rm top}$ is very complicated, its general variation is simple, and it is given in \cite[eq. 3.16]{Samtleben:2005bp}, where the fermionic terms and the supertransformations can be found as well. In particular the supertransformations of the fermions are given by
\bea
\delta \psi_\m^a &=& D_\m \e^a +\frac{1}{5{\sqrt 2}} \cF_{\m\n}^{ab} \left(\gamma^{\n\r}{}_\m +8 \gamma^\n\delta_\m^\r\right) \e_b
\nn\w2
&& +\frac{1}{15} \cH_{\n\r\lambda}^{ab}\left(\gamma^{\n\r\lambda}{}_\m +\frac92\gamma^{\n\r}\delta^\lambda_\m\right)\e_b + gA_1^{ab} \e_b\ ,
\nn\w2
\delta\chi^{abc} &=& -2 P_\m{}^{cd ab}\gamma^\m \e_d- {\sqrt 2} \gamma^{\m\n} \Big( \cF_{\m\n}^{c[a} \e^{b]} +\frac25 \Omega^{a(b}\cF_{\m\n}^{c)d} \e_d \Big)
\w2
&&-\frac16 \gamma^{\m\n\r} \Big( \cH_{\m\n\r}^{[ab]} \e^c +\frac15 \Omega^{ab} \cH_{\m\n\r}^{cd} \e_d +\frac45 \Omega^{c[a} \cH_{\m\n\r}^{b]d} \Big) \e_d -g A_2^{d,abc} \e_d\ ,
\nn
\eea
where
\bea
A_1^{ab} &=& -\frac{1}{4\sqrt2} \left(\frac14  B\Omega^{ab} +\frac15 C^{ab}\right)\ ,
\w2
A_2^{d,abc} &=& \frac{1}{2\sqrt2} \left[ C^{[ab](cd)}-B^{[ab][cd]}+\frac14 \left( C^{ab}\Omega^{cd} +\frac15 \Omega^{ab} C^{cd} +\frac45 \Omega^{c[a} C^{b]d}\right) \right]\ .
\nn
\eea
The degrees of freedom are correctly $128_B+128_F$, thanks to the (projected) duality equation. With fermion terms suppressed, this equation is given by \cite{Samtleben:2005bp} 
\be
Y_{MN} \Big( \cV^N_{ab} \cH^{\m\n\r ab}-\frac{1}{3!}
\ve^{\m\n\r \s_1...\s_4}  \cG_{\s_1...\s_4}{}^N \Big) = 0\ ,
\ee
which arises as a field equation of $C_{\m\n\r}{}^M$. We refer the reader to \cite{Samtleben:2005bp} for more details and several examples of gaugings based on different choices of the embedding tensor.

\subsection{ Half-maximal $7D$ gauged supergravity coupled to vector multiplets}

The $7D$ maximal supergravity for the multiplet \eq{7Dmax} admits a consistent truncation to half-maximal $7D$ multiplet with the field content $\big( e_\m{}^r, \phi, B_{\m\n}, 3A_\m ; \psi_\m^i, \chi^i \big)$, with $i=1,2$ labelling an $USp(2)$ doublet, coupled to three vector multiplets $(A_\m, 3\phi;\lambda)$, and the scalars parametrizing the coset $GL(4,\mR)/SO(4)$. This model with its $SO(4)\subset GL(4,\mR)$ gauging was constructed by Noether procedure in \cite{Salam:1983fa}. A dual version of half-maximal supergravity containing a massless field $A_{\m\n\r}$ was constructed in \cite{Townsend:1983kk}, and a two parameter deformation was found, one having to do with $SU(2)$ gauging, and the other one with a topological mass term. 
The bosonic part of the Lagrangian is given by \cite{Townsend:1983kk}\footnote{Typographical errors present in \cite{Townsend:1983kk} are corrected in \cite{Mezincescu:1984ta}, and the $F\wedge X$ term is corrected here.}
\bea
\e^{-1} \cL &=& -\frac12 R -\frac{1}{48} \sigma^{-4} F_{\m\n\r\s} F^{\m\n\r\s} -\frac14 \s^2 \tr (F_{\m\n} F^{\m\n}) -\frac12 \partial_\m\phi \partial^\m \phi
\nn\w2
&& +\frac{i}{24\sqrt 2} \ve^{\m\n\r\s\kappa\lambda\tau} F_{\m\n\r\s}\tr \big( X_{\kappa\lambda\tau} -\frac{2\sqrt 2}{3} h A_{\kappa\lambda\tau} \big)
\nn\w2
&& +g^2\sigma^{-2}+8\sqrt2 g h \s^2 -16h^2 \s^8\ ,
\eea
where $\sigma = {\rm exp} (-\phi/\sqrt5), \  \tr (FF)= F_i{}^j F_j{}^i,\  F_{\mu\nu\rho\sigma} = 4\partial_{[\mu} A_{\nu\rho\sigma]}\ ,\ F_i{}^j= dA_i{}^j +ig (A\wedge A)_i{}^j$, and the Chern-Simons form 
\be
X_{\m\n\r} =\tr( A_{[\m}\partial_\n A_{\r]}  +\frac23 ig A_{[\m} A_\n A_{\r]}\big)\ ,
\ee
and $g,h$ are arbitrary constants. The supertransformations of the fermions are
\bea
\delta \psi_{\m i} &=&  D_\m \e_i +\frac{i\s}{10\sqrt2} \big(\gamma_\m{}^{\r\s}-8\delta_\m^\r\gamma^\s\big) F_{\r\s i}{}^j \e_j
\w2
&& +\frac{1}{80\sqrt2} \s^{-2}\big( \gamma_\m{}^{\n_1...\n_4} -\frac83 \delta_\m^{\n_1} \gamma^{\n_2...\n_4}\big) F_{\n_1...\n_4} -\frac15 \left( \frac{1}{\sqrt2} g\s^{-1}+2h\s^4 \right) \gamma_\m \e_i\ ,
\nn\w2
\delta \chi_i &=& \frac12 \partial_\m \phi \gamma^\m \e_i -\frac{i}{2\sqrt{10}} \s \gamma^{\m\n} F_{\m\n i}{}^j \e_j +\frac{1}{24\sqrt{10}}\s^{-2} \gamma^{\m_1...\m_4} F_{\m_1...\m_4} \e_i
\nn\w2
&& +\frac{1}{\sqrt5} \left( \frac{1}{\sqrt2} g\s^{-1}-8 h\s^4 \right) \gamma_\m \e_i\ ,
\eea
where $D_\m \e_i = \nabla_\m \e_i +ig A_{\m i}{}^j \e_j$. It was shown in \cite{Mezincescu:1984ta} that if $h/g <0$ the potential has no critical points, and that if $h/g>0$, then  it has two extrema, one preserving the full $OSp(8^\star|2)$ supersymmetry of the $AdS_7$ background, while the other breaks the supersymmetry completely \cite{Mezincescu:1984ta}.

The most general gauging of half-maximal supergravity coupled to $n$ vector multiplets has not been carried out so far in the embedding tensor formalism to our best knowledge, though the special case of $n=3$ has been investigated in that framework \cite{Dibitetto:2015bia}. Nonetheless, the coupling to $n$ vector multiplets and the gauging of $(n+3)$ parameter group $G\subset SO(n,3)$ were constructed by Noether procedure in \cite{Bergshoeff:1985mr}, as we summarize below.

Combining $n$-copies of  Maxwell multiplet consisting of fields $(A_\mu, 3\phi, \lambda) $ with  the supergavity multiplet gives the field content
\be
\{ e_\m{}^a, \phi, B_{\m\n}, \cV_M{}^A, A_\m^M ;\,\psi_\m, \chi^i, \lambda^{ai} \}\ ,
\ee
where $M,A=1,...,n+3,\ a=1,..,n$, the scalar $\phi$ is the dilaton, the scalars $\cV_M{}^A= \left(\cV_M{}^{ij}, \cV_M{}^a\right)$, with $i,j=1,2$ labelling $USp(2)$ doublet and $\cV_M{}^{ij}\Omega_{ij}=0$, parametrize the $SO(n,3)/SO(n)\times SO(3)$ coset, and the fermions are symplectic-Majorana. Writing $A_\m^M = \left(A_\m^m, A_\m^\a\right)$ with $m=1,2,3$ and $\a=1,...,n$, the vector fields $A_\m^m$ belong to the supergravity multiplet.
The $SO(n,3)$ invariant tensor $\eta={\rm diag} (-1,-1,-1,+1,...,+1)$ and the positive definite scalar matrix $\cM$ are given by
\be
\eta_{MN} = -\cV_I{}^i{}_j \cV_J{}^j{}_i + \cV_M{}^a \cV_N{}^a \ , \qquad \cM_{MN} = \cV_I{}^i{}_j \cV_J{}^j{}_i + \cV_M{}^a \cV_N{}^a\ .
\ee
The inverse of $\cV_M{}^A$ is $\cV^M{}_A$, satisfying  the relations $\cV_M{}^i{}_j \cV^{Mk}{}_\ell = -\delta^i_\ell \delta^k_j +\frac12 \delta^i_j \delta^k_\ell\ ,\cV_M{}^a \cV^M{}_b =\delta^a_b$ and $\cV_M{}^a \cV^M{}_{ij}=0$.
Using  the $(n+3)$ vector fields to gauge a $(n+3)$ parameter semisimple group $G \subset SO(n,3)$, we introduce  the field strengths
\bea
\cF_{\m\n}^M &=& 2\partial_{[\m} A_{\n]}^M +f_{NP}{}^M A_\m^N A_\n^P\ ,
\nn\w2
\cH_{\m\n\r} &=& 3 \del_{[\m} B_{\n\r]}-\frac{3}{\sqrt2} \eta_{MN} \cF_{[\m\n}^M A_{\r]}^N  - \frac{1}{\sqrt2}f_{MN}{}^P A_{[\m}^M A_\n^N A_{\r]P}\ ,
\eea
with $f_{MN}{}^P$ representing structure constants of $G$. We also need the coset currents and $SO(n)\times SO(3)$ composite connections
\bea
\cP_{\m a}{}^i{}_j &=& \cV^M_a \left( \partial_\m \delta_M^N +f_{MK}{}^N A_\m^K\right) \cV_N{}^i{}_j\ ,
\nn\w2
\cQ_\m{}^i{}_j  &=& \cV^{Mi}{}_k \left( \partial_\m \delta_M^N +f_{MK}{}^N A_\m^K\right) \cV_N{}^k{}_j\ ,
\nn\w2
\cQ_{\m ab} &=& \cV^M{}_a \left( \partial_\m \delta_M^N +f_{MK}{}^N A_\m^K\right) \cV_{Nb}\ .
\eea
Using these building blocks, the Lagrangian takes the form \cite{Salam:1983fa}
\bea
e^{-1} \cL &=& \frac12 R -\frac14 e^\phi \cM_{MN} \cF_{\m\n}^M \cF^{\m\n N} -\frac{1}{12} e^{2\varphi} \cH_{\m\n\r} \cH^{\m\n\r} +\frac58 \del_\m \phi \del^\m\phi 
\nn\w2
&& -\frac12 \cP_\m^{ai}{}_j  \cP^\m{}_{ai}{}^j  -\frac14 e^{-\phi} \left( C^{ai}{}_j C_{ai}{}^j -\frac19 C^2\right)\ ,
\label{hm7}
\eea
where 
\be
C = if_{MN}{}^P \cV^{Mi}{}_k \cV^{Nj}{}_i  \cV_{P}{}^k{}_j\ , \qquad C^{ai}{}_j = if_{MN}{}^P \cV^{Mi}{}_k \cV^{Nk}{}_j \cV_P{}^a\ ,
\ee
and the supertransformations of the fermions are
\bea
\delta\psi_\m^i &=& 2 D_\m \e^i -\frac{1}{60} e^\phi \left( \gamma_\m\gamma^{\r\s\tau} +5\gamma^{\r\s\t} \gamma_\m\right) \cH_{\r\s\tau} \e 
\nn\w2
&& +\frac{i}{10\sqrt2} e^{\phi/2} \left( 3\gamma_\m\gamma^{\r\s} - 5\gamma^{\r\s} \gamma_\m \right) \cF_{\r\s}^M \cV_M{}^i{}_j \e^j-\frac{\sqrt2}{30} e^{-\phi/2} C\gamma_\m \e^i\ ,
\nn\w2
\delta\chi^i &=& -\frac12 \gamma^\m \del_\m \phi + \frac{1}{5\sqrt2} e^{\phi/2} \gamma^{\m\n} \cF_{\m\n}^M \cV_M{}^i{}_j \e^j -\frac{1}{15\sqrt2} e^\phi \gamma^{\m\n\r} \cH_{\m\n\r} \e^i +\frac{\sqrt2}{30}e^{-\phi/2}  C\e^i\ ,
\nn\w2
\delta\lambda^{ai} &=&  -\frac12 e^{\phi/2} \gamma^{\m\n} \cF_{\m\n}^M \cV_M{}^a \e^i -{\sqrt2} i\gamma^\m \cP_\m{}^{ai}{}_j \e^j+ie^{-\phi/2} C^{ai}{}_j \e^j\ .
\eea
Turning to the nature of the gauge group $G$ with the structure constants $f_{MN}{}^P$, 
it has been observed that it has the form $ G_0\times H \subset SO(3,n)$ where $G_0$ is one of the following \cite{Bergshoeff:2005pq} 
\be
SO(2,1),\quad SO(3,1)\ ,\quad SO(2,2)\ ,
\quad SO(2,2)\times SO(2,1),\quad SL(3,R),\quad SO(3)\ ,
\label{gg7D}
\ee
and $H$ is $(n+3-{\rm dim}\,G_0)$ dimensional gauge group. The question of whether this provides the most general general gaugings, the possibility of further deformations, and the question of which one of these theories can be embedded in $10D$ or $11D$ supergravities are remain to be investigated. For a reduction of $N=1,10D$ heterotic supergravity on a group manifold, in particular, see \cite{Lu:2006ah}. The supersymmetric double field theory and exceptional field theory approaches, both briefly reviewed in section 12, may also shed light on the embedding questions. In particular, it is worth noting that the Lagrangian \eq{Lag8D} agrees with the result in \cite[eq. (3.43)]{Baron:2017dvb} obtained in a certain reduction scheme of the (gauged) double field theory.

\section{D=6}

\subsection{ $(2,2)$ gauged supergravity in $6D$}

The ungauged $(2,2)$ theory was constructed in \cite{Tanii:1984zk} and it has the duality symmetry $SO(5,5)$. Using the embedding tensor formalism, the general gaugings of a subgroup $G\subset SO(5,5)$ that utilizes the 16 vector fields of the supergravity multiplet was carried out in \cite{Bergshoeff:2007ef}, as we summarize next. The  maximal gauged supergravity is built out of the  following multiplet of fields 
\be
\big( e_\m{}^r, V_A{}^{a\ad}, A_\m^A, B_{\m\n\,M},  C_{\m\n\r\,A}\, ;\, \psi_{+\m\a}, \psi_{-\m\ad}, \chi_{+a\ad}, \chi_{{-\dot a}\a} \big)\ ,
\ee
where $M=1,...,10$ labels the fundamental and  $A=1,...,16$ labels the spinor representation of the duality group $SO(5,5)$. The indices $\a,\ad=1,...4$ label the spinor representations of $SO(5)\times SO(5)$, and $V_A^{\a\ad}$ is the $SO(5,5)/SO(5)\times SO(5)$ coset representative. The spinors are symplectic-Majorana-Weyl, $a,{\dot a}=1,...5$ are the vectors indices of $SO(5)\times SO(5)$ and $\pm$ refer to chirality under $\gamma_7$. The two-form potential $B_{\m\n M} = \big(B_{\m\n m}, B_{\m\n}{}^m \big)$ consists of the electric and magnetic two-forms $B_m$ and $B^m$ transforming as $5$ and $5'$ of $GL(5)$ respectively, and combining into ${\bf 10}$ of $SO(5,5)$. The three-form $C_{\m\n\r\,A}$ is introduced as on-shell dual of the vector fields $A_\m^A$. Their properly covariantized field strengths will be related to each other via a duality relation, which arises, under a projection, as an equation of motion. 

All of  the $16$ gauge fields, or a subset of them, may be used to gauge a suitable subgroup of $S0(5,5)$. This gauging is encoded in a real embedding tensor $\theta_A{}^{MN}= \theta_A{}^{[MN]}$, which determines the generators of the gauge group $G$ among the $SO(5,5)$ generators $t_{MN}$ and thus the covariant derivative
\be
X_A = \theta_A{}^{MN} t_{MN}\ ,\qquad   D_\m=\nabla_\m -g A_\m^A X_A\ .
\ee
Supersymmetry  imposes a linear constraint on  the embedding tensor such that in the product ${\bf 16}\otimes {\bf 45}$, only the representations ${\bf 144}_c$ survives, and therefore it can be parametrized as
\be
\theta_A{}^{MN} = -\theta^{B[M} \gamma^{N]}_{BA}\ ,
\ee
where $\gamma^M_{AB}$ are  chirally projected gamma-matrices of $SO(5,5)$. The closure of the gauge algebra imposes the quadratic constraints
\be
\theta^{AM}\theta^{BN} \eta_{MN}=0\ ,\qquad \theta^{AM} \theta^{B[N} \gamma^{P]}_{AB}=0\ ,
\label{QC}
\ee
where $\eta_{MN}$ is the invariant tensor of $SO(5,5)$. The covariant field strengths are given by (from here on we absorb $g$ into the definition of the embedding tensor)
\bea
\cF_{\mu\nu}^A &\equiv&
2\,\partial_{[\mu} A_{\nu]}{}^A
+ X_{[BC]}{}^A \,A_\mu{}^B A_\nu{}^C- {\sqrt 2}\, \theta^{AM} \,B_{\mu\nu\,M}
\;,
\nn\w2
\cH_{\mu\nu\rho\,M} &\equiv&
3\, D_{[\mu} B_{\nu\rho]\,M} +
3{\sqrt 2}\,(\gamma_{M})_{AB}\,A_{[\mu}{}^A\,\Big(\partial_{\nu} A_{\rho]}{}^B+
\ft13  X_{[CD]}{}^B A_{\nu}{}^C A_{\rho]}{}^D\Big)
\nn\\
&&{}
- {\sqrt 2} \theta^{AN}\,\eta_{MN}\, C_{\mu\nu\rho\,A} \;,
\nn\w2
\cG_{\mu\nu\rho\lambda\,A} &\equiv&  4D_{[\mu} C_{\nu\rho\lambda]\,A} -(\c^M)_{AB} \Big( 6{\sqrt 2}\, B_{[\mu\nu M} \cF_{\rho\lambda]}^B+6\theta^{BN} B_{[\mu\nu M} B_{\rho\lambda] N}
\label{covAH}\w2
&& +8(\c_M)_{CD} A_{[\mu}^B A_\nu^C \partial_\rho A_{\lambda]}^D +2 (\c_M)_{CF} X_{DE}{}^F A_{[\mu}{}^B A_\nu{}^C A_\rho{}^D A_{\lambda]}{}^E\Big)\ ,
\nn
\eea
where $X_{AB}{}^C = \left(\gamma^M\theta^N\right)_A \left(\gamma_{MN}\right)_B{}^C$. Next, we define the coset currents and the $SO(5)$ composite connection in terms of the $SO(5,5)$ valued $16\times 16$ matrix $V_\alpha^{\a\ad}$ as
\be
P_\m^{a\adot} =\frac14 {\bar V}\gamma^a\gamma^\adot \cD_\m V\ ,\qquad  Q_\m^{ab}= \frac18 {\bar V}\gamma^{ab} \cD_\m V\ ,
\ee
where $\cD_\m V= D_\m(Q) V - \left( {\bar A}_\m\gamma^M\theta^N\right) \gamma_{MN} V$.
To construct the $T$-tensors, we first introduce  $10\times 10$ scalar matrix $\cV$ that is an element of $SO(5,5)$. For convenience explained in detail in ~\cite{Bergshoeff:2007ef}, we choose $\cV$ such that it satisfies $\cV^T \eta\cV =\eta_{\rm diag}$ where $\eta= \begin{pmatrix} 0 & 1\\1&0 \end{pmatrix}$, and $\eta_{\rm diag} = \begin{pmatrix} 1 & 0\\0&-1 \end{pmatrix}$. Writing
\be
\cV_M{}^A =\begin{pmatrix} \cV_m{}^a & \cV_m{}^\ad\\
\cV^{ma} & \cV^{m\ad}\end{pmatrix}\ ,
\ee
we find the relations
\be
\cV_M{}^a = \frac{1}{16} {\bar V}\gamma_M\gamma^a V\ ,\qquad \cV_M{}^\ad = -\frac{1}{16} {\bar V}\gamma_M \gamma^\ad V\ .
\ee
The $T$-tensors are then defined as
\be
T^a = \cV_M{}^a \theta^{AM} V_A\ ,\qquad T^\ad = -\cV_M{}^\ad \theta^{AM} V_A\ .
\ee
Note that the $\a\ad$ indices of $V_A^{\a\ad}$ are suppressed. Finally, we  define the following hyper-matrix which is key for writing the appropriate kinetic term for the two-form potential:
\be
K^{mn}=\cV^{ma} \left(\cV_n{}^a\right)^{-1} P_+ -\cV^{m\ad} \left(\cV_n{}^{\ad}\right)^{-1} P_-\ ,
\ee
where $P_\pm = \frac12 \left(1\pm j\right)$ and $j$ acts on a given the three-form as
$ j\omega = {\widetilde \omega}$ with the definition ${\widetilde\omega}= \frac{1}{3!} \ve_{\m\n\r\s\lambda\tau} \omega^{\s\lambda\tau}$. The bosonic part of the Lagrangian takes the form \cite{Bergshoeff:2007ef}
\bea
e^{-1} \cL_B &=&  \ft14 R  -\ft1{12} \cH_m \cdot K^{mn} \cH_n
-\ft14 \cM_{AB} \cF_{\mu\nu}^A \cF^{\mu\nu B} -\ft1{16} \cP_\mu^{a\adot} \cP^\mu_{a\adot} 
\nn\w2
&&  + \left( \tr\,T^a{\tilde T}^a -\ft12\,\tr\,T{\tilde T} \right)  +e^{-1} \cL_{top} \ ,
\eea
where $``\sim "$ denotes transposition, $M_{AB}= V_A^{\a\ad} V_{B\a\bd}$ and $\cL_{\rm top}$ is the topological part of the Lagrangian given explicitly in \cite{Bergshoeff:2007ef}. It is a complicated expression but its general variation takes a simple form, which makes it straightforward to derive the following equations of motion for the three-form potential $C_A$ and the ``magnetic" two-form potentials $B_M$:
\be
\theta_m^A \left(\cH^m -j K^{mn} \cH_n\right) =0\ ,\qquad 
\theta_m^A \left( \cG_{\m\n\r\s A} +\frac12 \ve_{\m\n\r\s\lambda\tau} \cM_{AB} \cF^{\lambda\tau B} \right)=0\ .
\ee
These first equation furnishes a duality relation between the electric and magnetic two forms,combined as $\cH_M = (\cH_m, \cH^m)$, and the second equation between the three-form potentials and the vector fields. Finally, the supersymmetry transformations of the fermionic field are given by \cite{Bergshoeff:2007ef}
\bea
\delta\psi_{\mu +} &=& \cD_\mu \epsilon_+ -\ft1{24}\cH_{\rho\sigma\kappa}^a\c^a\c^{\rho\sigma\kappa}\c_\mu\epsilon_+ +\ft18 \left(\c_\mu{}^{\nu\rho}-6\d_\mu^\nu\c^\rho\right) \cF_{\nu\rho}^\cA V_A\, \epsilon_-
+ \ft14 \c_\mu T\epsilon_-\ ,
\nn\w2
\delta\psi_{\mu -}&=& \cD_\mu \epsilon_- -\ft1{24}\cH_{\rho\sigma\kappa}^\adot\c^\adot\c^{\rho\sigma\kappa}\c_\mu\epsilon_-
+\ft18 \left(\c_\mu{}^{\nu\rho}-6\d_\mu^\nu\c^\rho\right) \cF_{\nu\rho}^\cA {\tilde V}_A\, \epsilon_+
-\ft14 \c_\mu {\tilde T}\epsilon_+\ ,
\nn\w2
\delta\chi^\adot &=& \ft14 \cP_\mu^{a\adot} \c^a \c^\mu \epsilon +\ft1{12} \cH_{\mu\nu\rho}^\adot \c^{\mu\nu\rho}\epsilon +\ft14 \cF_{\mu\nu}^\cA V_A \c^\adot\c^{\mu\nu}\,\epsilon + 2 T^\adot \epsilon
+ \ft12 T\c^\adot \epsilon\ ,
\nn\w2
\delta\chi^a &=& \ft14 \cP_\mu^{a\adot} \c^\adot \c^\mu \epsilon +\ft1{12} \cH_{\mu\nu\rho}^a \c^{\mu\nu\rho}\epsilon +\ft14 \cF_{\mu\nu}^\cA {\tilde V}_A \c^a\c^{\mu\nu}\,\epsilon + 2{\tilde T}^a \epsilon - \ft12 {\tilde T}\c^a \epsilon\ .
\label{susy2}
\eea
Possible solutions of the embedding constraint \eq{QC} and the identification of the resulting theories was provided in \cite{Bergshoeff:2007ef}. In particular, decomposing $\theta^{AM}$ under $GL(5)\subset SO(5,5)$ allows identification of possible $7D$ origins, as well as a possible origin in $11D$, in which context $GL(5)$ is associated to the five-torus on which the reduction is performed. Gaugings of half-maximal $6D$ supergravity coupled to $4$ vector multiplets (see the subsection below), which has the duality group $ R^+ \times SO(4,4)$ can also be obtained by decomposing $\theta^{AM}$ under this duality group and performing a truncation to $N=(1,1)$ supersymmetry \cite{Dibitetto:2019odu}. For the details of the classifications of gaugings along these lines, see \cite{Bergshoeff:2007ef}.

\subsection{ $(1,1)$ supergravity coupled to vector multiplets in $6D$}

Pure $(1,1), 6D$ supergravity is based on the Poincar\'e superalgebra $F(4)$. Its $SU(2)$ gauged version coupled to a single vector multiplet was obtained in \cite{Giani:1984dw} by circle reduction of $SU(2)$ gauged half-maximal $7D$ supergravity with a topological mass term \cite{Townsend:1983kk}, summarized above in section 6.2\footnote{
A generalized dimensional reduction of half-maximal $7D$ supergravity on a circle in which the scale and trombone symmetries are utilized was obtained 
in \cite{Kerimo:2003am} giving rise to a $(1,1)$ theory in $6D$ with four abelian vectors, and whose equations of motion cannot be derived from a Lagrangian}. The $6D$ theories obtained in this way have no stable ground states with maximal spacetime symmetry.  It was shown \cite{Romans:1985tw} that for $g>0, h=0$ with $SU(2)_{\rm diag} \subset SO(4)_R \approx SU(2) \times SU(2)$ gauged, the theory can be generalized by introducing a mass parameter $m$ for the two-form tensor $B_{\m\n}$, just as in type IIA supergravity in $10D$. (The parameters $(g,h, m)$ are defined in section 6.2.) Furthermore it was shown \cite{Romans:1985tw} that this generalized theory for $g>0, m>0$ admits a ground state which exhibits the full AdS supersymmetry $F(4)$.

The coupling to $n$ vector multiplets, which has the duality group $R^+ \times SO(n,4)/SO(n)\times SO(4)$, and gauging of $SU(2)_{\rm diag} \times G$, where $G$ is an $n$ dimensional subgroup of $SO(n)$, was carried out in \cite{Andrianopoli:2001rs}. More general gauging inherited from the truncation of the maximal $N=(2,2)$ supergravity described in the previous subsection to half-maximal $N=(1,1)$ supergravity, the resulting potential and its extrema were studied in \cite{Dibitetto:2019odu}. However, the most general $6D, N=(1,1)$ theory in the embedding tensor framework remains to be spelled out. Here we shall  summarize the bosonic action and supertransformations obtained in \cite{Andrianopoli:2001rs}, bearing in mind that it provides a particular gauging. 

Combining $n$-copies of  Maxwell multiplet consisting of  fields $(A_\mu, 4\phi, \lambda) $ with the supergavity multiplet gives the field content
\be
\{ e_\m{}^r, \varphi, B_{\m\n}, \cV_M{}^A, A_\m^M \,;\,\psi_\m^i, \chi^i, \lambda^{ai} \}\ ,
\ee
where $M,A=1,...,n+4,\ a=1,..,n$, the scalar $\phi$ is  dilaton, and $\cV_M{}^A$ is the $SO(n,4)/SO(n)\times SO(4)$ coset representative parametrized by $4n$ scalars. The fermions are symplectic-Majorana, with $i=1,2$ labeling an $SU(2)$ doublet. Denoting $SU(2) $ generators by $T^I$, {note that $T^I P_\pm$, where $P_\pm = \frac12 (1\pm\gamma_7)$, act independently on spinors, yielding the isomorphism group $SU(2)\times SU(2)$ of the supersymmetry algebra. Writing $\cV_M{}^A= \big( \cV_M{}^m, \cV_M{}^a \big)\, (m=1,...,4)$, the vectors $A_\m^m$ belong to the supergravity multiplet. In a $1+3$ split we have, $\cV_M{}^m (\s^m)_{ij} = \cV_M{}^0 \e_{ij} + \cV_M{}^I (\s^I)_{ij}$, with $I=1,2,3$.  In gauging the group $G \subset SO(2)_{\rm diag} \times SO(n)$, we need the scalar currents\footnote{Our $\cV_M{}^A=(\cV_M{}^0, \cV_M{}^I, \cV_M{}^a)$ and $\psi_\mu^i$ with $M, A=1,...,n+4,\ I=1,2,3,\ a=1,...,n$ and $i=1,2$ correspond to $L_\Lambda{}^\Sigma = (L_\Lambda{}^0, L_\Lambda{}^r, L_\Lambda{}^I)$ and $\psi_\m^A$ in \cite{Andrianopoli:2001rs}.}
\bea
\cP_\m^{a0} &=& \cV^{Ma} D_\m \cV_M{}^0\ ,\qquad \cP_\m^{aI} = \cV^{Ma} D_\m \cV_M{}^I\ , 
\nn\w2
\cQ_\m^{IJ} &=& \cV^{M[I} D_\m \cV_M{}^{J]}\ ,\qquad \cQ_\m^{ab} = \cV^{M[a} D_\m \cV_M{}^{b]}\ ,
\eea
where 
\be
D_\m \cV_M{}^A = \partial_\m \cV_M{}^A + f_{MP}{}^N A_\m^P \cV_N{}^A\ ,\qquad f_{MNP}= \{ g\,\e_{IJK},\ g'f_{abc} \}\ ,
\ee
where $f_{abc}$ are the structure constants of $n$-dimensional subgroup of $SO(n)$. Further needed definitions are those of the `boosted structure constants',
\bea
C &=& f_{MNP} \cV^M_I \cV^N_J \cV^P_K\,\e^{IJK}\ , \qquad C^I=  f_{MNP} \cV^M_J \cV^N_K \cV^P_0\,\e^{IJK}\ ,
\nn\w2
C_{1\,a}{}^I &=&  f_{MNP} \cV^M{}_a \cV^N_J \cV^P_K\,\e^{IJK}\ ,\qquad  C_{2\,aI}= f_{MNP} \cV^M_0 \cV^N_I \cV^P{}_a \ .
\eea
The bosonic part of the  Lagrangian is given by \cite{Andrianopoli:2001rs}
\bea
e^{-1} \cL &=& -\frac14 R -\frac18 e^{-2\varphi} \cM_{MN} \cF_{\m\n}^M \cF^{\m\n N} +\frac{3}{64} e^{4\phi} H_{\m\n\r} H^{\m\n\r} + \del_\m\phi\partial^\m \phi -\frac14 \cP_\m^{a0} \cP^\m_{a0} -\frac14 \cP_\m^{aI} \cP^\m_{aI}
\nn\w2
&& +\frac14 e^{-2\phi}\Big( \frac19 C^2 +C^I C^I +C_1^{aI} C_1^{aI} + C_2^{aI} C_2^{ aI} \Big)  -m^2 e^{-6\phi} \cM_{00} + 2m^2 e^{-2\phi} \big( C\cV_{00} -3C^I \cV_{0I} \big)
\nn\w2
&& -\frac{1}{64} \varepsilon^{\m\n\r\s\lambda\tau} B_{\m\n} \Big( \eta_{MN} \cF_{\r\s}^M \cF_{\lambda\tau}^N + m B_{\r\s} \cF^0_{\lambda\tau} +\frac13 m^2 B_{\r\s} B_{\lambda\tau}\Big)\ ,
\eea
where $m$ is an arbitrary mass parameter and
\bea
\cM_{MN} &=& \cV_M{}^m \cV_N{}^m  + \cV_M{}^a \cV_N{}^a\ ,\qquad H_{\m\n\r} = 3\del_{[\m} B_{\n\r]}\ ,
\nn\w2
\cF_{\m\n}^M &=& 2\del_{[\m} A_{\n]}^M + f_{NP}{}^M A_\m^N A_\n^P- m\delta^M_0 B_{\m\n}\ .
\eea
The supertransformations of the fermionic fields are given by \cite{Andrianopoli:2001rs}
\bea
\delta \psi_\m^i &=& D_\m\e^i -\frac{1}{16} e^{-\varphi} \Big[\cF_{\r\s}^M \cV_M{}^0 \e^{ij} -\cF_{\r\s}^M \cV_M{}^I  (\s^I)^{ij} \Big]\left(\gamma_\m{}^{\r\s} -6\delta_\m^\r \gamma^\s \right)  \e_j 
\nn\w2
&& +\frac{i}{32} e^{2\phi} H_{\r\s\tau} \gamma_7 \left(\gamma_\m{}^{\r\s\tau} - 3 \delta_\m^\r \gamma^{\s\tau}\right) \e^i -S^{ij} \e_j\ ,
\nn\w2
\delta\chi^i &=&  \frac{i}{2} \gamma^\m \partial_\m \phi \e^i + \frac{i}{16} \Big[ \cF_{\m\n}^M \cV_M{}^0 \e^{ij} -\cF_{\m\n}^M\cV_M{}^I  (\s^I)^{ij} \Big]\gamma^{\m\n} \e_j
\nn\w2
&&-\frac{1}{32} e^{2\phi} \gamma^{\m\n\r} H_{\m\n\r} \gamma_7 \e^i -N^{ij} \e_j\ ,
\nn\w2
\delta\lambda^{ai} &=& i \cP_\m^{aI} (\s^I)^{ij} \gamma^\m \e_j +i \cP^{a0} \e^{ij} \gamma^\m \e_j +\frac{i}{2} \cF_{\m\n}^M \cV_M{}^a \e^i -M^{aij} \e_j\ ,
\eea
where $D\e^i = D_\m(\omega, \cQ) \e^i$ and the shift functions are 
\bea
S_{ij} &=& \frac{i}{24} e^\phi \big(C \e_{ij} -3 C^I (\s^I)_{ij} \big) +\frac{i}{4} m e^{-3\phi} \big( \cV_{00} \e_{ij} + \cV_{I0} (\s^I)_{ij} \gamma_7 \big)\ ,
\nn\w2
N_{ij} &=& \frac{1}{24} e^\phi \big(C \e_{ij} +3 C^I (\s^I)_{ij} \big)  -\frac34 m e^{-3\phi} \big( \cV_{00} \e_{ij} - \cV_{I0} (\s^I)_{ij} \gamma_7 \big)\ ,
\nn\w2
M^a_{ij} &=& e^\phi \left( C_1^{aI} + 2i\gamma_7 C_2^{aI} \right) (\s^I)_{ij} -2m e^{-2\phi} \cV^a{}_0 \e_{ij}\ .
\eea
%
%

\subsection{ $(2,0)$ supergravity coupled to tensor multiplets in $6D$}

The $(2,0), 6D$ supergravity multiplet has the field content $ \{ e_\m^r, \psi_{\m+}^i, 5 B_{\m\n +} \}$ with the symplectic Majorana-Weyl gravitini in the 4-plet, and the  two-form potential with anti-self-dual field strength in the 5-plet of the R-symmetry group $USp(4)_R$. Combining this with $n$ copies of  the tensor multiplets, each containing the  fields $\{ B_{\m\n -}, \chi_{-}, \phi^{ab} \}$, where the scalars in the 5-plet of $USp(4)_R$, we get the multiplet
\be
\{ e_\m{}^r, B_{\m\n}^I, \cV_M{}^A ;\,\psi_{\m +}^i, \chi_{-}^{ai} \}\ ,
\ee
where $I=1,...,n+5$ labels  the fundamental of $SO(n,5),\ a=1,..,n,\ i=1,...,4$ and $L_I{}^A$ is a representative of the $SO(n,5)/SO(n)\times SO(5)$ coset. This theory was constructed in \cite{Romans:1986er,Riccioni:1997np} by means of Noether procedure, and in \cite{Bergshoeff:1999db} by using the superconformal tensor calculus. The coset representatives obey the defining relations
\be
\cV_M{}^a \cV^M{}_{ij}=0\ , \quad \cV_M{}^a \cV^M{}_b =\delta^a_b\ , \quad \cV_M{}^{ij} \cV^M{}_{k\ell} = -\delta^{[i}_{[k} \delta^{j]}_{\ell]} +\frac14 \Omega^{ij} \Omega_{k\ell}\ .
\ee
In terms of these elements, the scalar current, the composite connections and  the metric can be written as
\bea
P_\m^{aij} &=& \cV^{Ia}\del_\m \cV_I{}^{ij}\ ,\quad 
Q_\m^{ab} = \cV^{Ma}\del_\m \cV_M{}^b\ , \quad
Q_{\m i}{}^j = 2\cV^M{}_{ik} \del_\m \cV_M{}^{jk}\ , 
\nn\w2
\cM_{MN} &=& \cV_M^{ij} \cV_{Nij} + \cV_M{}^a \cV_{N a}\ .
\eea
The $USp(4)$ indices are raised and lowered with the symplectic metric, and  the indices $M,N$ by the $SO(n,5)$ invariant metric $\eta_{MN}=  -\cV_M^{ij} \cV_{Nij} + \cV_M{}^a \cV_{N a} ={\rm diag} (-1,-1,-1,-1,-1, +1,...,+1)$. The bosonic part of the pseudo-Lagrangian is simply
\be
e^{-1} \cL = -\frac14 R +\frac{1}{12} \cM_{IJ} H_{\m\n\r}^I H^{\m\n\r J} + P_\m^{aij} P^\m_{ij}\ ,
\ee
where $H_{\m\n\r}^I = 3\partial_{[\m} B_{\n\r]}^I$. The supersymmetry transformations of the fermions are 
\be
\delta \psi_\m^i = D_\m\e^i - \frac{1}{12}
H^{ij-}_{\r\s\tau}\c^{\r\s\tau}\c_\m\e_j \ ,\qquad 
\delta \chi^{ai} = \frac{1}{48} H^{+a}_{\m\n\r}\c^{\m\n\r} \e^i
+\frac14 P_\m^{a ij} \e_j \ ,
\ee
where $ H^{ij} := H^M\cV_M^{ij},\ H^a := H^M \cV_M^a$ and $D_\m \e^i= \nabla_\m \e^i + Q_\m^{ij} \e_j$. We recall that the correct equations of motion are obtained by imposing the duality equations $H_{\m\n\r}^{ij -}=0$ and $H_{\m\n\r}^{a+}=0$ by hand, {\it after} varying the action with respect to all fields. The $(2,0)$ supergravity coupled to $21$ tensor multiplet follows from Type IIB supergravity on $K3$, and it is anomaly free \cite{Townsend:1983xt}.

\subsection{ $(1,0)$ supergravity coupled to vector, tensor and hyper multiplets in 6D}

\subsubsection{Generalities}

Starting with $N=(1,0), 6D$ supersymmetry, we begin to see the appearance of off-shell supergravity and matter multiplets, thanks to having only 8 fermionic generators. In particular, there are two versions the off-shell $(1,0), 6D$ supergravity. One of them is obtained from the coupling of the standard Weyl multiplet to a linear multiplet and fixing the dilatations, conformal boost and special supersymmetry transformations, and fixing a gauge that breaks $Sp(1)_R$  down to $U(1)_R$. The resulting off-shell multiplet containing $48_B+48_F$  degrees of freedom is described by the fields \cite{Bergshoeff:1985mz}
\begin{equation}e_{\mu}{}^a\ (15)\ ,\ \ V^\prime_{\mu}{}^{ij}\
(12)\ ,\ \ V_\mu\ (5)\,,\ \ B_{\mu\nu}\
(10)\ ,\ \ \sigma\ (1)\ ,\ \ E_{\mu\nu\rho\sigma}\ (5)\ ;\ \ \psi_{\mu}{}^i\ (40)\ ,\ \ \psi^i \
(8)\ ,\label{m1}
\end{equation}
where $i=1,2$, the vector $V_\mu^{'ij}$ is symmetric and traceless, $E$ is a 4-form potential and ${\cal V}_\mu$ is the gauge field of the surviving $U(1)_R$ gauge symmetry. The fermions (the last two in the list) are symplectic Majorana-Weyl. An alternative off-shell multiplet is obtained by coupling the dilaton Weyl multiplet to a linear multiplet and fixing the symmetries mentioned above in a slightly different way.  This yields an off-shell multiplet in which $\sigma$ is replaced by the trace of the linear multiplet scalars, $\delta^{ij} L_{ij}$, thus resulting again with $48_B+48_F$ degrees of freedom. 
The 6D off-shell ${\cal N}=(1,0)$ supergravity was constructed in \cite{Bergshoeff:1985mz,Coomans:2011ih}. The off-shell formulation is very useful in constructing the higher derivative superinvariants. As our focus is on two-derivative supergravities here, we will set the auxiliary field to zero, by using their algebraic equations of motion, and review the resulting off-shell supergravities and their matter couplings. 

On-shell the $(1,0)$ super-Poincar\'e algebra in $6D$ admits the following multiplets\footnote{There also exists a linear multiplet which has a triplet of scalars $L_{ij}$, and a 4-form potential which is on-shell dual to a scalar field. In addition, there is a non-linear multiplet similar to the linear multiplet but in which the the 3 scalars form an element of $SU(2)$. In both of these multiplets, the fermions are doublets of the $R$-symmetry group. See \cite{Bergshoeff:1985mz} for a discussion of the off-shell versions of these multiplets, and \cite{Bergshoeff:1985mz,Atli:2020ejw} for some of their couplings.}

\be
\underbrace{ \left(e_\m^m, \psi_{\m +}^A, B_{\m\n}^{-} \right)}_{\rm graviton}\ ,\qquad \underbrace{\left(B_{\m\n}^{+}, \chi_{-}^A,\varphi  \right)}_{\rm tensor}\ ,\qquad \underbrace{\left(A_\m, \lambda_{+}^i\right)}_{\rm vector}\ ,\qquad \underbrace{\left(4\phi,\psi_{-} \right)}_{\rm hyper}\ .
\ee
The two-form potentials, $B_{\m\n}^{\pm}$, have (anti)selfdual field strengths. The spinors are symplectic Majorana-Weyl, $A=1,2$ labels the doublet of the $R$ symmetry group $Sp(1)_R$, and  chiralities of the fermions are denoted by $\pm$. 

\subsubsection{Hypersector and quaternionic K\"ahler manifolds}

The couplings of two-derivative $(1,0)$ supergravity in $6D$ to a single tensor multiplet, $n_V$ vector multiplets and $n_H$ hypermultiplets was given completely in \cite{Nishino:1984gk}. As was first shown in \cite{Bagger:1983tt}, the  locally supersymmetric coupling of  hypermultiplets to supergravity requires that  hyperscalars parametrize a quaternionic Kahler (QK) manifolds of negative scalar curvature. Such manifolds are typically noncompact \cite{Bagger:1983tt}. All such spaces are necessarily Einstein. Those which are symmetric are exhausted by noncompact Wolf spaces.
These spaces, listed in table 6 under $3D, N=4$ supergravities, are
\bea
&& \frac{Sp(n,1)}{Sp(1)\times Sp(1)}\ ,\quad \frac{SU(n,2)}{SU(n)\times SU(2)\times U(1)}\ ,\quad \frac{SO(n,4)}{SO(n)\times SO(4)}\ ,\quad \frac{E_{8(-24)}}{E_7\times Sp(1)}\ ,
\nn\w2
&& \frac{E_{7(-5)}}{SO(12)\times Sp(1)}\ ,\quad\frac{E_{6(2)}}{SU(6)\times Sp(1)}\ ,\quad\frac{F_{4(4)}}{Sp(3)\times Sp(1)}\ ,\quad \frac{G_{2(2)}}{SO(4)}\ .
\eea
The comments in table 6 are about how they are related to particular scalar manifold geometries in matter coupled $4D, N=2$ supergravity, known as (very) special K\"ahler, and those geometries that arise in $5D,N=2$ matter coupled supergravities, known as very special geometries; see for example,  \cite{Lauria:2020rhc}. We shall comment further on these geometries where these $4D$ and $5D$ models are reviewed. 

There also exists homogeneous but non-symmetric Alekseevsky spaces \cite{Alekseevsky:1975} whose classification was completed in \cite{deWit:1991nm}\footnote{For non-homogeneous QK manifolds, see \cite{LeBrun:1991,Cortes:2020klb}.}. These are coset spaces $G/H$ where $G$ is a parabolic group and $H$ is its maximal compact subgroup. The Lie algebra $\sg$ of the group $G$ is a semi-direct sum 
\def\cS{{\cal S}}
\def\mR{{\mathbb{R}}}
\def\mC{{\mathbb{C}}}
\def\mH{{\mathbb{H}}}
\def\mF{{\mathbb{F}}}

\bea
\sg &=& \sg_0 \oplus \sg_1 \oplus \sg_2\ ,
\nn\w2
\sg_0 &=& so(1,1) \oplus so(q+3,3)\oplus {\cal S}_q(P,Q)\ ,
\nn\w2
\sg_1 &=& \big( SO(q+3,3)\, \mbox{spinor}, {\cal S}_q(P,Q)\ \mbox{vector} \big)_1 \ ,
\nn\w2
\sg_2 &=& \big( SO(q+3,3)\, \mbox{vector}, {\cal S}_q(P,Q)\ \mbox{singlet} \big)_2\ ,
\eea
where $q, P, Q$ are positive integers or zero, the subscripts denote the $so(1,1)$ weights and the group ${\cal S}_q(P,Q)$ is explained and given in table 1 below \cite{deWit:1991nm}. The isotropy group $H$ is

\be
H= SO(q+3)\otimes SU(2) \otimes {\cal S}_q(P,Q)\ .
\ee
For the symmetric manifolds, $\sg_{-1}$ and $\sg_{-2}$ are included. For example, taking $q=8, P=Q=0$, the generator count for $\sg_{-2} \oplus\sg_{-1} \oplus \sg_0 \oplus \sg_{+1}\oplus \sg_{+2}$ is $14+ 64_c+ (1+91)+ 64_s+ 14$ giving a total of $248$ generators of the isometry group of the symmetric space $ E_{8(-24)}/(E_7\times Sp(1))$. This is a very special QK manifold. For more details, see \cite{deWit:1991nm}.

\begin{table}[H]
 \small{\begin{tabular}{|c|c|c|l|}
\hline
$q$ & ${\cal C}(q+1,0)$ & ${\cal D}_{q+1}$ &\  ${\cal S}_q(P,Q)$ \\
\hline
-1 & $\mathbb{R}$ & 1 & \ $SO(P)$\\
0 & $\mR\oplus \mR$ & 1 & \  $SO(P)\otimes SO(Q)$ \\
1 & $\mR(2)$ & 2 & \ $SO(P)$ \\
2 & $\mC (2)$ & 4 & \ $U(P)$\\
3 & $\mH(2)$ & 8 & \ $USp(2P)$ \\
4 & $\mH(2) \oplus \mH(2)$ & 8 & \  $USp(2P)\otimes USp(2Q)$ \\
5 & $\mH(4)$ & 16 & \ $USp(2P)$ \\
6 & $\mC(8)$ & 16 & \ $U(P)$ \\
7 & $\mR(16)$ & 16 & \ $SO(P)$ \\
$n+8$ &\ $\mR(16) \otimes {\cal C}(n+1,0)$ \ & 16 ${\cal D}_n $ & \ \mbox{as for} $q=n$\\
\hline
\end{tabular}}
\caption{ \footnotesize  Real Clifford algebras ${\cal C}(q+1,0)$. Here $\mF (n)$ stands for $n\times n$ matrices with entries over the field $\mF$, and ${\cal D}_{q+1}$ denotes the real dimensions of an irreducible representation of the Clifford algebra. ${\cal S}_q (P,Q)$ is the metric preserving group in the centralizer of the Clifford algebra in the $(P+Q){\cal D}_{q+1}$ dimensional representation.}
\end{table}
The QK manifolds have  tangent space group $Sp(n_H)\times Sp(1)_R$, and one can introduce  vielbeins $V_X^{ri}$ and their inverse $V^X_{ri}$ satisfying\footnote{Note that the third relation in eq. (2) of \cite{Bagger:1983tt}, which also appeared in several papers that followed, holds only for $n=1$, namely the case of a single hypermultiplet.}
\bea
g_{\a\b} V^\a_{aA} V^\b_{bB} =\Omega_{ab}\eps_{AB}\ ,
\qquad
V^\a_{aA} V^{\b bB} + \ ( \a \leftrightarrow \b ) = g^{\a\b} \d_A^B\ ,
\eea
where $g_{\a\b}$ is the metric. According to \cite{Bagger:1983tt} it is not known if there are negatively curved QK manifolds which do not admit globally defined vielbeins $V_a^{aA}$. An $Sp(n_H)\times Sp(1)_R$ valued connection is defined through the vanishing torsion condition
\bea
\partial_\a V_{\b aA} + A_{\a a}{}^b V_{\b bA} +A_{\a A}{}^B V_{\b aB}\
-( \a \leftrightarrow \b) =0\ .
\eea
From the fact that  the vielbein $V^\a_{aA}$ is covariantly constant, one derives that
\bea
R_{\a\b\gamma\delta} V^\delta_{aA} V^\gamma_{bB} &=& \epsilon_{AB}\,F_{\a\b}{}_{ab}+\Omega_{ab}\,F_{\a\b}{}_{aB}\ ,
\label{Riemann}
\eea
where $F_{AB}$ and $F_{ab}$ are the  curvature two-forms of  $Sp(1)_R$ and $Sp(n_H)$ connection, respectively.
The manifold has a quaternionic K\"ahler structure characterized by three locally defined $(1,1)$ tensors $J^r{}_\a{}^\b$ $(r=1,2,3)$ satisfying the quaternion algebra
\bea
J^r{}_\a{}^\b J^{s}{}_\b{}^\gamma = -\delta^{rs}\delta_\a^\gamma +\eps^{rst} J^t{}_\a{}^\gamma\ .
\eea
These tensors can be expressed as 
$J^r{}_\a{}^\b = -i (\sigma^r)_A{}^B\,V_\a^{aA} V^\b_{aB}$, 
where $\sigma^r$ are the Pauli matrices. We can define a triplet of two-forms $J^r_{\a\b}=J^r{}_\a{}^\gamma g_{\gamma\b}$, and these are covariantly constant, namely, $
\nabla_\a J^r_{\b\gamma} +\eps^{rst} A_\a^s J^t_{\b\gamma}=0$, 
with $A_\a^r \equiv \frac{i}2  (\sigma^r)_A{}^B A_{\a B}{}^A$.
For $n_H >1$, quaternionic K\"ahler manifolds are Einstein spaces,
i.e.\ $R_{\a\b}=\lambda g_{\a\b}$. Using the covariant constancy of $J^r_{\a\b}$, one finds that \cite{Galicki:1985qv}
\bea
F^r_{\a\b} = \frac{\lambda}{n_H+1} J^r_{\a\b}\ .
\label{ks}
\eea
Local supersymmetry relates $\lambda$ to the gravitational coupling constant (which we set to 1), and requires that $\lambda <0$ \cite{Bagger:1983tt}, explicitly $\lambda=-(n_H+1)$. For $n_H=1$ all Riemannian 4-manifolds are quaternionic K\"ahler\footnote{Sometimes (\ref{ks}) is used to extend the definition of the quaternionic K\"ahler to $4D$, which restricts  the manifold to be Einstein and self-dual \cite{Galicki:1985qv}.} Using this value, substitution of \eq{ks} into \eq{Riemann} and the use of the curvature cyclic identity gives \cite{Bagger:1983tt}
\be
F_{\a\b ab} = V_{[\a}^{cA} V^d_{\b]A} \left( -2\Omega_{ca}\Omega_{db} +\Omega_{abcd}\right) \ ,
\ee
where $\Omega_{abcd}$ is a totally symmetric tensor defined by this equation. For symmetric QK manifolds, $\Omega_{abcd}=0$. 
 
\subsubsection{The case of single tensor multiplet}

In the case of quaternionic projective space $Hp(n_H)=Sp(n_H,1)/Sp(n_H)\times Sp(1)_R$,  the maximal compact subgroup of its isometry group was gauged in \cite{Nishino:1984gk}, where couplings to Yang-Mills and hypermultiplets were completely determined. The full multiplet content in this case is given by
\be 
(e_\mu{}^m, \psi_\mu^A, B_{\mu\nu}, \chi^A, \varphi)\ ,\quad   (A_\m^\hI, \lambda^{\hI A})\ ,\quad  (\phi^\a, \psi^a)\ , \quad \hI= (I, i)\ ,\quad i=1,2,3, \quad  a=1,..., n_H\ ,
\ee
where $I$ labels the adjoint representation of $Sp(n_T)$, and $\a=1,...,4n_H$ labels the coordinates of $Hp(n_H)$. Here we have combined the $(1,0)$ supermultiplet with the single tensor multiplet, resulting in what we may think of as ``reducible supergravity multiplet". Further building blocks needed are defined as follows \cite{Nishino:1984gk},
\bea
\cP_\mu^{aA} &=& D_\mu \phi^\a\ V_\a^{aA}\ ,
\quad
\cQ_{\mu}^{AB} = D_\mu \phi^\a A_\a^{AB}+ A_\m^{AB} \ ,\quad
\cQ_{\mu}^{ab} = D_\m \phi^\a A_\a^{ab}  + A_\m^{ab} \ ,
\nn\w2
\cH_{\m\n\r} &=& 3\del_{[\m} B_{\n\r]} +6 v_z \tr_z \big( A_{[\m}\partial_\n A_{\r]} + \frac23A_{[\m}A_\n A_{\r]}\big)\ ,
\nn\w2
D_\m \phi^\a &=& \partial_\m \phi^\a - \tr (A_\m K^\a)\ ,
\nn\w2
C^i & \equiv & C^{i \hI} T^\hI\ ,\qquad C^{iI}= A_\a^i K^{\a I}\ , \qquad 
C^{ij} = A_\a^i K^{\a j} -\delta^{ij}\ ,
\eea
where $F=F^\hI T^\hI,\ \tr\, (T^\hI T^\hJ)=\delta^{\hI\hJ}$, the Killing vectors $K^\a = K^{\a\hI} T^\hI$, and $v_z=(v_1, v_2)$ with $v_1=1/g^2$ and $v_2=1/g'^2$ representing the coupling constants of $Sp(1)$ and $Sp(n_H)_R$, respectively. The bosonic part of the complete Lagrangian is given by \cite{Nishino:1984gk}\footnote{We use the conventions of \cite{Nishino:1984gk} with the replacements  $\eta_{mn}\to -\eta_{mn}, \gamma^m\to i\gamma^m, \varphi \to \varphi/{\sqrt 2},\ \lambda^i \to {\sqrt 2}\lambda^i/g,\ \lambda^I \to {\sqrt 2}\lambda^I/g', \ A_\mu \to A_\mu/g$.} Note also that while a coupling of the form $\Omega_{abcd} \bpsi^a\psi^b \bpsi^c\psi^d$ is given in \cite{Nishino:1984gk}, it is present only for non-symmetric QK manifolds.}
\begin{align}
e^{-1} \cL &= \frac14 R -\frac14 \partial_\m\varphi \partial^\m \varphi -\frac{1}{12} e^{2\varphi} \cH_{\mu\nu\rho} \cH^{\mu\nu\rho} -\frac14 e^\varphi v_z \tr_z \left(F_{\m\n} F^{\m\n}\right) -\frac12 \cP_\m^{aA} \cP^\m_{aA}
\nn\w2
& -\frac18 e^{-\varphi}  v_z^{-1} \tr_z \left(C^i C^j\right) \ .
\end{align}
and the local supersymmetry transformations are given by
\bea
\delta \psi_\mu^A &=& D_\mu \eps^A + \frac{1}{24} e^{\varphi} \cH_{\rho\sigma\tau}
\c^{\rho\sigma\tau} \c_\mu  \eps^A\ ,
\nn\w2
\delta \chi^A&=& \frac12 \c^\mu \eps^A \partial_\m \varphi - \frac{1}{12} e^\varphi \cH_{\mu\nu\rho}\gamma^{\mu\nu\rho} \eps^A\ ,
\nn\w2
\delta\lambda^{iA}&=&  \frac14 e^{\varphi/2} F^i_{\mu\nu}  \gamma^{\mu\nu} \eps^A   -\ft12  g e^{-\varphi/2} C^{iAB}\,\epsilon_B \ ,
\nn\w2
\delta\lambda^{IA}&=&  \frac14 e^{\varphi/2} F^I_{\mu\nu}  \gamma^{\mu\nu} \eps^A   -\ft12  g' e^{-\varphi/2} C^{IAB}\,\epsilon_B \ ,
\nn\w2
\delta\psi^a &=& -\cP_\mu^{aA} \gamma^\mu\epsilon_A\ ,
\label{susy3}
\eea
where $D_\m \epsilon^A = \nabla_\m \e^A + \cQ_\m^{AB} \e_B,\ C^{iAB}= C^{ij} (T^j)^{AB}$ and $C^{IAB}= C^{iI} (T^i)^{AB}$. Passing to a `string frame' entails the following steps \cite{Chang:2023pss}
\bea
&& e_\mu{}^m \to e^{\varphi/2} e_\mu{}^m\ ,\quad \psi_\mu \to e^{\varphi/4} (\psi_\mu +\frac12 \gamma_\mu \chi)\ ,\quad \left( \chi, \lambda, \psi^a \right) \to e^{-\varphi/4} \left( \chi, \lambda, \psi^a \right)\ , 
\nn\w2
&& \epsilon \to e^{\varphi/4}\epsilon\ ,\quad \delta_\epsilon +\delta_\Lambda \to \delta_\lambda\ ,\quad \lambda^m{}_n =\frac12 {\bar\eps}\gamma^m{}_n \chi\ ,
\eea
where $\delta_\eps$ and $\delta_\Lambda$ are the supersymmetry and local Lorentz transformations,  and gives
\begin{align}
\cL &= e e^{2\varphi} \Big[ \frac14 R -\frac14 \partial_\m\varphi \partial^\m \varphi -\frac{1}{12}H_{\mu\nu\rho} H^{\mu\nu\rho} -\frac14 v_z \tr_z \left(F_{\m\n} F^{\m\n}\right) -\frac12 P_\m^{aA} P^\m_{aA}
\nn\w2
& -\frac18 v_z^{-1} \tr_z \left(C^i C^j\right)\Big] \ ,
\end{align}
with the supertransformations of the fermionic fields given by
\bea
\delta \psi_\mu^A &=& D_\mu(\omega_{-}) \eps^A\ ,
\nn\w2
\delta \chi^A&=& \frac12 \c^\mu \eps^A \partial_\m \varphi - \frac{1}{12} H_{\mu\nu\rho}\gamma^{\mu\nu\rho} \eps^A\ ,
\nn\w2
\delta\lambda^{i A}&=&  \frac14  F^i_{\mu\nu}  \gamma^{\mu\nu} \eps^A   -\ft12 g C^{i AB}\,\epsilon_B \ ,
\nn\w2
\delta\lambda^{I A}&=&  \frac14  F^I_{\mu\nu}  \gamma^{\mu\nu} \eps^A   -\ft12 g' C^{I AB}\,\epsilon_B \ ,
\nn\w2
\delta\psi^a &=& -P_\mu^{aA} \gamma^\mu\epsilon_A\ ,
\label{susy1}
\eea
where $C^{\hat I AB}= C^{i \hat I} (T^i)^{AB}$ and the the field strength $H$ has been absorbed into the definition of the spin connection as torsion so that $\omega_{-\mu ab} := \omega_{\mu ab} -H_{\mu ab}$. 

\subsubsection{$U(1)_R$ gauged Einstein-Maxwell supergravity }

The special case in which the hypermultiplets are left out and only a single vector multiplet is kept to gauge $U(1)_R \subset Sp(1)_R$ \cite{Salam:1984cj} has attracted much interest, finding applications in cosmology and phenomenology \cite{Maeda:1984gq,Maeda:1984gq,Gibbons:1987ps,Aghababaie:2003wz}. In this case the bosonic part of the Lagrangian, in the conventions of \cite{Salam:1984cj},  is given by
\be
e^{-1} \cL = \frac{1}{4\kappa^2} R-\frac14  \partial_\m\varphi\partial^\m \varphi -\frac{1}{12} e^{2\kappa \varphi} \cH_{\m\n\r} \cH^{\m\n\r} -\frac14 e^{\kappa\varphi} F_{\m\n} F^{\m\n} -\frac12 g^2\kappa^{-4} e^{-\kappa\varphi}\ ,
\ee
and the supertransformations of the fermions are
\bea
\delta\psi_\m &=& \kappa^{-1} D_\m \e +\frac{1}{24} e^{\kappa \varphi} \cH_{\n\r\s} \gamma^{\n\r\s}\gamma_\m\e\ ,
\nn\w2
\delta\chi &=& -\frac12 \partial_\m \vp \gamma^\m \e +\frac{1}{12} e^{\kappa\varphi} \cH_{\m\n\r} \gamma^{\m\n\r} \e\ , 
\nn\w2
\delta \lambda &=& \frac{1}{2\sqrt2} e^{\kappa\varphi/2} F_{\m\n}\gamma^{\m\n} \e -\frac{i}{\sqrt2}g e^{\kappa\varphi/2}\e\ ,
\eea
where the fermions are Weyl, $F=dA$ and $\cH_{\m\n\r}=3\partial_{[\m} B_{\n\r]} +3\kappa F_{[\m\n} A_{\r]}$. The ${\rm Minkowski}\times S^2$ compactification was found in \cite{Salam:1984cj}, and dyonic string solution in \cite{Gueven:2003uw}.

The $U(1)_R$ gauged $N=(1,0),6D$ supergravity by itself is anomalous. The simplest way to remedy this is to couple to 245 hypermultiplets. Anomaly freedom by more general couplings to a suitable set of vector, hyper and tensor multiplets, turns out to be rare if one insists on gauge groups other than $U(1)$ and $SU(2)$. It is still not known if the few such models obtained so far \cite{Randjbar-Daemi:1985tdc,Avramis:2005qt,Avramis:2005hc,Becker:2023zyb} can be embedded in string/M theory. As a progress towards this end, it was shown in \cite{Cvetic:2003xr} that a dimensional reduction of the (anomalous) pure $N=1, 10D$ supergravity on a noncompact 3-manifold $H_{2,2}$ which can be embedded in a plane with $(2,2)$ signature and which has $U(1)\times U(1)$ isometry, yields an $SO(2,2)$ gauged supergravity in $7D$, whose reduction on a circle followed by a chiral truncation yields the $U(1)_R$ gauged supergravity in $6D$. It has also been shown that the $SO(2,1)$ and $SO(3,1)$ gauged half-maximal supergravities in $7D$, reviewed in section 6.2 here, also admit circle reduction followed by chiral truncation that yield the $U(1)_R$ gauged supergravity in $6D$ \cite{Bergshoeff:2005pq}. It has also been shown that the $N=(1,1), 6D$ model obtained from a generalized dimensional reduction of pure half-maximal $7D$ supergravity, and whose equations of motion cannot be derived from a Lagrangian, has a truncation, albeit nonsupersymmetric one, that yields the equations of motion of the $U(1)_R$ gauge Maxwell-Einstein theory \cite{Kerimo:2003am}. 

\subsubsection{The case of multi-tensor multiplets }

Coupling an arbitrary number of tensor multiplets to $N=(1,0), 6D$ supergravity brings in a number of new features and subtleties. In that case, the scalar fields of the tensor multiplets parametrize the coset $SO(n_T,1)/SO(n_T)$, and the chiral two-form of supergravity multiplet together with the $n_T$ anti-chiral two-forms coming from the tensor multiplets transform a vector of $SO(n_T+1)$. The field equations of multi-tensors coupled to $(1,0)$ supergravity in leading order in fermions were obtained in \cite{Romans:1986er}. The vector fields were included in \cite{Ferrara:1996wv}, and the hypermultiplet couplings as well in \cite{Nishino:1997ff}, where the complete supertransformations were also found. Subsequently, the complete field equations without hypermultiplets were found in \cite{Ferrara:1997gh}. Finally, the results of \cite{Nishino:1997ff,Ferrara:1997gh} were completed to include the higher order fermion terms in \cite{Riccioni:2001bg}, where a pseudo-Lagrangian was given as well. 

In the rest of this subsection, we leave out the hypermultiplets and combine $n_T$ copies of the tensor multiplet consisting of the fields $(B_{\m\n}, \varphi, \lambda)$ with the pure supergavity multiplet, and consider Yang-Mills sector with gauge a semi-simple gauge group $G=\Pi_z G_z$, without $R$-symmetry gauging. For the field content, we introduce the notation
\be
\{ e_\m{}^m, B^I_{\m\n}, L_I{}^A ;\,\psi^i_\m, \chi^{r}\}\ ,\qquad \{ A_\m, \lambda\}_z\ ,
\ee
where $I, A=0,1,..,n_T,\ r=1,..,n_T$, and $L_M^A= \left(\L_I{}^0, L_I{}^r \right)$ is the $SO(n_T,1)/SO(n_T)$ coset representative parametrized by $n_T$ tensor multiplet scalars. The fermions are symplectic-Majorana-Weyl. The $SO(n_T,1)$ invariant tensor $\eta={\rm diag} (-1,+1,...,+1)$ and the positive definite scalar matrix $\cM$ are given by
\be
\eta_{IJ} =- L_I{}^0 L_J{}^0 + L_I{}^r L_J{}^r\ ,\qquad \cM_{IJ} = L_I{}^0 L_J{}^0 + L_I{}^r L_J{}^r\ .
\ee
Suppressing the fermionic field dependence, the self-duality equation is 
\be
\cM_{IJ} H^{J \m\n\r} = \frac16 \eta_{IJ} \ve^{\m\n\r\s\lambda\tau} H^J_{\s\lambda\tau}\ ,
\label{de2}
\ee
where
\be
H^I_{\m\n\r} = 3\del_{[\m} B_{\n\r]} +6 c^{Iz} \tr_z \big( A_{[\m}\partial_\n A_{\r]} + \frac23A_{[\m}A_\n A_{\r]}\big)\ .
\ee
The Yang-Mills equation obtained from supersymmetry considerations, and suppressing the fermionic terms, is given by
\be
D_\m \big( c^{Iz} L_I F_z^{\m\n} \big) = \frac12 c^{Iz} \cM_{IJ} H^{J\n\rho\sigma} F_{\rho\sigma}^z\ ,
\label{YM1}
\ee
where $c^{Iz}$ are coupling constants and we have defined $L_I{}^0 \equiv L_I$. 
Writing this equation as $D_\m \left( c^{Iz}L_I F_z^{\m\n} \right) = J_z^\n$, and taking the covariant divergence of both sides, and using the duality equation, one finds that the current $J_z^\m$ fails to be conserved \cite{Ferrara:1996wv}, since \footnote{In presence of hypermultiplets, there will be a contribution to the current coming from the hypermultiplet scalars which is conserved.}
\be 
D_\m  J_z^\mu = \frac{1}{16} \ve^{\m\n\r\s\lambda\tau} \eta_{IJ} c^{Iz} c^{Jz'} F_{z \m\n}\, \tr_{z'} \left(F_{\r\s} F_{\lambda\tau} \right) \ .
\ee
In view of this (covariant) anomaly, the Yang-Mills equation of motion cannot be derived from an action, pseudo or not.  The presence of the anomaly in the gauge symmetry also implies, by supersymmetry algebra, that it must imply the presence of supersymmetry anomaly as well. A useful way to characterize both supersymmetry and gauge anomalies is to integrate all the field equations, except those of the Yang-Mills sector, into a pseudo action and add the Wess-Zumino term of the form $B(\wedge \tr F\wedge F)$ which will modify the Yang-Mills equation \eq{YM1} such that it yields the so-called consistent anomaly. The gauge anomaly thus manifest itself as the non-vanishing of the gauge variation of the Wess-Zumino term, and the expected supersymmetry anomaly, as its non-vanishing variation under supersymmetry \cite{Ferrara:1997gh}. Moreover, these anomalies have been shown \cite{Ferrara:1997gh} to obey the Wess-Zumino consistency conditions
\bea
&& \delta_{\Lambda_1} \cA_{\Lambda_2} - \delta_{\Lambda_2} \cA_{\Lambda_1} =\cA_{\Lambda_3}\ , \quad  \delta_\Lambda \cA_\e = \delta_\e \cA_\Lambda \ ,
\nn\w2
&& \delta_{\e_1} \cA_{\e_2} - \delta_{\e_2} \cA_{\e_1} = \cA_\Lambda + \cA_{\e_3}\ ,
\eea
where $\cA_\Lambda$ and $\cA_\e$ denote the gauge and supersymmetry anomalies, respectively. 

The bosonic part of the pseudo-Lagrangian is given by \cite{Ferrara:1997gh}
\footnote{ Here we go from the conventions of \cite{Ferrara:1997gh} to those of \cite{Nishino:1997ff} by letting $ c_z^I \to -c_z^I/2, H_{\m\n\r}\to H_{\m\n\r}/2,  \eta_{\m\n} \to -\eta_{\m\n}, \eta_{IJ} \to -\eta_{IJ}, \gamma^\m \to i\gamma^\m$ and $\lambda \to \lambda/{\sqrt 2}$. In the expression for the potential given in eq. (2.19) of \cite{Riccioni:2001bg},  $\cA_\alpha^A{}_B \cA_\beta^B{}_A \xi^{\alpha i}\xi^{\alpha j}$ should be replaced by $(\cA_\alpha^A{}_B \xi^{\alpha i} -\delta^A_B)  (\cA_\beta^B{}_A \xi^i-\delta^B_A)$, and in eq. (2.20) for $\delta\lambda^{iA}$, the expression $\cA_\alpha^A{}_B \xi^{\alpha i}$ should be replaced by $\cA_\alpha^A{}_B \xi^{\alpha i}-\delta^A_B$. }
\bea
e^{-1} \cL &=& \frac14 R-\frac{1}{48} \cM_{IJ} H^{I \m\n\r} H^J_{\m\n\r} - \frac14 \partial_\m L_I \partial^\m L^I -\frac14 c_z^I L_I \tr_z \left( F_{\m\n} F^{\m\n}\right)
\nn\w2
&& -\frac{1}{32} \ve^{\m\n\r\s\lambda\tau} \eta_{IJ} c^{Iz} B^J_{\m\n} \tr_z \left(F_{\r\s} F_{\lambda\tau}\right) \ .
\eea
The positivity of the Yang-Mills kinetic term requires that $c_z^I L_I >0$, which must hold at least in a region of the moduli space, if the theory is to be physically acceptable. As to the significance of the point at which it vanishes as a phase transition point, see \cite{Duff:1996cf}. 

The Yang-Mills field equation resulting from the action above, upon the use of the duality equation \eq{de2}, is now given by 
\begin{align}
D_\m \big( c^{Iz} L_I F_z^{\m\n} \big) &=  \frac{1}{12} \ve^{\n\r\s\lambda\tau\kappa} \eta_{IJ} c_z^I \Big( H_{\rho\sigma\lambda} F^z_{\tau\kappa} - \frac12  c_{z'}^J  F^{z'}_{\r\s} \omega^{z'}_{\lambda\tau\kappa}-\frac34  c^{J z'} A_\r \tr_{z'} \big( F_{\s\lambda} F_{\tau\kappa}\big)\,\Big) \ .
\end{align}
Writing this equation as $D_\m \left( c^{Iz}L_I F_z^{\m\n} \right) = {\hat J}_z^\n$, and taking the covariant divergence of both sides, one now finds the (consistent) anomaly \cite{Ferrara:1996wv}
\be 
D_\m  {\hat J}_z^\mu = \frac{1}{16} \ve^{\m\n\r\s\lambda\tau} \eta_{IJ}c_z^I c_{z'}^I \big(\partial_\m A_\n \big)\, \tr_{z'} \left(F_{\r\s} F_{\lambda\tau} \right) \ .
\ee
For a description of the resulting supersymmetry anomaly, and how it satisfies the Wess-Zumino consistency conditions, see \cite{Ferrara:1996wv}. At the one-loop level, generically there will be gauge, gravitational and mixed anomalies. An analysis of the effective action including one loop effects, and potential counterterms is needed to determine the fate of these anomalies. Global anomalies can also arise. Much work has been done on anomalies in $(1,0), 6D$ supergravities, and the review of this vast subject goes beyond the scope of this survey. See for example, \cite{Avramis:2005hc,Kumar:2009ae,Monnier:2017oqd} and references therein.  

Turning to the supertransformations of the fermionic fields in leading order in fermion terms, which are given by
\bea
\delta \psi_\m &=& D_\m \e +\frac{1}{48} L_I H^I_{\r\s\tau} \gamma^{\r\sigma\tau} \gamma_\m \e \ ,
\nn\w2
\delta \chi^r &=& \frac12 \big(L_I{}^r \partial_\m L^I \big)\gamma^\m \e +\frac{1}{24} L_I{}^r H^I_{\m\n\r} \gamma^{\m\n\r} \e\ ,
\nn\w2
\delta \lambda_z &=& -\frac14 F_{z \m\n} \gamma^{\m\n} \e\ .
\eea
We shall next describe special cases of the matter couplings of $(1,0)$ supergravity known as the magical supergravities.

\subsection{ $(1,0)$ magical supergravities in $6D$}
 
There exists a special class of supergravity theories in $D=3,4,5,6$, known as magical supergravities \cite{Gunaydin:1983rk,Gunaydin:1983bi} whose remarkable geometries and symmetries correspond to those of the Magic Square of Freudenthal, Rozenfeld and Tits. The scalar manifolds arising in all magical supergravities are collected are tabulated in appendix C.  The magical theories in $6D$ are parent theories from which all magical supergravities in $D=3,4,5$ can be obtained by dimensional reduction, see table 2 below. The geometries arising in $D=3,4,5$ \cite{Gunaydin:1983rk} were later referred to as very special quaternionic K\"ahler, very special K\"ahler and very special real, respectively.  See~\cite{VanProeyen:2001wr} for a review of these geometries, their relation to $6D$ theories and a more complete list of  references.
%
Stringy origins and constructions of some of the magical supergravity theories in various dimensions, with or without  additional hypermultiplet couplings, are known \cite{Sen:1995ff,Dolivet:2007sz,Bianchi:2007va}.
Gaugings of magical supergravities have been investigated in $D=5$ (see \cite{Gunaydin:2003yx} and references therein) as well as in $4$ and $3$ dimensions~\cite{Andrianopoli:1996cm,Gunaydin:2005df,deWit:1992up,deWit:2003ja}, and finally in $6D$ \cite{Gunaydin:2010fi}, where coupling to hypermultiplets was constructed as well. Here we shall briefly review the key results of \cite{Gunaydin:2010fi}.

\TableE

In $(1,0), 6D$ magical  supergravities, the  global symmetry groups, representation content of the vectors and two-forms are displayed in table 3. Note that the global symmetry for $n_T=5$ and $n_T=3$ has additional factor $USp(2)$ and $U(1)$, respectively, exclusively acting on the hypermultiplets. 
\TableD 
The structure of  magical supergravity Lagrangian is in many ways similar to  the one described in the previous section but it also differs in interesting ways, in particular relying on the existence of the gamma-matrix identity,
\be 
\Gamma_{I(AB} \Gamma^I_{C)D}=0\ .
\label{md}
\ee
Given that there is a fixed number of vector fields in magical supergravities,  the embedding tensor formalism is the most convenient method for determining their gaugings. In this framework, the  general covariant derivative is
\be
D_\m =\partial_\m -A_\m^A X_A\ ,
\ee
where the  representation carried by the vector field is as given in table 3, and 
\be
X_A =\theta_A{}^{IJ} t_{IJ} + \theta_A{}^\chi t_\chi + \Theta_A{}^\cA t_\cA\ ,
\label{gmagic}
\ee
where $t_{IJ}$ are the generators of $SO(n_T,1)$, the  generators $t_\chi$ span the additional symmetries $USp(2)$ and $U(1)$ for $n_T=5$ and $n_T=3$, respectively, and $t_\cA$ are  generators of the isometries of the quaternionic K\"ahler manifold\footnote{Note  that the $R$-symmetry group $Sp(1)_R$ is contained in $t_\cA$.}.  The generators $X_A$ are required to obey the closed algebra 
\be
[X_A, X_B] =-X_{AB}{}^C X_C\ ,\qquad X_{(AB)}{}^C X_C =0\ .
\ee
As shown in \cite{Gunaydin:2010fi}, the consistency of  gauge algebra on  vector fields imposes the constraint 
\be
X_{(BC)}{}^A = \Gamma^I_{BC} \theta_I^A\ ,
\ee
where $\theta_I^A$ is a constant tensor. Consequently, the hierarchy of the covariant field strength takes the  form
\bea
\cF_{\m\n}^A &=& 2\del_{[\m} A_{\n]}^A +X_{[BC]}{}^A A_\m^B A_\n^C + B_{\m\n}^I \theta_I^A\ ,
\nn\w2
\cH^I_{\mu\nu\rho} &=& 3\, D_{[\mu} B^I_{\nu\rho]} +
6\,\Gamma^I_{AB}\,A_{[\mu}{}^A\,\Big(\partial_{\nu} A_{\rho]}{}^B+
\ft13  X_{[CD]}{}^B A_{\nu}{}^C A_{\rho]}{}^D\Big) +\,\theta^{IA}\, C_{\mu\nu\rho\,A} \ ,
\nn\w2
\cG_{\mu\nu\rho\sigma\,A} &=&
4D_{[\mu} C_{\nu\rho\sigma]\,A}
-\Gamma_{I AB} \Big( 6 B^I_{[\mu\nu} {\cF}_{\rho\sigma]}^B+6 \theta^{BJ} B^I_{[\mu\nu}
B^{\vphantom{I}}_{\rho\sigma] J}
\nn\w2
&& {}+8\Gamma^I_{CD} A_{[\mu}^B A_\nu^C \partial_\rho A_{\sigma]}^D +2 \Gamma^I_{CD} X_{EF}{}^{D} A_{[\mu}{}^B A_\nu{}^C A_\rho{}^E A_{\sigma]}{}^F\Big)\ .
\eea
The gauge transformations and the Bianchi identities can be found in \cite{Gunaydin:2010fi}. The construction so far is entirely off-shell, but the (anti-) self-duality condition \eq{de2} will be imposed by hand, and the field equation of $B_{\m\n}^I$, up to fermion terms, 
obtained from the  pseudo-action will give rise to the first order equation
\bea
\theta_I^A  \Big( \cG^{\mu\nu\rho\sigma}_{A} +\frac12\, \,\ve^{\mu\nu\rho\sigma\lambda\tau}\,m_{AB}\,\cF_{\lambda\tau}^B \Big)=0\ ,
\label{de}
\eea
where 
\be
m_{AB} \equiv L_I \Gamma^I_{AB}\ .
\ee
The constraints on the embedding tensor, linear and quadratic, turn out to be highly constraining, and it was shown that they are solved such that
\be
\theta_A{}^{IJ} = -\Gamma^{[I}_{AB} \theta^{J]B}\ , \qquad \theta^{IA}= \Gamma^I_{BC} \zeta^A \zeta^B \zeta^C\ ,\qquad \theta^{IA} \Theta_A{}^\cA  =0\ ,
\ee
where $\zeta^A$ is an unconstrained constant (commuting) spinor of $SO(n_T,1)$. All different choices of the spinor $\zeta^A$ lead to equivalent gaugings \cite{Gunaydin:2010fi}.

In the hypermultiplet sector, the generators $t_\cA$ act on the hyperscalar manifold by a Killing vector field $K_\cA{}^X$ as
\be
t_\cA \cdot \phi^X = K^X_\cA (\phi)\ ,
\ee
and the gauge covariant derivative of the hyperscalars is given by
\be
D_\m\phi^X = \partial_\m \phi^X -g A_\m^A \cK_A^X\ ,\qquad \cK_A^X \equiv \theta_A{}^\cA K_\cA^X\ .
\ee
The covariant derivative of the fermionic fields requires the following connections
\be
Q_\m^{ab} = L^{I[a} D_\m L_I^{b]}\ ,
\qquad
Q_\m^{ij} = \partial_\m\phi^X +A_\m^A C_A{}^{ij}\ ,
\qquad
Q_\m^{rs} = \partial_\m \phi^X +A_\m^A C_A{}^{rs}\ ,
\ee
where
\be
C_{A i}{}^j = -\frac{1}{n_H} V^X_{ri} V^{rj}_Y \nabla_X \cK^Y_A\ ,\qquad 
C_{A r}{}^s = -\frac12 V^X_{ri} V_Y^{si} \nabla_X \cK^Y_A\ .
\ee
Putting together  the above ingredients, and the following the standard Noether procedure, we find that the pseudo-action for the gauged magical supergravity is given by
\bea
e^{-1} \cL &&= R    -\frac{1}{12}  g_{IJ} {\cal H}_{\mu\nu\rho}^I {\cal H}^{\mu\nu\rho J} -\frac14 m_{AB} \cF_{\mu\nu}^A \cF^{\mu\nu B}  -\frac14 P_\mu^a P^{\mu a}
-\frac12 P_\mu^{ri} P_{\mu ri}
\nn\w2
&& -\frac14 \left( \theta^{IA} \theta^{JB} m_{AB} g_{IJ}
+  m^{AB}\,C_A{}_{ij} C_B^{ij}  \right) +\cL_{\rm top}\ ,
\label{action-1}
\eea
where ${\cal L}_{\rm top}$ is the gauge invariant completion of  the $B\wedge \cF\wedge \cF$ term and its variation is given in \cite[eq. (4.42)]{Gunaydin:2010fi}.
Using that variation, it is easy to see the equation of motion for $B_{\m\n}^I$, up to fermion terms, gives the (projected) duality equation \eq{de}. The supertransformations of the fermions terms are
\be
\begin{split}
\delta \psi_\mu^i &= {\cal D}_\mu \eps^i + \frac{1}{48}
\c^{\rho\sigma\tau} \c_\mu  \eps^i {\cal H}_{\rho\sigma\tau}\ ,
\nn\w2
\delta \chi^a_i &= \frac12 \c^\mu \eps_i P_\mu^a - \frac{1}{24} \c^{\mu\nu\rho} \eps_i
\cH^a_{\mu\nu\rho}\ ,
\nn\w2
\delta\lambda^A_i &= - \frac14 \c^{\mu\nu} \eps_i \cF^A_{\mu\nu}  -\ft12 \theta^{IA} L_{I}\, \epsilon_i
-\ft12  m^{AB} C_{B}{}_{ij}\,\epsilon^j \ ,
\nn\w2
\delta\psi^r &= P_\mu^{ri} \gamma^\mu\epsilon_i\ ,
\label{t7-1}
\end{split}
\ee
where ${\cal H}_{\mu\nu\rho} := \cH_{\m\n\r}^I\,L_I$ and $\cH_{\m\n\r}^a := \cH_{\mu\nu\rho}^I\,L_I{}^a$. In establishing the supersymmetry of the action, it is important to recall that 
the duality equations $\cH_{\m\n\r}^+=0$ and $\cH_{\m\n\r}^{a-}=0$ are to be used after varying  action. 

To describe the gauge group, consider the case of $n_T=9$, from which all the lower magical theories can be obtained by truncation. In this case, there exists the 3-graded decomposition $so(9,1) \rightarrow N^{-}_{(8)} \oplus so(8) \oplus so(1,1)\oplus N^{+}_{(8)}$, where the subscripts refer to the $so(1,1)$ charges, and $N^\pm_{(8)}$ are nilpotent generators. Accordingly, splitting the spinor index of $so(9,1)$ as $A=(\alpha,t,0)$, where $\alpha=1,...,8, \ t=1,...,7$, in this basis the gauge algebra is found to be non-semisimple nilpotent with commutation rules taking the form
\be
[X_\alpha, X_\beta] = -g \gamma^t_{\alpha\beta} X_t\ ,\qquad [X_\alpha,X_t]=0=[X_t,X_u]\ ,\quad X_0=0\ .
\label{vtg}
\ee
The generators $X^t$ act as central extensions of the algebra which vanish when evaluated on vector and tensor fields but can act nontrivially on the hypermultiplet scalars. More specifically, $X_\alpha \phi^X = \theta_\alpha^{IJ} t_{IJ} \phi^X + \Theta_\alpha{}^\cA K_\cA^X$ and $X_t \phi^X = \Theta_t{}^\cA K_\cA^X$, where $\phi^X$ denotes the hypermultiplet scalars, and $K_\cA^X$ are the Killing vectors supported by the quaternionic K\"ahler manifold generating an algebra isomorphic to the one in \eq{vtg}. For various embeddings of this group into the isometries of the hyperscalar manifolds, and further details, see \cite{Gunaydin:2010fi}. 

In the case of generic couplings of Yang-Mills and multi-tensor multiplets, the Yang-Mills field equations $D_\mu F^{\mu\nu}=J^\mu$ displays a gauge anomaly in the form of $D_\mu J^\mu =\cA$. In the case of magic supergravities whose tensor-Yang-Mills content is fixed as given in table 3, one finds that $D_\mu J^\mu=0$ by using the gamma-matrix identity \eq{md}. This holds also in presence of the hypermultiplet couplings. Nonetheless, only in the ungauged theory, and by introducing $n_H= 273+n_V-29n_T$ hypermultiplets, and in the case of $n_T=2,3,5$ by employing Green-Schwarz-Sagnotti mechanism \cite{Green:1984sg,Sagnotti:1992qw} as well, that the gravitational anomaly can be removed. In presence of gaugings, gravitational, gauge and mixed anomalies are generically expected to arise at the quantum level. Their analysis and possible removal remains to be investigated.

\subsection{ Comments on other supergravities in $6D$ }

\subsection*{$(3,1)$ and $(4,0)$ exotic supergravities and anomalous $(2,1)$ supergravity}

In $6D$, in addition to the  $(2,2)$ Poincar\'e superalgebra, the $(3,1), (4,0)$ and $(2,1)$ versions also exist. In the first two cases, the lowest dimensional representations do not contain the graviton, and hence  the terminology of `exotic supergravities'. Such supergravities have been conjectured to exist and to describe the strong coupling limits of maximal supergravity in $5D$ \cite{Hull:2000zn,Hull:2000rr}. Evidence has been presented for an exceptional field theory which can possibly accommodate these theories, or the one with $(2,2)$ supersymmetry, depending on  solutions of the embedding constraints. However, nonlinear theories for exotic supergravities have not been constructed as yet.

In the case of $(3,1)$, the $R$-symmetry group is $USp(6)\times USp(2)$ and  the lowest dimensional representation has the field content 
\be
(3,1) :  \quad \{ C_{\m\n,\r},\ B_{\m\n +}^{ai},\ A_\m^{ab},\ \phi^{ab i} ;\ \psi_{\m +}^a,\ \psi_{\m\n -}^i,\ \chi_+^{ab},\ \chi_-^{ab i} \}\ ,
\ee
where $a=1,...,6$ and $i=1,2$ label the fundamentals of $USp(6)$ and $USp(2))$, the pairs of $USp(6)$ indices are symplectic traceless,and the exotic field has  the hook symmetry, $C_{\m\n,\r}=-C_{\n\m,\r},\ C_{[\m\n,\r]}=0$.  The two-form bosonic field, and  the exotic two-form fermion, have self-dual field strengths, and $\pm$ indicate  chiralities of the fermions.  The field equations for this multiplet exhibit curious features \cite{Bertrand:2020nob,Henneaux:2018rub}. 
%

In  the case of $(4,0)$,  the $R$-symmetry group is $USp(8)$ and the lowest dimensional representation has the field content 
\be
(4,0) :  \quad \{ C_{\m\n,\r\s},\ B_{\m\n +}^{ab},\ \phi^{abcd} ;\ \psi_{\m +}^a,\ 
\chi_{-}^{abc} \}\ , 
\ee
where $a=1,...,8$ labels the $USp(8)$ fundamental, the exotic bosonic field has the window symmetry, $C_{\m\n,\r\s}= C_{\r\s,\m\n}= -C_{\n\m,\r\s},\ C_{[\m\n,\r\s]}=0$. The  two-form boson and  spin $1/2$ fermions are in  the $27$ and $42$ dimensional representations of $Sp(4)$, respectively. As usual, $\pm$ indicate  chiralities of the  fermions.  The yet to be constructed interacting $(4,0)$ theory has been conjectured to have a global $E_{6(6)}$ symmetry with  scalars parametrizing the $E_{6(6)}/USp(8)$ coset\cite{Hull:2000zn}. Based on  its field content, the $(4,0)$ theory has gravitational anomalies, which, however, can be cancelled by adding $21$ tensor multiplets \cite{Seiberg:1988pf,DAuria:1997caz}. For further interesting aspects of $(4,0), 6D$ supergravity, see \cite{Henneaux:2017xsb,Gunaydin:2020mod}. 

Finally, let us recall that there also exists a $(2,1)$ multiplet with the $R$ symmetry group $Sp(2)\times Sp(1)$ that has the field content
\begin{align}
(2,1): && \{\ e_\m^a \ , && B_{\m\n +}\ , &&  B_{\m\n-}\ , && A_\m\ , && \phi\ , && \psi_{\m +}\ , && \psi_{\mu -}\ , && \chi_+\ , && \chi_-\ \}\ , 
\nn\\
 && (1,1)  &&  (1,1) &&  (4,1) &&   (4,2) &&  (5,1) && (1,2)&&  (4,2) &&  (5,2) &&  (4,1) && 
\end{align}
where $(p,q)$ in the second row denote the  $USp(6)\times USp(2)$ representation content. Supergravity with this field content has been studied in \cite{DAuria:1997caz,Bergshoeff:1986hv}. The five scalars parametrize the coset $SO(5,1)/SO(5)$\cite{Hawking:1981bu}, and the theory has a bosonic sector which is identical to that of $(1,0)$ supergravity coupled to eight vector multiplets and five tensor multiplets. These are referred to as {\it twin supergravities} \cite{Roest:2009sn}, as their bosonic sectors coincide, while their fermionic sectors differ. Both have gravitational anomalies \cite{Bergshoeff:1986hv} but while $(1,0)$ supergravity can be coupled to vector, tensor and hypermultiplets to cure its anomalies (see next section), $(2,1)$ supergravity does not have such matter multiplets available to remove its anomalies\footnote{Adding one gravitino multiplet of suitable chirality removes this anomaly, but  result is just the $(2,2)$ theory.}.

\section{D=5}

We shall summarize the results for maximal $N=8$ supergravity, half-maximal $N=4$ supergravity coupled to vector multiplets, and $N=2$ supergravity coupled to vector, tensor and hypermultiplets. In the last case, some of the vectors, the number of which depends on the gauged symmetries, are dualized to two-form potentials. The gaugings in the embedding tensor formalism will be summarized for $N=8,4$. The scalar manifold geometries involved are tabulated in appendix C. 

\subsection{$N=8$ gauged supergravity in $5D$}

Maximal supergravity $5D$ was constructed long ago in \cite{Cremmer:1979uq}, and certain gaugings of it were obtained in \cite{Pernici:1985ju,Gunaydin:1984qu}. Using the embedding tensor formalism, the most general gaugings were constructed in \cite{deWit:2004nw}, and it is this work which we summarize briefly below. In the embedding formalism framework, which brings in a tensor hierarchy, the maximal gauged supergravity is built out of the  following multiplet of fields 
\be
\big(\, e_\m{}^m, \cV_M{}^{ij}, A_\m^M, B_{\m\n\,M} ; \psi_\m^i, \chi^{ijk}\,\big)\ ,
\ee
where $i=1,...,8$ labels the fundamental representation of  the $R$-symmetry group $USp(8)$, $\cV_M{}^{ij}$ is the $E_{6(6)}/USp(8)$ coset representative, and $M=1,...27$ labels the  fundamental representation of $E_{6(6)}$. The spinors are symplectic-Majorana, $\chi^{ijk}=\chi^{[ijk]}$ and $\Omega_{ij}\chi^{ijk}=0$, thus in  the $48$-plet of $USp(8)$. The two-form potential $B_{\m\n\,M}$, in ${\bf 27}$ of $E_{6(6)}$, is introduced as on-shell dual of the vector fields $A_\m^M$. Their covariantized field strengths will be related to each or via a duality equation, which arises, under a projection, as an equation of motion. 
All of the $27$ gauge fields, or a subset of them, may be used to gauge a suitable subgroup of $E_{6(6)}$. This gauging is encoded in a real embedding tensor $\theta_M{}^\alpha$ which determines the  gauge group $G_0$ among the $E_{6(6)}$ generators $t_\alpha$ with $\alpha=1,...,78$, as
\be
X_M = \theta_M{}^\alpha t_\alpha \ .
\ee
Thus, the covariant derivatives are given by $D_\m=\nabla_\m -g A_\m^M X_M$. Supersymmetry requirement imposes a linear constraint on the embedding tensor such that in the product ${\bf 27}\otimes {\bf 78}$, only  the representations ${\bf 351}$ survives. This implies the conditions
\be
t_{\alpha M}{}^N \theta_N{}^\alpha =0\ ,\qquad \left(t_\b t_\a \right)_M{}^N \theta_N{}^\b =0\ .
\ee
The closure of  gauge algebra, on the other hand, imposes the quadratic constraints
\be
f_{\beta\gamma}{}^\alpha \theta_M{}^\b \theta_N{}^\gamma + t_{\beta N}{}^P\theta_M{}^\beta \theta_P{}^\alpha=0\ .
\ee
As a consequence, it is shown in \cite{deWit:2004nw} that
\be
X_{(MN)}{}^P =d_{MNQ} Z^{PQ}\ ,\qquad X_{[MN]}{}^Q= 10 d_{MQS}\, d_{NRT}\, d^{PQR}\, Z^{ST}\ ,
\ee
where $d_{MNP}$ is the totally symmetric invariant tensor of $E_6$ and $Z_{MN} = Z_{[MN]}$ is defined by these relations.
In  the scalar field sector, the $E_{6(6)}$ valued Maurer-Cartan decomposes as
\be
\cV_{ij}{}^M \Big( \partial_\m \cV_M{}^{k\ell} -g A_\m{}^P X_{PM}{}^N \cV_N{}^{k\ell} \Big)  = P_{\m ij}{}^{k\ell} + 2Q_{\m[i}{}^{[k} \delta_{j]}^{\ell]}\ .
\ee
In the conventions of \cite{deWit:2004nw}, $[X_m,X_N] - X_{MN}{}^P X_P$. The bosonic Lagrangian is given by \cite{deWit:2004nw}
\bea
e^{-1} \cL &=& -\frac12 R -\frac{1}{16} \cF_{\m\n}^{ij} \cF^{\m\n}_{ij} -\frac{1}{12} |P_\m^{ijk\ell}|^2 
+ 3g^2|A_1^{ij}|^2 -\frac13 g^2 |A_2^{i,jk\ell}|^2 
\nn\w2
&& + e^{-1}\cL_{\rm top}\ ,
\eea
where 
\be
\cF_{\m\n}^{ij} = \cV_M^{ij} \left( 2\partial_{[\mu} A_{\nu]}^M +g X_{[NP]}{}^M A_\mu{}^N A_\nu{}^P  + g Z^{MN} B_{\mu\nu N}\right)\ ,
\ee
and  $ (A_1^{ij}, A_2^{ijk\ell} )$ are defined by%
\bea
 X_{MN}{}^P \cV_P{}^{[k\ell} \cV^{mn] N} \cV_{ij}{}^M &=& 12A_2^{q,[k\ell m}\delta^{n]}{}_{[i} \Omega_{j]q} +9 A_2^{p,q[k\ell}\Omega^{mn]} \Omega_{p[i} \Omega_{j]q}\ ,
\nn\w2
X_{MN}{}^P \cV_{P im} \cV^{jm\,N} \cV^{k\ell\,M} &=& 3\Omega_{im} A_2^{(m,j)k\ell} +3\Omega_{im}\big(\Omega^{m[k} A_1^{\ell]j} +\Omega^{j[k} A_1^{\ell]m}
\nn\w2
&& +\frac14 \Omega^{k\ell} A_1^{mj} \big)\ ,
\eea
and $\cL_{\rm top}$, which has a complicated form but simple general variation, can be found explicitly in \cite{deWit:2004nw}. Using that variation, it is straightforward to find that the field equation of $B_{\m\n M}$ is, modulo the fermion terms, the (projected) duality equation \cite{deWit:2004nw}
\be
Z^{MN} \Big( \cH^{\m\n\r}{}_ N - \frac{i}{\sqrt 5} \ve^{\m\n\r\s\tau}  \cM_{NP} \cF_{\s\t}^P \Big)=0\ ,
\ee
where the `internal' metric $\cM_{MN} =\cV_M{}^{ij} \cV_{N ij}$, and the field strength $\cH_{\mu\nu\rho M}$ is defined by $3 D_{[\mu} \cF_{\nu\rho]}^M= g Z^{MN} \cH_{\mu\nu\rho N}$.

Finally, the  supertransformations are given by \cite{deWit:2004nw} 
\bea
\delta \psi_\m^i &=& D_\m \e^i -\frac{1}{12} i\left(\gamma_\m{}^{\n\r} -4\delta_\m^\n \gamma^\r \right) \cF_{\n\r}{}^{ij} \e_j +ig A_1{}^{ij} \e_j\ ,
\nn\w2
\delta\chi^{ijk} &=& -\frac12 i \gamma^\m P_\m^{ijk\ell} \e_\ell -\frac{3}{16}\gamma^{\m\n} \left( \cF_{\m\n}{}^{[ij} \e^{k]} +\frac13 \Omega^{[ij} \cF_{\m\n}{}^{k]\ell}\e_\ell \right) -g A_2^{\ell,ijk} \e_\ell\ , 
\eea
where $D_\m \e^i = \left( \partial_\m\delta^i_j -\frac14 \omega_{\m ab} \gamma_{ab}\delta^i_j -Q_{\m j}{}^i \right)\e^j$. 

\subsection{$N=6$ supergravity in $5D$}

The field content of $N=6$ supergravity is 
\be
\{ e_\m^r,\  A_\m^{ij}, A_\m, \cV_\a{}^{ij};\ \psi_\m^i, \chi^i, \chi^{ijk} \}\ ,
\ee
where $i=1,...,6$ is an $USp(6)$ index, the fermions are symplectic Majorana, $A_\m^{ij}$ and $\chi^{ijk}$ are in ${\bf 14}$ and ${\bf 14'}$ representations of $USp(6)$ and the scalars parametrize the coset $SU^\star(6)/USp(6)$ \cite{Cremmer:1983kc}. Ungauged or gauged, the Lagrangians and supersymmetry transformations of the $N=6$ theory have not been spelled out entirely in the literature so far. However, it is straightforward to obtain them from a consistent reduction of the $N=8$ theory, which requires the removal of a single gravitino multiplet. 

The $N=6, 5D$ supergravity has the same bosonic sector as that of $N=2, 5D$ supergravity, where the coset space parametrized by the scalars, $SU^\star(6)/USp(6)$, arises as one of the very special real (VSR) manifolds in the magic family. As such they are referred to as ``twins". We shall say more on this in section 9.2. For various other aspects of the $N=6, 5D$ supergravity, see \cite{Ferrara:1998zt}.

\subsection{$N=4$ supergravity coupled to vector multiplets in $5D$}

We combine the $N=4$ supergravity multiplet with $n$ copies of the vector multiplet $(A_\m, 5 \phi; \lambda)$, and introduce the notation 
\be
\{ e_\m^r,\  A_\m^M, \phi,\ \cV_\a{}^A;\ \psi_\m^i, \chi^i, \lambda^{ia} \}\ ,\quad A_\m^M=\{ A_\m^0, A_\m^\a\}\ ,  
\ee
where $\a =1,...,n+5\ , a=1,...,n$ and $\cV_\a{}^A= \big(\cV_\a{}^m, \cV_\a{}^a\big)$ with $m=1,...,5$, is a representative of the coset $SO(n,5)/SO(n)\times SO(5)$ parametrized by the scalar fields, and $\phi$ is the dilaton. The vectors $( A_\m^0, A_\m^m)$ belong to the supergravity multiplet. Note that the abelian vector gauge fields form one vector  $A_\mu^\alpha$ and one singlet $A_\mu^0$ under $SO(n,5)$. The index $i=1,...,4$ labels the fundamental representation of $USp(4)_R \approx SO(5)$, and the fermions are symplectic-Majorana. The undeformed theory has the global symmetry $SO(1,1)\times SO(n,5)$. The gauging of (on-shell) trombone symmetry will not be considered here. 
For gauging of such symmetries, see \cite{LeDiffon:2008sh}. 

Pure $N=4$ supergravity was constructed by Noether procedure in \cite{Romans:1985tw} where a subset of the vector fields were used to gauge $SU(2)\times U(1)$. Coupling $n$ vector multiplets to supergravity, in which a particular $SU(2)$ is gauged, was obtained in \cite{Awada:1985ep}. A more general gauging with the gauge group taking the form $K_A \times K_S$ where $K_A$ is an abelian and $K_S$ is a semisimple subgroup of $SO(n,5)$ was achieved in \cite{DallAgata:2001wgl}, where it was shown that $K_A$ has to be $SO(2)$ or $SO(1,1)$ gauged by $a_\m$. These results were generalized further in \cite{Schon:2006kz} to account for most general gaugings, using the embedding tensor formalism, which we summarize next. 
The $SO(5,n)$ invariant tensor $\eta={\rm diag} (-1,-1,-1,-1,-1,+1,...,+1)$ and the positive definite scalar matrix $\cM$ are given by
\be
\eta_{\a\b} =- \cV_\a{}^m \cV_\b{}^m + \cV_\a{}^a \cV_\b{}^a\ ,\qquad \cM_{\a\b} = \cV_\a{}^m \cV_\b{}^m + \cV_\a{}^a \cV_\b{}^a\ ,
\ee
We also define $\cV_\a^{ij} \equiv \cV_\a{}^m \left(\Gamma_m \right)^{ij}$ where $\Gamma_m^{ij}=\Gamma_m^{[ij]}$ are the $SO(5)$ gamma-matrices\footnote{The conventions are:  $\Omega^{ij}\Omega_{jk}=-\delta^i_k$, and $X^i= \Omega^{ik} X_k$ and $X_i= X^k \Omega_{ki}$.}. Taking into account the fact that $A_\m^\a$ and $A_\m^0$ have $SO(1,1)$ charges $1/2$ and $-1$ respectively, the gauge group generators $(X_M)_N{}^P$ in the covariant derivative $D_\m=\nabla_\m +g A_\m^M X_M$, are parametrized as \cite{Schon:2006kz}
\be
X_{\a\b}{}^\gamma=-f_{\a\b}{}^\gamma -\frac12 \eta_{\a\b} \xi^\gamma +\delta_{[a}^\gamma \xi_{\b]}\ ,\qquad X_{\alpha 0}{}^0 =\xi_\a\ ,\qquad X_{0 \alpha}{}^\b = \xi_\a{}^\b\ .
\ee
Thus the general gauging  is parametrized by three real tensors, $f_{\a\b\gamma}, \xi_{\a\b}$ and $\xi_\a$. The closure of the gauge algebra imposes the following quadratic constraints \cite{Schon:2006kz}
\begin{align}
\x_\a \xi^\a & = 0\ , &  \xi_{\a\b} \xi^\b &=0\ , & f_{\a\b\gamma} \xi^\gamma=0\ ,
\nn\w2
3 f_{\kappa [\a\b} f_{\gamma\delta]}{}^\kappa &= 2 f_{[\a\b\gamma} \xi_{\delta]}\ , & \xi_\a{}^\kappa f_{\kappa \b\gamma} &= \xi_\a \xi_{\b\gamma} -\xi_{[\b} \xi_{\gamma]\a}\ .
\end{align}
The covariant field strengths are given by
\bea
\cF_{\m\n}^M &=& 2\partial_{[\m} A_{\n]}^M + g X_{NP}{}^M A_\m^N A_\n{}^P + g Z^{MN} B_{\m\n N}\ ,
\w2
Z^{MN} \cH_{\m\n\r N} &=& Z^{MN} \Big[ 3 D_{[\m} B_{\n\r]N} +6 gd_{NPQ} A_{[\m}^P \big( \partial_\n A_\r^Q\big) +\frac13 g X_{RS}{}^QA_\n^R A_{\r]}^S \big) \Big]\ ,
\nn
\eea
where $d_{MNP}= d_{(MNP)}$ and $Z_{MN}=Z_{[AB]}$ are defined by
\bea
&& d_{0\a\b}=d_{\a 0 \b}=d_{\a\b 0}= \eta_{\a\b}\ ,\qquad \mbox{all other components zero}\ ,
\nn\w2
&& Z^{\a\b} =\frac12 \xi^{\a\b}\ , \qquad Z^{0\a}=-Z^{\a 0}= \frac12 \zeta^\a\ .
\eea
We will also need the coset current and composite $SO(5) \approx USp(4)$ connection
\be
\cP_\m^{a ij} = \cV^{\alpha a} D_\m \cV_\alpha{}^{ij}\ ,\qquad \cQ_{\m i}{}^j= 2 \cV^\a_{ik} D_\m \cV_\a{}^{jk}\ .
\ee
We are now ready to give the Lagrangian \cite{Schon:2006kz}
\bea
e^{-1} \cL &=& \frac12 R- \frac14 e^{-\phi} \cM_{\a\b} \cF_{\m\n}^\a \cF^{\m\n \b} -\frac14  e^{2\phi} \cF_{\m\n}^0 \cF^{0 \m\n} -\frac38 \partial_\m\phi \partial^\m\phi + \cP_\m^{a ij} \cP^\m_{a ij} 
\nn\w2
&& -4 \left( A_1^{ij} A_{1 ij} -A_2^{ ij} A_{2 ij} -A_2^{a ij} A_{2 ij}^a \right)  +e^{-1}\cL_{\rm top}\ ,
\eea
where the shift functions are given by
\bea
A_1^{ij} &=& \frac{1}{\sqrt6} \big(-\zeta^{ij}+2\r^{ij}\big)\ ,\qquad A_2^{ij} = \frac{1}{\sqrt6} \big(\zeta^{ij} +\r^{ij}+\frac32 \tau^{ij}\big)\ ,
\nn\w2
A_2^{a ij} &=& \frac12 \big( -\zeta^{aij} +\r^{a ij} -\frac{\sqrt2}{4} \tau^a \Omega^{ij}\big)\ ,
\eea
with the scalar field dependent tensors defined as
\bea
\tau^{ij} &=& e^{\phi/2} \cV_\a{}^{ij} \zeta^\a\ , \qquad \tau^a = e^{\phi/2} \cV_\a{}^a \zeta^\a\ , 
\nn\w2
\zeta^{ij} &=& \sqrt2 e^{-\phi} \cV_\a{}^i{}_k \cV_\b{}^{jk} \zeta^{\a\b}\ ,
\qquad \zeta^{a ij} = e^{\phi} \cV_\a{}^a \cV_\b{}^{ij} \zeta^{\a\b}\ , 
\w2
\r^{ij} &=& -\frac23 e^{\phi/2} \cV_\a^{ik} \cV_\b^{j\ell} \cV^\gamma{}_{k\ell} f^{\a\b}{}_\gamma\ , \quad \r^{aij} = \sqrt2 e^{\phi/2} \cV_\a{}^a \cV_\beta{}^i{}_k \cV_\gamma{}^{jk} f^{\a\b\gamma}\ .
\nn
\eea
The tensors $(\zeta^{ij}, \r^{ij}, \r^{aij} )$ are symmetric, and  $(\lambda^{ij}, \zeta^{a ij})$ are anti-symmetric, and therefore  $A_1^{ij}=A_1^{(ij)}$. The topological Lagrangian $\cL_{\rm top}$ is given in \cite{Schon:2006kz} where its general variation is also provided,
\be
\delta \cL_{\rm top} = \frac{1}{4\sqrt2} e^{\m\n\r\s\lambda} \Big( \frac13 Z^{MN} \cH_{\m\n\r M} \Delta B_{\s\lambda N} + d_{MNP} \cF_{\m\n}^M \cF_{\r\s}^N \delta A_\lambda^P \Big)\ ,
\ee
up to total derivatives, and with $\Delta B_{\m\n M} \equiv Z^{MN} \left( \delta B_{\m\n M} -2d_{NPQ} A_{[\m}^P \delta A_{\n]}^Q \right)$. It follows that the equation of motion for the two-form field is the (projected) duality equation
\be
Z^{MN} \left( \frac{1}{6\sqrt2} \e_{\m\n\r\s\lambda} \cH^{\r\s\lambda}{}_N -\cM_{NP} \cF_{\m\n}^P \right)=0\ ,
\ee
where $\cM_{MN} \equiv {\rm diag} \left( e^{-4\s}, e^{2\s} M_{\a\b} \right)$. The Yukawa couplings and the following supertransformations are given by \cite{Schon:2006kz}:
\begin{align}
\delta \psi_\m^i &= D_\m\e^i +\frac{i}{6} \left( e^{-\phi/2} \cV_\a{}^{ij}\cF_{\n\r}^\a +\frac{\sqrt2}{4} \Omega^{ij} \cF_{\n\rh}^0 \right) \left(\gamma_\m{}^{\n\r} -4\delta_\m^\n\gamma^\r \right) \e_j -\frac{1}{\sqrt6} A_1^{ij} \gamma_\m \e_j\ ,
\nn\w2
\delta\chi^i &= -\frac{\sqrt3 i}{4} \partial_\m \phi \gamma^\m \e^i +\frac{\sqrt3}{6} \big( e^{-\phi/2} \cV_M{}^{ij} \cF_{\m\n}^M -e^\phi \Omega^{ij} \cF_{\m\n}^0\big) \gamma^{\m\n}\e_j -\sqrt2 A_2^{ij}\e_j\ ,
\nn\w2
\delta\lambda^{ia} &= \cP_\m^{a ij} \gamma^\m \e_j -\frac14 e^{-\phi/2} \cV_M{}^a \cF_{\m\n}^M \gamma^{\m\n} \e^i +A_2^{aij} \e_j\ ,
\end{align}
where $D_\m \e^i = D_\m (\omega, \cQ)\e^i$.

\subsection{$N=2$ supergravity coupled to vector, tensor and hypermultiplets in $5D$ and very special real manifolds}

The  $N=2, 5D$ Poincar\'e superalgebra in $5D$ admits the  following multiplets\footnote{For the off-shell multiplets, see \cite{Bergshoeff:2001hc,Coomans:2012cf}.}
\be
\underbrace{ \left(e_\m^A, \psi_\m^i, A_\m \right)}_{\rm graviton}\ ,\qquad  \underbrace{\left(A_\m, \sigma,\ \lambda^i \right)}_{\rm vector} \qquad \underbrace{\left(B_{\m\n}, \varphi,\ \chi^i \right)}_{\rm tensor}\ ,\qquad \underbrace{\left(4\phi,\psi \right)}_{\rm hyper}\ .
\ee
The spinors are symplectic Majorana, $i=1,2$ labels  the doublet of the $R$ symmetry group $USp(2)$. General couplings of two-derivative $N=2$ supergravity in $5D$ coupled to $n_V$ vector multiplets, $n_T$ tensor multiplets and $n_H$ hypermultiplets, including  gaugings of possible isometries of the  scalar manifolds, based largely on previous work, e.g. \cite{Gunaydin:1983bi,Gunaydin:1999zx,Ceresole:2000jd}, was given completely in \cite{Bergshoeff:2004kh}, where a more complete list of references can be found. Here we shall follow the book \cite{Lauria:2020rhc} to give a brief account of these couplings. The conventions are largely those of \cite{Bergshoeff:2004kh}. 

To begin with,  we group together  field of the supergravity multiplet with those of $n_V$ vector multiplets and $n_T$ tensor multiplets and use the notation
\be
\{ e_\m^a, \psi_\m^i, A_\m^I, B_{\m\n}^r, \varphi^x ;\ \psi_\m^i, \lambda^{ix} \},\qquad \{\phi^X, \psi^A \}\ ,
\ee
where 
\bea
&& I=0,1,...,n_V,\qquad r=1,...,n_T,\qquad  x=1,...,n_V+n_T\ ,
\nn\\
&& X=1,...,4n_H\ , \qquad A=1,...,2n_H\ ,\qquad i=1,2\ .
\eea
The hyperscalars parametrize a quaternionic K\'ahler manifold $\cM_{QK}$, while  $n_V+n_T$ real scalars $\varphi^x$ parametrize a {\it very\, special\ real\, manifold} $\cM_{VSR}$. To define $\cM_{VSR}$, introduce $n_V+n_T+1$ scalars $h^\tI$ and impose the constraint
\be
C_{\tI\tJ\tK} h^\tI h^\tJ h^\tK =1\ ,\quad \tI=0,1,...,n_V+n_T\ ,
\label{VSR}
\ee
where $C_{\tI\tJ\tK}$ is a totally symmetric constant real tensor. The index $\tI$ is split as $\tI=(I,r)$ with $I=0,1,...,n_V$ and $r=n_{V+1},...,n_V+n_T$. The  following relations fields can then be defined
\bea
&& a_{\tI\tJ}= -2C_{\tI\tJ\tK} h^\tK +3 h_\tI h_\tJ\ , \qquad g_{xy}= a_{\tI\tJ}\,h^\tI_x h^\tJ_y\ ,
\nn\w2
&& h_\tI h^\tI =1\ ,\qquad  h_\tI = a_{\tI\tJ} h^J\ , \qquad h^\tI_x = -\sqrt{\frac32}\, h^\tI_{\ ,x}\ ,
\eea
where $a_{\tI\tJ}$ is the metric in the ambient space, and $g_{xy}$ is the metric on $\cM_{VSR}$. For several useful identities, see for example \cite{Gunaydin:1983bi,Bergshoeff:2004kh}. In particular, all cubic polynomials whose invariance group acts transitively on the corresponding real spaces have been classified in \cite{deWit:1991nm}. The symmetric VSR manifolds are listed in table 5, under $D=5, N=2$. 

Next, consider the  gauge transformations \cite{Lauria:2020rhc}
\bea
\delta A_\m^I &=& \del_\m\Lambda^I -\Lambda^J f_{JK}{}^I A_\m^K\ , \qquad \delta B_{\m\n}^r = -\Lambda^J t_{J\tI}{}^r H_{\m\n}^\tI\ ,
\nn\w2
\delta \varphi^X &=& \Lambda^I k_I{}^X\ ,\qquad \delta \phi^x =\Lambda^I k_I{}^x\ , \qquad k_I{}^x := \sqrt{\frac32}\, t_{I\tJ}{}^\tK \,h^\tJ h_\tK{}^x\ ,
\eea
where $\Lambda^I(x)$ is the gauge parameter,  $k_I{}^X (\phi)$ and $k_I{}^x(\varphi)$ are the Killing vectors on the manifolds parametrized by the scalars of the  vector and hypermultiplets, respectively, and they separately generate the gauge algebra with structure constants $f_{JK}{}^I$. Furthermore, the generators $t_I$ in the $(n_V+n_T+1)$ dimensional representation are given by
\be
(t_r)_\tJ{}^\tK =0\ , \qquad (t_I)_\tJ{}^\tK = \begin{pmatrix} f_{IJ}{}^K & t_{IJ}{}^r \\ 0 & t_{Ir}{}^s \end{pmatrix}\ ,
\label{r1}
\ee
obeying  gauge algebra $[t_I,t_J]= f_{IJ}{}^K t_K$. The following relations must hold \cite{Lauria:2020rhc}
\be
C_{r\tJ\tK}= \sqrt{\frac38} \, t_{\tJ\tK}{}^s\Omega_{sr}\ ,\qquad t_{I[r}{}^t \Omega_{s]t}=0\ ,\qquad t_{I(\tJ}{}^{\widetilde M} C_{{\widetilde K}{\widetilde L}]\widetilde M} =0\ .
\label{r2}
\ee
With these basic ingredients at hand, the bosonic part is given by \cite{Bergshoeff:2004kh}\footnote{In the absence of hypermultiplet couplings, and for $t_{IJ}{}^r=0$, this result agrees with that of \cite{Gunaydin:1999zx} where $t_{IJ}{}^r$ was set to zero.}
\bea
e^{-1} \cL &=& \frac12 R -\frac14 a_{\tI\tJ} H_{\m\n}^\tI H^{\tJ\m\n} -\frac12 g_{xy} D_\m\varphi^x D^\mu \varphi^y -\frac12 g_{XY} D_\m\phi^X D^\m\phi^Y -V
\nn\w2
&& +\frac{1}{16 g} \,\varepsilon^{\m\n\r\s\tau} B_{\m\n\,r} \Big( \del_\r B_{\s\tau}^r + t_{Is}{}^r A_\r^I B_{\s\tau}^s + 2\, t_{IJ}{}^r A_\m^I F_{\r\s\tau}^J \Big)
\nn\w2
&& +\frac{1}{6\sqrt6} \varepsilon^{\m\n\lambda\r\s} C_{IJK} A_\m^I 
\Big[ F_{\n\lambda}^J F_{\r\s}^K +f_{FG}{}^J A_\n^F A_\lambda^G \big( -\frac12 F_{\r\s}^K +\frac{1}{10} f_{HL}{}^K A_\r^H A_\s^L \big) \Big] 
\nn\w2
&& -\frac18 \varepsilon^{\m\n\lambda\r\s} \Omega_{rs} t_{IK}{}^r t_{FG}{}^s A_\m^I A_\n^F A_\lambda^G \Big( -\frac12 F_{\r\s}^K +\frac{1}{10} f_{HL}{}^K A_\r^H A_\s^L \Big)\ ,
\eea
where $\Omega_{rs}$ is the constant symplectic invariant matrix, we have set the coupling constant equal to one, and 
\bea
H_{\m\n}^\tI &=& \left( F_{\m\n}^I, B_{\m\n}^r \right)\ , \quad D_\m \varphi^x = \del_\m \varphi^x-A_\m^I k_I{}^x\ ,
\nn\w2
D_\m \phi^X &=& \del_\m \phi^X -A_\m^I k_I{}^X\ ,\qquad F_{\m\n}^I= 2\partial_{[\m} A_{\n]}^I + f_{JK}{}^I A_\m^J A_\n^K\ .
\eea
The potential is given by
\be
V= {\bf P}\cdot {\bf P} -\frac12 {\bf P}^x \cdot {\bf P}_x -2 W_x W^x -2 \cN^{iA} \cN_{iA}\ ,
\ee
where
\bea
&& W^x = -\frac{\sqrt6}{4} h^I k_I{}^x\ ,\ \ \cN^{iA} = -\frac{\sqrt6}{4}h^I k_I{}^X \cV_X^{iA}\ ,
\nn\w2
&& \ \ {\bf P} = h^I {\bf P}_I\ ,\ \ {\bf P}_x = h^I{}_x {\bf P}_I\ ,\ \  {\bf P}_I = \frac{1}{4n_H} {\bf J}_X{}^Y \nabla_Y k_I{}^X\ . 
\eea
Here ${\bf J}_X{}^Y$ are the three quaternionic K\"ahler structures, $\cV_X^{iA}$ is the vielbein on $\cM_{QK}$, and  the moment maps ${\bf P}_I$ are defined by the relation (see section 3.3.2 in \cite{Bergshoeff:2002qk} for a detailed discussion)
\be
\partial_X {\bf P}_I = -\frac12 {\bf J}_{XY} k_I{}^Y\ .
\ee
The supertransformations of the fermions are given by 
\bea
\delta \psi_\m^i &=& D_\m \e^i +\frac{i}{4\sqrt6} \Big[h_\tI H^{\tI \n\r} \left(\gamma_{\m\n\r}-4g_{\m\n}\gamma_\r \right) \e^i +2 P_j{}^i \gamma_\m \e^j \Big]\ ,
\nn\w2
\delta \lambda^{xi} &=& -\frac{i}{2} \gamma^\m D_\m\varphi^x \e^i +\frac14 \gamma\cdot H^\tI h_\tI{}^x \e^i +P^x{}_j{}^i \e^j +\frac12 W^x \e^i\ ,
\nn\w2
\delta \psi^A &=& \frac{i}{2} \gamma^\m D_\m \phi^X V_X^{iA} \e_i-\cN^{iA} \e_i\ ,
\eea
where
\be
D_\m \e^i = \nabla_\m \e^i +\del_\m \phi^X \omega_{X j}{}^i \e^j +\frac12 A_\m^I P_{Ij}{}^i \e^j\ ,
\ee
with the $USp(2)$ connection $\omega_{X j}{}^i$ and the definition $ P_{Ii}{}^j := {\bf P}_I \cdot ({\bm \sigma})_j{}^i$ used. The specifics of the gauge group $K\subset G$, and its representations carried by the tensor multiplets, of course, depend on the solutions of the  constraints \eq{r1} and \eq{r2}. In the case of last four entries for $5D, N=2$ in the table provided in appendix B, the following gauge groups $K$, and their irreps carried by the $n_T$ tensors, are given in \cite{Gunaydin:1999zx}:
\begin{align}
 & K= SO^\star (6) \subset E_{6(-26)}\ , & &n_T= 6+6\ ,
 \nn\w2
 &  K= SU(3)\times U(1)_R \subset E_{6(-26)}\ , & &n_T= 3+\bar3 +3 + \bar3 + 3 + \bar3\ ,
 \nn\w2
 & K= SO^\star(6) \subset SU^\star (6) \ , & &n_T= 0\ ,
 \nn\w2
 & K= SU(3)\times U(1)_R  \subset SU^\star (6) \ , & &n_T= 3+\bar3\ ,
\nn\w2
& K= SU(3)\times U(1)_R  \subset SL(3, \mathbb{C})\ , & &n_T= 0\ ,
\nn\w2
& K= SU(2)\times U(1)\times U(1)_R  \subset SL(3, \mathbb{C})\ , & &n_T= 2+\bar2\ ,
\nn\w2
& K= SL(2,\mathbb{R})\times U(1)_R  \subset SL(3, \mathbb{C})\ , & &n_T= 2\ .
\end{align}
For other gaugings, including the gauging of the full $R$-symmetry group $SU(2)_R$, see \cite{Gunaydin:1999zx,Gunaydin:2000ph}. It is natural to expect that these results, as well as a systematic search for all possibilities, are best obtained by employing the embedding tensor formalism.

Finally, it is worth noting that dimensional reduction chain $5D \to 4D \to 3D$ applied to the $N=2, 5D$ supergravity coupled to vector multiplets, with very special real (VSR) scalar manifold, and referred to as $r$-maps \cite{deWit:1992cr} and $c$-maps \cite{Cecotti:1988qn}, produce $4D, N=2$ supergravity with very special K\"ahler (VSK), and $N=4, 4D$ supergravity with special quaternionic K\"ahler (SQK) manifolds, as can be seen in the tables in appendix C:
\be
VSR\ (5D)\ \xrightarrow{\rm r-map}\ VSK\ (4D)\ \xrightarrow{\rm c-map}\ SQK\  (3D)\ .
\ee
See, for example, the books \cite{Freedman:2012zz,Lauria:2020rhc} for details, and further references. 

\section{D=4}

Gauged supergravities in $4D$ have been reviewed extensively in \cite{Trigiante:2016mnt}, and a full treatment of $N=4, 4D$ supergravity has appeared recently in \cite{DallAgata:2023ahj}. Their scalar manifolds and duality symmetries are tabulated in appendix C. For an earlier review of $N=8$ supergravity, see \cite{deWit:2002vz}. The case of $N=1,2$ has been treated in great detail in the book \cite{Freedman:2012zz}. For completeness, we shall give a brief survey of these supergravities.  

\subsection{ $N=8$ gauged supergravity in $4D$}

The  maximal supergravity in $4D$ was constructed long ago in \cite{Cremmer:1979up}. As is well known, this theory has on-shell global $E_{7(7)}$ symmetry. The gauging of $SU(8)\subset E_{7(7)}$ was achieved in \cite{deWit:1981sst}. A systematic analysis of electric gaugings was provided in \cite{Cordaro:1998tx}, and a formulation which accommodates magnetic duals and the attendant more general gaugings were given in an embedding tensor formalism in \cite{deWit:2007kvg}. We shall review the last construction here. In this framework, the $N=8$ supergravity multiplet of fields, together with the two-form fields that are on-shell related to the scalars, are  
\be
\big( e_\m{}^m, \cV_M{}^A, A_\m^M, B_{\m\n\,\a} ; \psi_\m^i, \chi^{ijk}  \big)\ ,
\ee
where $i=1,...,8$ labels  the fundamental of the  $R$-symmetry group $SU(8)$, $M=1,...,56$  labels the  fundamental representation of $E_{7(7)}$, and $\cV_M{}^A= (\cV_M{}^{ij}, \cV_{M ij} )$, with antisymmetry in $ij$, is the vielbein on the coset $E_{7(7)}/SU(8)$. The vector fields consist of electric and magnetic potentials, $A_\m^M =(A_\m^\Lambda, A_{\m\n \Lambda})$, with no raising and lowering of the index $\Lambda=1,...,28$. The spinors are Majorana, and $\chi^{ijk}=\chi^{[ijk]}$, thus in the  $56$-plet of $SU(8)$. The two-form potential $B_{\m\n\,\a}$ is in  the adjoint representation of $E_{7(7)}$. Their properly covariantized field strengths will be related to the scalar current by a duality relation, which arises, under a projection, as an equation of motion. 
All of the $56$ gauge fields, or a subset of them, may be used gauge a suitable subgroup of $E_{7(7)}$. This gauging is encoded in a real embedding tensor $\theta_M{}^\alpha$ which determines the  gauge group $G_0$ among  $E_{7(7)}$ generators $t_\alpha$ with $\alpha=1,...,133$, as
\be
X_M = \theta_M{}^\alpha t_\alpha \ .
\ee
Thus,  the covariant derivatives are given by $D_\m=\nabla_\m -g A_\m^M X_M$. Supersymmetry requirement imposes a linear constraint on the embedding tensor such that in  the product ${\bf 56}\otimes {\bf 133}$, only the representation ${\bf 912}$ survives. This implies  the conditions
\be
t_{\alpha M}{}^N \theta_N{}^\alpha =0\ ,\qquad \left(t_\b t_\a \right)_M{}^N \theta_N{}^\b = -\frac12 \theta_M{}^\a\ .
\label{LC1}
\ee
The closure of the gauge algebra, on the other hand, imposes the quadratic constraints equivalent to 
\be
\Omega^{MN} \theta_M{}^\alpha \theta_N{}^\beta=0\ ,
\ee 
provided that the constraints \eq{LC1} are used. Here $\Omega^{IJ}$ is the $E_7$ invariant constant tensor satisfying $\Omega^{MN}\Omega_{NP}= -\delta^M_P$. 
There are only two $N=8$ supersymmetric solutions of the full set of constraints on the embedding tensor. One of them is the $SU(8)$ gauged supergravity constructed long ago \cite{deWit:1981sst}, and the other one is the the so-called $\omega$-deformed $N=8$ supergravity which was realized in \cite{DallAgata:2012mfj}.
We are not aware of a complete classification of all solutions to the embedding constraints. However, all electric gaugings have been found, albeit by different methods, in \cite{Hull:1984qz,Cordaro:1998tx}.

The scalar current $P_{\m ijk\ell}$ and the  composite $SU(8)$ connection are defined as
\bea
P_{\m ijk\ell} &=& \cV_{ij}{}^M \Big( \partial_\m \cV_M{}^{k\ell} -g A_\m{}^P X_{PM}{}^N \cV_{N\,k\ell} \Big)   \ ,
\nn\w2
Q_{\m\,i}{}^j  &=& \frac23 i \cV_{ik}{}^M \Big( \partial_\m \cV_M{}^{jk} -g A_\m{}^P X_{PM}{}^N \cV_N{}^{kj} \Big) \ ,
\eea
and the bosonic part of the gauged maximal supergravity Lagrangian is given by
\bea
e^{-1} \cL &=& -\frac12 R -\frac14  \Big( i\cN_{\Lambda\Sigma} \cF^+_{\m\n}{}^\Lambda \cF^{+\m\n \Sigma} +h.c. \Big) -\frac{1}{12} |P_\m^{ijk\ell}|^2 
\nn\w2
&& +\frac34 g^2|A_1^{ij}|^2 -\frac{1}{24} g^2 |A_{2\,i}{}^{jk\ell}|^2 + e^{-1}\cL_{\rm top}\ ,
\eea
where a $56$-plet of $E_7$ is decomposed as $V^M =(V^\Lambda, V_\Lambda)$ with no raising and lowering of the index $\Lambda=1,...,28$, and
\be
\cF_{\m\n}^M =  2\partial_{[\mu} A_{\nu]}^M +g X_{[NP]}{}^M A_\mu{}^N A_\nu{}^P +g Z^{M\alpha} B_\alpha\ ,\qquad Z^{M\alpha} := \frac12 \Omega^{MN} \theta_N{}^\alpha\ ,
\ee
The complex (anti)self dual projections $F_{\mu\nu}^\pm$ are normalized such that $F_{\mu\nu} = F_{\mu\nu}^+ + F_{\mu\nu}^-$. Furthermore $A_1^{ij}, A_2^{ijk\ell},\cN_{\Lambda\Sigma}$ are defined by
\bea
 && \cV^N{}_{kn} X_{MN}{}^P \cV_P{}^{\ell n} \cV^{M ij} = -3A_1^{\ell[i} \delta_k^{j]} -\frac32 A_{2\,k}{}^{\ell ij}\ ,
 \nn\w2
 && \cV^{\Sigma ij} \cN_{\Lambda\Sigma} = -\cV_\Lambda{}^{ij}\ ,
\eea
with $A_1^{ij} = A^{[ij]}, A_{2\,\ell}{}^{ijk} = A_{2\,\ell}{}^{[ijk]}$ and $A_{2\,k}{}^{ijk}=0$. As to the Lagrangian $\cL_{\rm top}$, it has a complicated form but a simple general variation \cite{deWit:2007kvg},
\be
e^{-1} \delta\cL_{\rm top} = i\cF^{+\m\n\Lambda} D_\m \delta A_{\n \Lambda} +\frac14 ig \cF^{+\m\n}{}_\Lambda \theta^{\lambda\a} \Delta B_{\m\n \Lambda} + h.c.\ ,
\label{4Dtop}
\ee
where the covariant variation $\Delta B_{\m\n\Lambda} = \delta B_{\m\n\Lambda} -2(t^\a)_M{}^P\Omega_{NP} A_\m^M \delta A_\n^N$. The two-form field is related to the scalar fields by the (projected) duality equation that follows from the Lagrangian as a field equation given by \cite{deWit:2007kvg} 
\be
 \theta^{\Lambda \a} \ve^{\m\n\r\s} \cH_{\n\r\s \a} = 2i P^\Lambda{}_{ijk\ell} P^{\m ijk\ell}\ ,
 \label{4DDE}
\ee
 where $P^\Lambda{}_{ijk\ell} =i\cV^M{}_{ij} X^\Lambda{}_{MN} \cV^N{}_{k\ell}$, and the field strength $\cH_{\mu\nu\rho \alpha}$ is defined by  $3 D_{[\mu} \cF_{\nu\rho]}^M= g Z^{M\alpha} H_{\mu\nu\rho \alpha}$.
Finally, the supertransformations of the  fermions are
\bea
\delta \psi_\m^i &=& 2 D_\m \e^i -\frac{1}{2\sqrt{2}} \cF_{\r\s}{}^{ij} \gamma^{\r\s}\gamma_\m \e_j +{\sqrt 2} g A_1^{ij} \gamma_\m \e_j\ ,
\nn\w2
\delta\chi^{ijk} &=&- 2{\sqrt 2} P_\m^{ijk\ell} \gamma^\m  \e_\ell +\frac32 \cF^-_{\m\n}{}^{[ij} \gamma^{\m\n} \e^{k]} -2 g A_{\,\ell}{}^{ijk} \e^\ell\ , 
\eea
where $D_\m \e^i = \left( \partial_\m\delta^i_j -\frac14 \omega_\m{}^{ab} \gamma_{ab}\delta^i_j +\frac12 Q_\m{}^i{}_j \right)\e^j$.

\subsection{ $N=6$ supergravity in $4D$}

The ungauged $N=6$ theory can be obtained as a consistent truncation of maximal supergravity by branching $E_{7(7)}\to SO^\star(12)$ and $SU(8)\to U(6)\times SU(2)$, and retaining only the $SU(2)$ singlets \cite{Andrianopoli:2008ea,Roest:2009sn,Trigiante:2016mnt}. The resulting field content, consisting of $64_B+64_F$ degrees of freedom, is given by
\be
\big( e_\m{}^m, \cV_M{}^A, A_\m^M, B_{\m\n \alpha} ; \psi_{\m i}, \chi_{ijk}, \chi_i \big)\ ,
\ee
where $i=1,...,6$ labels  the fundamental representation of the  $R$-symmetry group $SU(5)\times U(1)$, $M,A=1,...32$  labels the  ${\bf 32_c}$ of $SO^\star(12)$, and $\cV_M{}^A$ is  the $G/H=SO^\star(12)/U(6)$ coset vielbein. 
The vectors consist of $16$ electric, and $16$ magnetic ones, and the duality symmetry $G$ is realized on-shell. The spinors are Weyl, the gravitino has $U(1)$ charge $+\frac12$, the dilatinos $\chi_{ijk}=\chi_{[ijk]}$ and $\chi_i$ are in the ${\bf 20+6}$ of $H$ with the same chirality, and  $U(1)$ charges $+\frac32$ and $-\frac52$, respectively. The two-form potential $B_{\m\n \alpha}$ is in the adjoint representation of $SO^\star(12)$. Its properly covariantized field strength will be related to the scalar current by a suitable duality relation. 
Various gaugings were considered in \cite{Andrianopoli:2008ea}. The most general gauging will be encoded as usual in an embedding tensor $\theta_M{}^\a$, where $\a=1,...,66$. The linear constraint on the embedding tensor requires that in the product ${\bf 32_c}\times {\bf 66}$, only the representation ${\bf 352_s}$ survives \cite{Roest:2009sn,Trigiante:2016mnt}. As for the quadratic constraints on the embedding tensor, it turns out that they set to zero the representations ${\bf 66}+{\bf 2079}+ {\bf 462_s}$ in the symmetric product $\left({\bf 352_s} \times {\bf 352_s}\right)_s $ \cite{Roest:2009sn}. 

The $N=6$ supergravity has some exceptional properties that have been discussed in detail in \cite{Andrianopoli:2008ea}. One of them is the fact that the underlying superalgebra $OSp(6|4)\times SO(2)$ has zero Cartan-Killing form defined as $Str (ad_X ad_Y)$. Furthermore, the zero-center module of the\footnote{The zero-center module of a superalgebra is a representation characterized by the vanishing of all super-Casimir operators.} coincides with the supergravity multiplet. Another noteworthy property is that $N=6$ supergravity has the same bosonic sector (but different fermionic sector) and scalar coset space as one of the $N=2$ magical supergravities in $4D$, with the special $8$ vector and $5$ tensor multiplet couplings. (See section 7.5 for a discussion of the magical supergravities in $D=3,4,5,6$.) This is referred to as  a ``dual relation'' in \cite{Andrianopoli:2008ea}, and ``twins'' in \cite{Roest:2009sn}. The two theories can be obtained from different truncations of the $N=8$ supergravity. However, this is not the case for the gauged cases \cite{Roest:2009sn}. In particular, the gauged versions of these twin theories can have different potentials \cite{Roest:2009sn}. For more details on $N=6$ supergravity see \cite{Andrianopoli:2008ea,Roest:2009sn,Trigiante:2016mnt}.

\subsection{ $N=5$ supergravity in $4D$}

The  $N=5$ gauged supergravity does not have matter multiplets, and its field content, which has $32_B+32_F$ degrees of freedom, is given by
\be
\big( e_\m{}^m, \cV_M{}^A, A_\m^M, B_{\m\n\a} ; \psi_{\m i}, \chi_{ijk}, \chi \big)\ ,
\ee
where $i=1,...,5$ labels  the fundamental of the  $R$-symmetry group $U(5)$, $M,A=1,...20$ labels the  ${\bf 20}$ of $SU(5,1)$, and $\cV_M{}^A$ is the $G/H=SU(5,1)/U(5)$ coset vielbein. The two-form potentials are in the adjoint representation of $SU(5,1)$. The vectors consist of $10$ electric, and $10$ magnetic ones, and the duality symmetry $G$ is realized on-shell. The spinors are Weyl, the gravitini have $U(1)$ charge $+\frac12$, the dilatinos $\chi_{ijk}=\chi_{[ijk]}$ and $\chi$ have opposite chirality, and they are in the $10+1$ representations of $H$ with $U(1)$ charges $+\frac32$ and $-\frac52$, respectively.   The two-form potential $B_{\m\n \alpha}$ is in the adjoint representation of $SU(5,1)$. Its properly covariantized field strengths will be related to the scalar current by a suitable duality relation. 

The most general gauging is encoded in an embedding tensor $\theta_M{}^\a$, where $\a=1,...,35$, and as a result of the linear constraint on it, in the product ${\bf 20}\times {\bf 35}$, only the $SU(5,1)$ representation ${\bf 70} + {\bar 70}$ survives. They are described by constant tensors $\theta_{mn,p}=\theta_{[mn],p}$ satisfying $\theta_{[mn,p]}=0$, and the complex conjugate \cite{Trigiante:2016mnt}. Of course, these must also satisfy the standard quadratic constraints \eq{qc}. See \cite{Trigiante:2016mnt} for further details.

\subsection{ $N=4$ supergravity coupled to vector multiplets in $4D$}

We combine the $N=4$ supergravity multiplet with $n$ copies of the vector multiplet$(A_\m, 6 \phi; \lambda)$, and introduce the notation 
\be
\{ e_\m^r, \tau, \cV_M{}^A,  A_\m^{M\a}, B_{\m\n}{}^{MN}, B_{\m\n}{}^{\a\b}\, ;\ \psi_{\m i}, \chi_{ijk}, \lambda_{ia} \}\ ,\quad A_\m^M=\{ A_\m^0, A_\m^\a\}\ ,    
\ee
where $M =1,...,n+6$ and $\cV_M{}^A= \big(\cV_M{}^m, \cV_\a{}^a\big)$ with $m=1,...,6,  a=1,...,n$, is a representative of the coset $SO(n,6)/SO(n)\times SO(6)$, and $\tau$ is the complex scalar that parametrizes the coset $SL(2,\mR)/SO(2)$. The index $\a=1,2$ labels the fundamental of $SL(2,\mR)$, and can be decomposed into the electric $A_\m^{M+}$ and magnetic $A_\m^{M-}$ vector fields, where $\pm$ denote the $SO(2)\subset SL(2,\mR)$ charges. The two-form potential needed by the tensor hierarchy are in the adjoint representation of $SO(n+6)\times SL(2,\mR)$. In the fermion sector, the $SU(4) \approx SO(6)$ index $i=1,...,4$, and the fermions are Weyl and $\chi_{ijk}=\chi_{[ijk]}$. The $SO(2)$ charges of $(\psi_{\m i}, \chi_{ijk}, \lambda_{ia})$ are $(+\frac12, \frac32, -\frac12)$. The field equations give a projected duality equations that relate the field strengths of the electric and magnetic field strengths, as well as the field strengths of the two-form field s and appropriate scalar currents, as we have already seen in many gauged supergravities surveyed so far above. Therefore the duality symmetry is on-shell.

The $N=4, 4D$ supergravity coupled to vector multiplets and its certain gaugings were  constructed long ago \cite{Bergshoeff:1985ms,deRoo:1985jh}, but general general gauging has been achieved in the in the embedding tensor formalism in \cite{Schon:2006kz}, and more recently in full generality in \cite{DallAgata:2023ahj}, where an extensive literature on previous works can be found. See also the review \cite{Trigiante:2016mnt}. A key result in the embedding tensor formalism is the characterization of the gauge group, which implies the following covariant derivative acting on representations of $SL(2)\times SO(6,n)$ \cite{Schon:2006kz}
\be
D_\m =\nabla_\m - g A_\m{}^{M\a}\big( f_{\a M}{}^{NP} -\xi_\a^N \delta_M^P\big) t_{NP} + g A_\m{}^{M \a} \xi^{\b}_M\,t_{\a\b}\ ,
\ee
where $ f_{\a MNP}= f_{\a[MNP]}$ and $\xi_M^\a$ are the components of the constant embedding tensor that govern the gauging, and $t_{MN}=t_{[MN]}$ and $t_{\a\b}=t_{(\a\b)}$ are the generators of $SO(6,n)$ and $SL(2,\mR)$, respectively. This form of the embedding tensor is dictated by the linear constraint. The quadratic constraints \eq{qc} give rise to five equations which can be found in \cite{Schon:2006kz}. The results of \cite{DallAgata:2023ahj} are the most general as they extend those of \cite{Schon:2006kz} as follows. Defining a symplectic frame as the choice of $(n+6)$ vector fields among the $2(n+6)$ vectors and dual vectors,  a \mbox{standard frame} was chosen and partial results for the Lagrangian and supertransformations were given in \cite{Schon:2006kz}. On the other hand, arbitrary symplectic frames were considered, and the full Lagrangian and supertransformations were provided in  \cite{DallAgata:2023ahj}.  For further details such as particular solutions of the constraints, and possible connections between the resulting gauge supergravities and type IIB flux compactifications, see \cite{Trigiante:2016mnt,DallAgata:2023ahj}, and references therein. Relation between gauge $N=4, 4D$ supergravity coupled to vector multiplets and the double field theory formulation of $N=1, 10D$ heterotic supergravity also offers an approach to its embedding to string theory \cite{Aldazabal:2011nj,Geissbuhler:2011mx}. 

\subsection{ $N=3$ supergravity coupled to vector multiplets in $4D$}

The complete $N=3$ supergravity coupled to vector multiplets was constructed in \cite{Castellani:1985ka}, using the group manifold approach. Combining the supergravity multiplet with $n$ vector multiplet $(A_\m, 6 \phi; \lambda_i, \lambda)$, where $i=1,2,3$, gives the field content
\be
\{ e_\m^r, \cV_M{}^A,  A_\m^M, B_{\m\n\a}\, ;\ \psi_{\m i}, \chi, \lambda_{ia}, \lambda \}\ ,
\ee
where $M =1,...,n+3$ and $\cV_M{}^A= \big(\cV_M{}^m, \cV_M{}^a\big)$ with $m=1,...,3,  a=1,...,n$, is a representative of the coset $SU(n,3)/SU(n)\times SU(3)$. The two-form potentials are in the adjoint representation of $SU(n,3)$. In the fermionic sector, the $SU(3)$ index $i=1,...,3$, and the fermions are Weyl. The $U(1)_R \subset U(3)_R$ charges of the fermions $(\psi_{\m i}, \chi, \lambda_{ia}, \lambda)$ are given by {\small{$( 1/2, 3/2, (n+6)/2n, 3(n+2)/2n )$}}, respectively. Subgroups of $SO(n,3) \subset SU(n,3)$ of dimensions $n+3$ were gauged in \cite{Castellani:1985ka}. As observed in \cite{Karndumri:2016miq}, there is a similarity with the allowed gauged groups in half-maximal $7D$ supergravity coupled to $n$ vector multiplet. As such, the gauge groups listed in \eq{gg7D} were found in the present case. More general gauging based on the embedding tensor formalism is described in \cite{Trigiante:2016mnt}, where it is shown that the embedding tensor is restricted by the linear constraint to transform as an irreducible $SU(n,3)$ tensor $\theta_{MN}{}^P = \theta_{[MN]}{}^P$ and its conjugate, $\theta^{MN}{}_P$. The quadratic constraints \eq{qc} need to be imposed as well. A systematic analysis of these constraints and a classification of all possible gaugings remain to be carried out.  

\subsection{ $N=2$ supergravity coupled to scalar and vector multiplets in $4D$}

Combining the $N=2, 4D$ supergravity multiplet, which contains a single vector, with $n_V$ vector multiplets each containing a complex scalar $z$, and $n_T$ hypermultiplets, upon introducing the magnetic vector potentials and two-form fields that are dual to the scalar currents, we get the field content\footnote{For the off-shell multiplets, see \cite{deWit:1980lyi,deWit:2006gn}.}
\be
\{ e_\m^r, A_\m^M, z^\a , B_{\m\n \a};\ \psi_\m^i, \lambda^{\a i} \},\qquad \{\phi^X, \psi^a \}\ ,
\nn
\ee
where
\bea
&& i=1,2,\ \ \a=1,...,n_V,\ \ X=1,...,4n_H,\ \ a=1,...,2n_H
\nn\w2
&& A_\m^M = (A_\m^\Lambda, A_{\m \Lambda}),\ \ \Lambda=0,1...,n_V\ . 
\eea
The $n_V$ complex scalars, $z^\a$, parametrize a \mbox{\it special K\"ahler (SK) manifold} $\cM_{SK}$, described below, and $\phi^X$ parametrize a quaternionic K\"ahler manifold $\cM_{QK}$, described in section 7.4, as required by supersymmetry. As a special case of SK geometries, the so-called VSK geometries arise from the dimensional reduction of the VSR geometries present in $5D$ supergravities via the $r$-map, as mentioned briefly at the end of section 8. See table 5 in appendix C for a list of supergravities in which SK and VSK geometries arise. These geometries and the relationships between them are treated in great detail for example, in \cite{Lauria:2020rhc,Trigiante:2016mnt}.  

A good source for original papers on $N=2, 4D$ supergravities coupled to vector and scalar multiplets is \cite{deWit:1984rvr}. The general electric gaugings of $N=2, 4D$ supergravity coupled to vector multiplets are given by using the group manifold approach in \cite{Andrianopoli:1996cm,Andrianopoli:2011zj}, and the duality covariant gaugings in \cite{deWit:2011gk} by means of superconformal tensor calculus. The duality covariant couplings were also obtained in \cite{Andrianopoli:2015rpa} by a direct computation. See also \cite{Lauria:2020rhc}. Here we shall summarize briefly the detailed account given in \cite{Trigiante:2016mnt}, which primarily follows \cite{Andrianopoli:2015rpa}.

Some of the key ingredients in the description of SK manifolds are the symplectic vectors and K\"ahler potentials which take the form
\bea
V^M  &=& \begin{pmatrix} X^\Lambda \\ F_\Lambda  \end{pmatrix}\ ,\qquad v^M(z) = \begin{pmatrix} L^\Lambda(z)\\ M_\Lambda (z) \end{pmatrix}\ ,\qquad v^M(z)= e^{-\cK/2} V^M
\nn\w2
\cK(z,\zb) &=& -\log \big[ i(\vb^T \Omega v)\big]\  ,\qquad \Omega=\begin{pmatrix} 0 & \mathbbm{1}\\ -\mathbbm{1} \end{pmatrix}\ .
\eea
It is also useful to define the K\"ahler covariant derivatives
\bea
U^M_\a &\equiv &  \begin{pmatrix} f_\a{}^\Lambda \\ h_{\a \Lambda}   \end{pmatrix} = \nabla_\a V^M =\partial_\a V^M + \frac12 (\partial_\a \cK) V^M\ .
\eea
For SK manifolds the following conditions must be satisfied 
\bea
&& \nabla_\a V^M = 0\ ,\qquad \nabla_\a U_\b^M = i C_{\a\b\g}\, g^{\gamma{\bar\gamma}}\, {\overline U}_{\bar\gamma}^M \ ,
\nn\w2
&& \nabla_\a {\overline U}_\bb = g_{\a\bb} {\overline V}^M\ ,\qquad V^M \Omega_{MN} U_\a^N =0\ ,
\eea
where $C_{\a\b\g}$ is a total symmetric tensor. To describe the Yang-Mills kinetic terms, it proves useful to define \cite{Lauria:2020rhc}
\be
\cN_{IJ} = \big(F_\Lambda\nabla_\ab {\overline F}_\Lambda\big) \big( X^\Sigma \nabla_\ab {\overline X}^\Sigma \big)^{-1}\ .
\ee
Next, consider the gauging of a subgroup of the isometry group $G_{SK} \times G_{QK}$ of the scalar manifold $\cM_{SK} \times \cM_{QK}$ with generators $(t_u, t_m)$, respectively. Thus, $u=1,...,{\rm dim}\,G_{SK}$ and $m=1,...,{\rm dim}\, G_{QK}$. The gauge generators $X_M$ are expressed as
\be
X_M = \theta_M{}^u t_u + \theta_M{}^m t_m\ ,\qquad X_{MN}{}^P = \theta_M{}^u (t_u)_N{}^P\ .
\ee
The resulting linear and quadratic constraints on the embedding tensors are \cite{Trigiante:2016mnt}
\bea
&& X_{(MNP)}=0\ ,\quad \theta_M{}^u \theta_N{}^v f_{uv}{}^w +X_{MN}{}^P \theta_P{}^w=0\ ,
\w2
&& \theta_M{}^m \theta_N{}^n f_{mn}{}^p +X_{MN}{}^P \theta_P{}^p=0\ ,\quad \theta^{MA} \theta_M{}^B=0\ ,\quad A=(u,m)\ .
\nn
\eea
The moment maps are given by \cite{Trigiante:2016mnt}
\bea
\cP_m{}^x &=& \frac{1}{2\lambda n_H} J^x{}_Y{}^X D_X k_m{}^Y \ ,\qquad x=1,2,3\ ,
\eea
where $J^x{}_Y{}^X$ are the quaternionic structures, $k_m{}^Y$ are the Killing vectors generating the algebra of the generators $t^m$ on $\cM_{QK}$, which has negative constant scalar curvature $R=8\lambda n_H (n_H+2)$ that defines the constant $\lambda$. 

With the above definitions at hand, the bosonic part of the Lagrangian is given by \cite{Lauria:2020rhc}
\bea
e^{-1} \cL &=& \frac12 R -g_{\a\bb} D_\m z^\a D^\m \zb^\bb -\frac12 g_{XY} D_\m\phi^X D^\m\phi^Y  
\nn\w2
&& +\frac{i}{4} \Big( \cN_{\Lambda\Sigma} \cF_{\m\n}^{+\Lambda} \cF^{+\m\n \Sigma} - h.c. \Big)+\cL_{\rm top} -V\ ,
\eea
where
\bea
D_\m z^\a &=& \partial_\m z^\a -A_\m^M k_M{}^\a\ ,\qquad D_\m \phi^X = \partial_\m \phi^X -A_\m^M k_M{}^X\ ,
\nn\w2
k_M{}^\a &=& \theta_M{}^u k_u{}^\a\ ,\qquad k_M{}^X =\theta_M{}^m k_m{}^X\ ,
\eea
and  $k_u{}^\a$ are the Killing vectors generating the algebra of the generators $t^u$ on $\cM_{SK}$. All Killing vectors preserve the complex and quaternionic structures. The field strength $\cF_{\m\n}^\Lambda$ is defined as
\be
\cF_{\m\n}^\Lambda = 2\partial_{[\m} A_{\n]}^\Lambda + g X_{[NP]}{}^\Lambda A_\m^N A_\n^P +\frac12 g \theta^{\Lambda A} B_{\m\n A}\ ,
\ee
where we recall that $A=(u,m)$. The potential which takes the form
\be
V= g^2 \Big[ \big(k_M{}^\a k_N{}^\bb g_{\a\bb} +4g_{XY}k_M{}^X k_M{}^Y \big) {\overline V}^M V^N + \big(U^{MN}-3V^M {\overline V}^N\big) \cP_N^x \cP_N^x \Big]\ ,
\ee
where $U_{MN} := g^{\a\bb} U_\a^M {\overline U}_\bb^N$.
The Lagrangian $\cL_{\rm top}$ has the same for as in \eq{4Dtop}, and the resulting duality equation which relates the three-form field strength (which can be deduced from the Bianchi identity for the two-form field strength defined above) to the dual of the scalar current has a form similar to \eq{4DDE}.

Finally, supertransformations of the fermions are given by \footnote{In this section we use the conventions of \cite{Trigiante:2016mnt}, where $X^M = -\Omega^{MN} X_N, \ X_M= X^N \Omega_{NM}$. Furthermore, the position of the index $i$ on a spinor is associated with its chirality, such that $\psi_{\m i}, \lambda_i^\a$ have positive chirality, while $\psi_\m^i, \lambda^{i\a}$ have negative chirality. The hyperino $\psi^a$ has positive chirality.}
\bea
\delta \psi_{\m i} &=& D_\m \e_i +\frac12 \e_{ij} H_{\m\n}^{-} \e^j  + ig \gamma_\m S_{ij}\e^j\ ,
\nn\w2
\delta \lambda^{\a i} &=& i \gamma^\m D_\m z^\a \gamma^\m \e^i+\frac{i}{4} H_{\m\n}^{-\a} \gamma^{\m\n} \e_j \e^{ij}  +g W^{\a ij} \e_j\ ,
\nn\w2
\delta \psi_a &=& i\cV_X{}^{bj} D_\m \phi^X \gamma^\m \e_{ij} \Omega_{ab} + gN_a{}^i \e_i\ ,
\eea
where $\cV_X{}^{ai}$ are the vielbeins on $\cM_{QK}$, and 
\bea
D_\m \e_i &=& \left(\nabla_\m -\frac{i}{2} Q_\m \right)\e_i -Q_{\m j}{}^i \e^j \ ,
\nn\w2
Q_\m &=& (\frac{i}{2} \partial_\a K \partial_\m z^\a -h.c. ) -A_\m^I P_I^0\ ,
\ \
Q_\m{}^{ij} = - \omega_X{}^{ij} \partial_\m \phi^X -\frac12 A_\m^M \cP_M{}^{ij}\ ,
\nn\w2
H_{\m\n}^{-} &=& 2i L^\Lambda \big({\rm Im}\, \cN_{\Lambda\Sigma}\big) \cH_{\m\n}^{\Sigma -}\ ,\quad 
H_{\m\n}^{-\a} = 2i g^{\a\bb} {\overline f}_\bb{}^\Lambda \big({\rm Im}\, \cN_{\Lambda\Sigma}\big) H_{\m\n}^{\Sigma -}\ ,
\eea
with $\omega_X{}^{ij}$ representing the composite $Sp(1)$ connection on $\cM_{QK}$, and $\cP_M{}^x := \Theta_M{}^m \cP_m{}^x$. The shift functions are given by
\bea
S_{ij} &=& \frac12 i (\s^x)_i{}^k \e_{jk} \cP_M^x V^M\ , \qquad N_a{}^i = -2\cV_X{}^i{}_a k_M{}^X {\overline V}\ ,
\nn\w2
W^{\a ij} &=& -\e^{ij} k_M{}^\a {\overline V}^M +i(\sigma^x)_k{}^j \e^{ki} \cP_M{}^x g^{\a\bb} {\overline U}_\bb^M\ .
\eea
For a wealth of information on the subject we have barely touched, see, for example, the book \cite{Lauria:2020rhc}, and references therein.

\subsection{ $N=1$ supergravity coupled to scalar and vector multiplets in $4D$}

It is natural that the general matter coupled  $N=1$ supergavity is the most studied one for many years. The general couplings were given long ago in \cite{Cremmer:1982en}. Its structure has been thoroughly treated in the excellent book by Freedman and van Proeyen \cite{Freedman:2012zz}, where some applications are discussed as well. See also the reviews \cite{Bagger:1984ge,DallAgata:2022nfr}. For the sake of completeness, we shall give the bosonic part of the matter coupled $N=1$ supergravity action and the supersymmetry transformations here. 

The supergravity, chiral and Yang-Mills multiplets have the fields $(e_\m^a, \psi_\m), (z^\a, \chi^\a_L)$ and $(A_\m^A, \lambda^A)$, with $\a=1,...,n$ and $A=1,..., {\rm dim}\, G$\footnote{For off-shell multiplets, see \cite{Stelle:1978ye,Ferrara:1978em,Sohnius:1981tp,Muller:1985vga}.} As is well known, the complex scalar fields of the $n$ chiral multiplets, denoted by $z^\alpha$ here, parametrize a Hodge-K\"ahler manifold ${\cal M}$. We shall assume that $\cM$ has an isometry group $G$ which is fully gauged by the introduction of the Yang-Mills multiplet \footnote{Gauging a subgroup of $G$ is straightforward, with minimal adjustment of notation.}. Prior to giving the bosonic part of the full action, it is useful to specify their building blocks.

The key building blocks are the K\"ahler potential $\cK(z,\zb)$, holomorphic superpotential $W(z)$, holomorphic gauge function  $f_{AB}(z)= f_{(AB)}(z)$, and holomorphic Killing vectors $k_A{}^\a(z)$.  The metric $g_{\a\bb}$ is invariant under the K\"ahler transformations $\cK(z,\zb) \to \cK(z,\zb)+ F(z) + {\bar F}(\zb)$, but under the gauge transformation that acts as $\delta z^\a = \theta^A (x)\, k_A{}^\a (z)$, where $\theta(x)^A$ is the gauge parameter, the K\"ahler potential $K$ is invariant up to a gauge transformation, viz.
\be
\delta_\theta \cK = \theta^A k_A{}^\a \partial_\a \cK + h.c. = \theta^A r_A(z) + h.c.\ ,
\ee
where $r_A(z)$ is defined by this equation.  Invariance of the action under the K\"ahler transformation also requires that $W \to e^{-F} W$. Further conditions that need to be satisfied for gauge and K\"ahler invariance are  \cite{Bagger:1984ge,Freedman:2012zz}
\bea
&& k_A{}^\a g_{\a\bb}k_B{}^{\bb} - A \leftrightarrow B = if_{AB}{}^C \cP_C\ , \ \ k_A{}^\a \nabla_\a W = i \cP_A\ ,
\nn\w2
&& k_C{}^\a  \partial_\a f_{AB}(z)=2f_{C(A}{}^D f_{B)D}(z) + iC_{AB,C}\ ,
\eea
where $C_{AB,C}=C_{(AB),C}$ are real constants, and the real scalar moment map
\bea
\cP_A (z,\zb) &=& i\left[ k_A{}^\a (z)\partial_\a \cK(z,\zb)  -r_A (z)\right]\ ,
\eea
which is defined up to arbitrary constant shifts, is related to the Fayet-Iliopoulos $D$-terms in the presence of abelian gauge group factors. It is also useful to specify the gauge covariant derivative $D_\m z^\a$, and the K\"ahler covariant derivative $\nabla_\a W$,
\be
D_\m z^\a =\partial_\m -A_\m^A k_A{}^\a\ , \qquad \nabla_\a W =\partial_\a W+(\partial_\a K)W\ .
\ee
With the above ingredients at hand, the bosonic part of the Lagrangian is given by \cite{Cremmer:1982en,Anastasopoulos:2006cz,DeRydt:2007vg}

\bea
e^{-1} \cL &=& \frac12 R -g_{\a\bb} D_\m z^\a D^\m z^\bb -\frac14 ({\rm Re} f_{AB}) F_{\m\n}^A F^{\m\n B} - i \frac14 ({\rm Im} f_{AB}) F_{\m\n}^A {\widetilde F}^{\m\n B}
\nn\w2
&& -e^K \Big( \nabla_\a W g^{\a\bb} {\bar\nabla}_\bb {\bar W} + 3 W{\bar W} - \frac12 ({\rm Re} f)^{-1 AB} \cP_A \cP_B \Big) + e^{-1} \cL_{CS}\ ,
\eea
where $F_{\m\n}^A = 2\partial_{[\m}A_{\n]}^A +g f_{BC}{}^A A_\m^B A_\n^C$ and the Chern-Simons Lagrangian is given by \cite{DeRydt:2007vg}
\be
\cL_{CS} = \frac12 \left(C_{AB,C} -C_{(AB,C)}\right) \ve^{\m\n\r\s} \Big( \frac13 A_\m^C A_\n^A F_{\r\s}^B +\frac14 f_{DE}{}^A A_\m^D A_\n^E A_\r^C A_\s^B \Big)\ .
\ee
If the theory has gauge anomalies, they will be encoded in a totally symmetric constant tensor $d_{ABC}$, and they can be removed by shifting the $C$-dependent factor in $\cL_{CS}$ by $d_{ABC}$ \cite{Anastasopoulos:2006cz,DeRydt:2007vg}.

The supertransformations of the fermions are
\bea
\delta \psi_{\m L} &=& D_\m\e_L +\frac12 e^{\cK/2} \gamma_\m \e_R\ ,
\nn\w2
\delta \chi^\a_L &=& \frac{1}{\sqrt2} \gamma^\m D_\m z^\a \e_R -\frac{1}{\sqrt2} e^{K/2} g^{\a\bb} {\overline\nabla}_\bb \overline{W} \e_L\ ,
\nn\w2
\delta \lambda^A &=& \frac14 \gamma^{\m\n} F_{\m\n}^A \e +\frac14 i (Re f)^{-1 AB} P_B \e\ .
\eea
where $D_\m \e_L= \big( \partial_\m +\frac14 \omega_\m{}^{ab} \gamma_{ab} -\frac32 i Q_\m \big)\e_L$ with 
\be
Q_\m = \frac{i}{6} \big(\partial_\m z^\a \partial_\a \cK - h.c.\big) -\frac13 A_\m^A \cP_A\ .
\ee
Invariance of the action under the K\"ahler transformation requires, in addition to $W\to e^{-F} W$, that 
\be
\psi_{\m L} \to e^{-iF/2} \psi_{\m L}\ ,\ \ \lambda^A_L \to e^{-iF/2} \lambda^A_L\ ,\ \ \chi^\a_L \to e^{iF/2} \chi^\a_L\ ,
\ee
While globally $N=1$ supersymmetric sigma models exist for all K\"ahler manifolds, only a subclass known as Hodge manifolds can be coupled in a globally well defined manner to supergravity. A Hodge manifold is a K\"ahler manifold on which it is possible to define a complex line bundle whose first Chern class is proportional to the K\"ahler form $J := (i/2) \big(\partial^2 K/\partial z^\a\partial z^{\bb}\big) dz^\a \wedge dz^\bb $. For topologically non-trivial Hodge manifolds this leads to the quantization of Newton's constant in terms of the scalar self-coupling \footnote{(For the noncompact case with nontrivial topology, the story is more complicated; see \cite{Witten:1982hu}.}. For more details, see \cite{Bagger:1984ge,Freedman:2012zz}. 
While scalar manifolds for $N>1$ are always non-compact, in the case of $N=1$ compact scalar manifolds are possible \cite{Freedman:2012zz}. Finally, we note that under certain circumstances, the K\"ahler symmetry may develop anomalies at the quantum level. For a discussion of these anomalies, see \cite{Freedman:2005up} and references therein.

\section{D=3}

The on-shell $N$-extended supergravity multiplet consist of the vielbein and gravitini, $(e_\mu{}^a, \psi_\mu^I)$, where $I=1,...,N$. The two-derivative supergravity theory for this multiplet is topological. An $N=8$  non-topological theory is obtained by coupling to $n$ scalar multiplets whose scalars parametrize the coset $SO(n,8)/(SO(n)\times SO(8))$, and an $N=16$ theory is obtained by coupling to $128$ scalar multiplets whose scalars parametrize the coset $E_{8(8)}/SO(16)$, as was shown in \cite{Marcus:1983hb}\footnote{The locally supersymmetric sigma model for $E_{8(8)}/SO(16)$ has also be derived from dimensional reduction of $11D$ supergravity \cite{Mizoguchi:1997si}.}. These results were generalized to obtain other general ungauged $N<16$ supergravities coupled to nonlinear sigma models in \cite{deWit:1992psp} where the geometries of the scalar manifolds were identified. A subset of the scalars can be dualized to vector fields, which in turn can be used to gauge  certain isometries of the scalar manifolds. In view of the duality relation, it was shown in \cite{Nicolai:2003bp} that there are two types of gaugings, known as Chern-Simons (CS) type and Yang-Mills type gaugings, depending on whether the scalars of the vectors carry the propagating degrees of freedom. Below we shall give a summary of the CS gaugings described in detail in \cite{deWit:2003ja}. Those which admit an AdS vacuum solution with $OSp(2|p)\otimes OSp(2|q)$ supersymmetry, upon truncation down to the topological theory, may be referred to as $N=(p,q)$ supergravity, constructed long ago in \cite{Achucarro:1987vz}. The topological theory is based on gauging the $N=(p,q)$ super AdS algebra and it exists by itself for any $N$.

\subsection{All gauged on-shell supergravities in $3D$ }

Supergravity in $3D$ coupled to $d$ real scalar fields has the field content~\cite{deWit:1992psp,deWit:2003ja}
\be
\{ e_\mu{}^a, \psi_\mu^i, \phi^\a, \chi^{\alpha i} \}\ ,
\ee
where $i=1,...,N$ and $\a =1,...,d$.  The fermions are Majorana, and the scalars parametrize a $d$ dimensional target space. We shall primarily follow \cite{deWit:2003ja} in our brief review here. Supersymmetry leads to stringent conditions on the target space, which can only be satisfied when the  number of supersymmetries is restricted to $N \le 16$, and it has to admit $N(N-1)/2$ complex structures $(f^{ij})_\alpha{}^\beta$. For $N>2$ the target space has to be Einstein with $ R_{\a\b}= \left( (N-2) +\frac{d}{8} \right) g_{\a\b} > 0$. In particular, for $N=3$ the  target space has to be quaternionic Kahler, while for $N=4$ it has to be a direct product of two quaternionic spaces of dimension $d_\pm$. For $N>4$,  the target space has to be homogeneous. There exists no theory with $N = 7$ supersymmetry and beyond $N = 8$ there are only four possible theories. They are $N = 9,10,12,16$ supersymmetric, and their target spaces, which  are unique, are respectively the symmetric spaces,
\be
\frac{F_{4(-20)}}{SO(9)}\ ,\qquad  \frac{E_{6(-14)}}{SO(10)\times SO(2)}\ ,\qquad \frac{ E_{7(7)}}{SO(12)\times SO(3)}\ ,\qquad \frac{E_{8(8)}}{SO(16)}\ .
\ee

A proper subgroup $G_0$ of the isometry group $G$ of the  scalar manifold can be gauged by introducing an embedding tensor $\theta_{MN}$ which specifies $G_0$ and it is invariant under it. This is expressed by
\be
\theta_{PL}\left(f^{KL}{}_M \theta_{NK} + f^{KL}{}_N \theta_{MK}\right) = 0\ ,
\ee
where $\theta_{PL} f^{KL}{}_M$ are the structure constants of $G_0$. Furthermore, the requirement of local supersymmetry of the action imposes the following  constraint linear in the embedding tensor,
\be
(N-2)( T^{ij,k\ell} -T^{[ij,k\ell]}) +\left( \delta^{i[k} T^{\ell]m,mj} - i \leftrightarrow j \right) +\frac{2\delta^{i[k} \delta^{\ell]j}}{(N-1)} T^{mn,mn}=0\ ,
\label{LC}
\ee
where 
\be
T^{ij,k\ell} = \cV^{Mij}\theta_{MN} \cV^{N k\ell}\ ,\qquad \theta_{MN}= \theta_{NM}\ ,
\ee
and $\cV^{M ij}$ are  moment maps associated with the Killing vector fields $X^{M \a}(\phi)$ that generate the isometry group satisfying the defining equation
\be
D_\a\cV^{M ij} (\phi, X) = \frac12 f^{ij}_{\a\b} (\phi) X^{M\b}(\phi)\ .
\ee
The  Lie derivative of the complex structures, the  Lie derivative of  $SO(N)$ connection and  the complex structures $f^{ij}_{\a\b}$ vanish up to a local $SO(N)$ rotation with parameter depending on $(X,\phi)$. As for  the condition \eq{LC}, it also holds for $N=1,2$ in which case it degenerates to an identity~\cite{deWit:2003ja}. 

The bosonic part of  Lagrangian is given by~\cite{deWit:2003ja}
\begin{eqnarray}
\cL &=& -\frac12\,\rmi \,\ve^{\mu\nu\rho}
e_\mu{}^a\,R_{\nu\rho a} -\frac12 e\,g_{\a\b} D_\mu \phi^\a\, D_\mu \phi^\b +\frac14\,\rmi g\,\ve^{\mu\nu\rho} \,
A_\mu^{M} \,\theta_{MN} \Big( \partial_{\nu} A_{\rho}^N - 
\frac13 g\, \widehat f_{PQ}{}^{N} \, A_\n^P A_\r^Q\Big)   
\nn\w2
&& + \frac{4\, e g^2}{N^2}  \left(N\,  A_1^{ij}A_1^{ij} -\frac12 
g^{\a\b}\, D_\a A_1^{ij} \,D_\b A_1^{ij} -2\,g^{\a\b} \, T^{ij}_\a\,T_\b^{ij}
\right)\ ,
\end{eqnarray}
where $R_{\mu\nu a} = \epsilon_{abc} R_{\mu\nu}{}^{bc}$, and $D^{~}_\a A_1^{ij} =\partial_\a A_1^{ij}  +2\, Q^{k[i}_\a\, A^{j]k}_{1}$ in which $Q_\a^{ij}$ is the $SO(N)$ target space connection, and 
\bea
D_\mu\phi^\a &=&\partial_\mu \phi^\a +g\, \theta_{MN} \,A^M_\mu \, X^{N\,\a}\ ,
\nn\w2
(N-2) A_1^{ij}
&=& \mu \, (N-2)\, \delta^{ij} - 4\,T^{i\ell,j\ell}  + \frac{2}{N-1}\, T^{mn,mn}\,\delta^{ij}\ , 
\nn\w2
T^{ij \a} &=& \cV^{M ij} \theta_{MN} X^{N\a}\ ,
\nn\w2
\widehat{f}_{MN}{}^P &=& \theta_{MQ} \,f^{PQ}{}_N \ .
\eea
The  supertransformations of the fermionic fields are 
\begin{eqnarray}
\delta \psi_\mu^i &=& D_\mu \epsilon^i\ ,
\nn\w2
\delta \chi^{\a i} &=& \frac12 \left( \delta^{ij} -f^{ij}\right)^\a{}_\b \gamma^\mu D_\mu \phi^\b \e^j\ ,
\end{eqnarray}
where
\be
D_\mu \epsilon^i = \left(\partial_\mu + \frac14 \omega_{\mu ab} \gamma^{ab} \right)\epsilon^i + \partial_\mu\phi^\a Q_\a^{ij} \epsilon^j + g\,\theta_{MN} A_\mu^M {\cal V}^{N ij} \epsilon^j\ .
\ee
Several properties of the scalar manifold as well as the solutions of the constraints on the embedding tensor $\theta_{PQ}$ for different values of $N$ have been described in \cite{deWit:2003ja}. It is noteworthy that the scalar manifold is an arbitrary Riemannian manifold for $N=1$, K\"ahler for $N=2$, quaternionic for $N=3$, locally a product of two quaternionic manifolds for $N=4$ and symmetric homogeneous space $G/H$ such that $d = {\rm dim}\,G-{\rm dim}\,H$ for $N>4$. The symmetric ones for $4\le N\le 16$ are listed in table 6.

\subsection{ Comments on off-shell supergravities in $3D$}

In $3D$, the off-shell $N=1$ supergravity has the fields $(e_\m^a, S)$, where the auxiliary field $S$  is a real scalar. The off-shell $N=(1,1)$ supergravity\footnote{The $N=(p,q)$ terminology in $3D$ refers to supergravities with the $R$-symmetry group $SO(p) \times SO(q)$ and admitting a vacuum solution with anti-de Sitter group, $OSp(2,p) \oplus OSp(2,q)$, symmetry.}, on the other hand, has the field content $\{ e_\mu{}^a, \psi_\mu, V_\mu, S \}$ where the gravitino is Dirac, the auxiliary vector is non-gauge, and the auxiliary scalar $S$ is complex. As to the off-shell  $N=(2,0)$ supergravity, the field content is $\{ e_\mu^a, \psi_\mu, C_\mu, V_\mu, D \}$ where the  gravitino is Dirac, $C_\mu$ is gauge field, the  auxiliary field $V_\mu$ is non-gauge and the auxiliary field $D$ is real; see, for example, \cite{Bergshoeff:2010mf,Alkac:2014hwa} for the Lagrangians and further details of these off-shell supergravities. Off-shell $N\le 4$ matter coupled supergravities were constructed in superspace formalism in \cite{Kuzenko:2011xg}, and in AdS $(p,q)$ superspace framework for $p+q\le 4$ in \cite{Kuzenko:2012bc}.

\section{Lower dimensions }

In one and two dimensions we get into the realm of infinite dimensional rigid symmetries and gaugings. 
It is technically quite a  complicated matter to render these symmetries manifest. As this would take us far afield, we shall be content with only providing some basic facts and references. 

\subsection{$D=2$}

In $2D$, there exists $(p,q)$ type of supersymmetry where $p$ and $q$ refer to the number of left- and right-handed supersymmetry generators. Couplings of scalar multiplets to $2D$ conformal supergravity were constructed long ago \cite{Deser:1976rb,Brink:1976sz,Bergshoeff:1985qr,Bergshoeff:1985gc,Pernici:1985dq,Bergshoeff:1986ys}. These are often referred to as locally supersymmetric sigma models. The geometry of the scalar manifold is Riemannian for $(1,0)$, hermitian with torsion for $(2,0)$, and hyperk\"ahler or quaternionic K\"ahler for $(4,0)$\footnote{As observed in \cite{Pernici:1985dq}, the possibility of hyperk\"ahler in the $(4,4)$, or $(4,0)$ cases is due to the fact that in $2D$ there is no gravitino kinetic term}. In the case of $N=(1,0)$, taking the scalars to correspond to the coordinates of spacetime, and by inclusion of Wess-Zumino term and heterotic fermions, i.e. fermions which are singlets under supersymmetry, the model describes a heterotic string in curved background. In the case of $N=(8,0)$, the algebraic constraints on the scalar multiplet fields that follow from the supergravity multiplet equations of motion are solved to obtain $N=(8,0)$ supergravity coupled to $8n$ scalars which parametrize the coset $SO(n,8)/(SO(n)\times SO(8))$ \cite{Bergshoeff:1986ys}. 

The simplest nonconformal coupling of supergravity to scalar multiplets has the action  $S=\int d^2x\,e\, \phi \left(R -\Lambda\right)$ as its bosonic part, and it is known as Jackiw-Teitelboim (JT) gravity which has been a subject of numerous studies in recent years owing to its being a solvable model of gravity, with remarkable properties including its dual relation to random matrix models in the boundary of $AdS_2$; see \cite{Mertens:2022irh} for a review. One generalization replaces $\Lambda$ by a function $W(\phi)$. A general coupling of $N=(2,2), 2D$ supergravity to scalar multiplets was given in \cite{Gates:2000fj}. 

The classical Lagrangian for the maximal $N=(8,8), 2D$ supergravity, often referred to as the $N=16$ supergravity, is easily obtained from a circle reduction of the maximal $3D$ supergravity in which the $128$ physical scalars parametrize the coset $E_8/SO(16)$. The equations of motion of the $N=16, 2D$ supergravity have been shown to be integrable in the sense that they follow from a linear system of equations \cite{Nicolai:1987kz,Nicolai:1988jb}. This theory has long been expected to have a hidden $E_9$ symmetry \cite{Julia:1980gr}. The group $E_9$ is based on the hyperbolic Kac-Moody algebra $\mathfrak{e_9}$, and it is the centrally extended loop group over $E_8$, extended with the Virasoro generator $L_0$. It turns out that there is an extension of this symmetry to ${\widehat E}_8 \rtimes {\rm Vir}_-$ based on an algebra with the generators \cite{Julia:1996nu}
\be
\underbrace{T^A_n\,(n\in \mathbb Z),\ K}_{{\widehat E}_8} \ , \underbrace{ L_n\, (n\le 0)}_{{\rm Vir}^-}\ ,\quad A=1,...,248\ .
\ee
For the algebra of these generators see, for example,  \cite[eqs. (2.5)-(2.7)]{Bossard:2023wgg}. The generators of ${\widehat E}_8$ together with $L_0$ form the algebra that underlines $E_9$. Denoting its maximal compact subgroup by $K(E_9)= K({\widehat E}_8)$, the bosonic fields of the $2D$ theory, all scalars, parametrize the coset \cite{Bossard:2021jix,Bossard:2023wgg}
\be
\frac{{\widehat E}_8 \rtimes {\rm Vir}^-}{K(E_9)}\ .
\ee
A building block for the description of the theory is the representative of this coset given by 
\be
V= \underbrace{\rho^{-L_0} e^{-\varphi_1 L_{-1}} e^{-\varphi_2 L_{-2}} \cdots}_{{\rm Vir}^-}\, \underbrace{e^{-\sigma K} V_0 e^{Y_{1A} T^A_{-1}} e^{Y_{2A} T^A_{-2}} \cdots}_{{\widehat E}_8}\ ,
\ee
where $\rho$ is the dilaton, $\varphi_n,\,n\ge 1,$ are scalars dual to the dilaton,  $\sigma$ is the conformal scalar coming from the choice of conformal gauge $g_{\mu\nu} =e^{2\sigma} \eta_{\mu\nu}$, $V_0$ is the representative of the coset $E_8/SO(16)$, and $Y_{nA}$ are dual to the physical scalars that parametrize the coset $E_8/SO(16)$. 

A systematic construction of the bosonic sector of gauged matter-coupled supergravity theories in $2D$ was initiated in \cite{Samtleben:2007an} where the embedding tensor is determined to transform in the basic representation of $E_9$. 
The complete bosonic dynamics of all gauged maximal supergravities
that admit a geometric uplift to $10D$ and $11D$, including the construction of the full potential,  was achieved recently in \cite{Bossard:2023wgg}, by starting from  the $E_9$ exceptional field theory based on the building blocks outlined above \cite{Bossard:2018utw,Bossard:2021jix}, and performing a generalized Scherk-Schwarz dimensional reduction. Denoting the generators of the algebra underlying ${\widehat E}_8 \rtimes {\rm Vir}^-$ by $T_\alpha$, an arbitrary gauging is described by an embedding tensor $\theta_M{}^\alpha$ which defines the generators, 
\be
X_M = \theta_M{}^\alpha T_\alpha\ ,
\ee
such that $ D_\mu =\partial_\mu -g A_\mu{}^M X_M$. In absence of a full supersymmetry analysis of $2D$ gauged supergravity, investigations of examples and analogy with higher-dimensional situations suggest that the embedding tensor $\theta_{M \alpha}$ is of the restricted form \cite{Samtleben:2007an,Bossard:2023wgg}
\be
\theta_{M\alpha} = - \eta_{-1 \alpha\beta} \theta_N T^{\beta N}{}_M\ , 
\ee
where $\eta_{-1\alpha\beta}$ is defined below in eq. \eq{etak}. Given the coset representative $V$ discussed above, the covariantized scalar current takes the form \cite{Bossard:2023wgg}
\be
P_\mu = \frac12 D_\mu V\, V^{-1} + \mbox{h.c.} = \frac12\Big(\partial_\mu V\,V^{-1} -A_\mu{}^M \theta_M{}^\alpha  V T_\alpha V^{-1}\Big) + \mbox{h.c.}\ ,
\ee 
The desired equations of motion are encoded in the duality equation 
\be
P_\mu^{(n)}= S_n(P_\mu) +\chi_{\mu n}\,K\ ,\quad n \in \mathbb Z\ ,
\label{e9de}
\ee
where $S_n, n\in \mathbb Z$ are the shift operators acting as 
\be
S_n (T_m^A) = T_{n+m}^A\ ,\quad S_n (L_m)= L_{n+m}\ ,\quad S_n(K)=0 \ ,
\ee
and $\chi_{\mu n}$ for $n>0$ are independent auxiliary fields, introduced to achieve $K(E_9)$ covariance of the shifted currents. A pseudo-Lagrangian which gives the duality equation \eq{e9de} under a suitable projection (hence the terminology of `pseudo-Lagrangian, since the full duality equation is to be imposed by hand), turns out to be \cite{Bossard:2023wgg} 
\be
\cL= \cL_{\rm top} -\frac{1}{2\rho^3} \langle\theta|M^{-1}|\theta\rangle -\frac{1}{2\rho}  \eta_{-2\alpha\beta} \langle\theta|T^\alpha M^{-1} T^\beta|\theta\rangle
\ee
where $M=V^\dagger V$, and ${\widehat E}_8$ invariant symmetric bilinear forms $\eta_k$ are defined by 
\be
\eta_{k\alpha\beta} T^\alpha \otimes T^\beta = \sum_{n\in \mathbb Z} \eta_{AB} T_n^A \otimes T_{k-n}^B - L_k\otimes K -K\otimes L_k\ .
\label{etak}
\ee
and $\cL_{\rm top}$ is the topological Lagrangian, which has a rather complicate form, and is given in \cite[eq. (4.51)]{Bossard:2023wgg}. The bra-ket notation here is simply denoting contraction of the matrix indices with vectors and their transpose. See \cite{Bossard:2023wgg} for further details.

Finally let us not that infinite dimensional hidden symmetries naturally arise also in $2D$ supergravity in less than maximal supersymmetry. For example, see \cite{Sen:1995qk} for a treatment of the $O(8,24)$ loop group arising as a symmetry of the equations of motion of half-maximal $2D$ supergravity, obtained from the toroidal compactification of heterotic string theory.

\subsection{D=1}

$N=1, 1D$ supergravity multiplet consists of $(e, \psi_t)$, where $e$ is the einbein and $\psi_t$ is the gravitino. Its coupling to $d$ scalar multiplets $(\phi^i, \chi^i), i=1,...,d$, where $\phi^i(\tau)$ and $\chi^i(\tau)$ are bosons and fermions, was constructed in \cite{Brink:1976sz} as an action for a spinning particle. Generalization of this model to $N$-extended locally supersymmetric model in which the global $SO(N)$ symmetry is gauged by introducing the gauge fields $f_{ab}=f_{[ab]},\ a=1,...,N$ was constructed in \cite{Howe:1988ft}, where it shown that upon quantization it provides a relativistic wave equation for spin $N/2$. For issues related to compatibility of  curved space background with local supersymmetry, see \cite{Howe:1988ft,Bonezzi:2020jjq}. 

$N=16, 1D$ supergravity coupled to scalar multiplets whose scalars parametrize the coset $G/H=SO(9,9)/(SO(9)\times SO(9))$ has been obtained by dimensional reduction of Type I supergravity on a torus $T^9$ \cite{Kleinschmidt:2004dy}. The $1D$ theory has the bosons $(e, \varphi, \cV_M^A)$ and the fermions $(\psi_{t\alpha}, \chi_\alpha, \chi_{\bi \alpha} )$, where $i,\bi =1,...,9$ label the $SO(N)\times SO(N)$ vectors, $\alpha=1,...,16$ is the $SO(9)$ spinor index, and $\cV_M^A= (\cV_M^i, \cV_M^\bi)$ is the representative of the coset $G/H$. The scalar currents are defined by 
\be
\cV^{-1} \partial_t \cV= \frac12 Q^{ij}X^{ij}+\frac12 Q_{\bi\bj} X^{\bi\bj} + P_{i\bi} Y^{i\bi}\ ,
\ee
where $(Q^{ij},Q_{\bi\bj})$ is the $SO(9)\times SO(9)$ composite gauge fields, $X_{ij}$ and $X_{\bi\bj}$ are the $SO(9)\times SO(9)$ generators, and $Y_{i\bi}$ are the $G/H$ coset generators. The bosonic part of the Lagrangian is given by \cite{Kleinschmidt:2004dy}
\be
\cL= \frac{1}{4}e^{-1}\big( P_{i\bj}P_{i\bj} -\dot\vp^2 \big)\ ,
\ee
and the supertransformations by
\be
\delta\psi_{t\alpha}= D_t\eps_\alpha\ ,\qquad 
\delta \chi_\alpha = -\frac12 e^{-1} \eps_\alpha \dot\varphi\ ,\qquad
\chi_{\bj\alpha} = -\frac12 e^{-1}\left(\gamma_i \eps\right)_\alpha P_{i\bj}\ .
\ee

The dimensional reduction of $11D$ supergravity to $1D$ was conjectured long ago  to have an infinite dimensional $E_{10}$ symmetry \cite{Julia:1985ams}, which is a hyperbolic Kac-Moody algebra. This reduction was carried out in \cite{Mizoguchi:1997si} for the bosonic sector, first by a toroidal reduction to $3D$, then a circle reduction to $2D$ followed by a null reduction to $1D$.  A nonlinear $\sigma$-model in $1D$ based on a coset $E_{10}/K(E_{10})$, where $K(E_{10})$ is the maximal compact subgroup of $E_{10}$, was proposed in \cite{Damour:2002cu}. The model was extended to include the fermions in \cite{Damour:2006xu}, where relation to $11D$ supergravity was studied as well. For a detailed treatment of the relation between the $E_{10}/K(E_{10})$ coset model, which requires an infinite set of fields most conveniently described in an $SO(9)\times SO(9)$ basis, and the $1D$ supergravity action discussed above, see \cite{Kleinschmidt:2004dy}. See also the review \cite{Kleinschmidt:2006ks}, and reference therein. 

\section{Supergravities in extended geometry framework} 

It is well known that the $U$-duality groups $E_{n(n)}$ arise in maximal supergravities in $D$ dimensions, where $n=11-D$, as an enhancement of the manifest $GL(n,\mathbb{R})$ symmetry in the reduction of $11D$ supergravity on $n$-torus. In the case of $N=1, 10D$ supergravity coupled to $n_V$ abelian vector multiplets, the enhanced symmetry is $O(n,n+n_V)$ which generates the $T$-duality group. In the case of non-abelian vectors, $O(n,n+n_V)$ symmetry is broken but its covariance is maintained, as in gauged supergravities. An extended geometrical framework has been developed in which the above mentioned symmetries/covariance are manifest prior to any reduction. Such theories in the case of $E_n$ symmetries are called Exceptional Field Theories (ExFT), and in the case of $O(n,n+n_V)$ symmetry they are known as Double Field Theories (DFT).  See the reviews \cite{Aldazabal:2013sca,Hohm:2013bwa,Berman:2013eva} and references therein. To achieve these symmetries, extra internal space coordinates are introduced in ExFTs, and the spacetime doubling occurs as well in DFTs, together with the so-called section constraints to ensure the closure of the generalized diffeomorphisms. Depending on how these constraints are solved, the ExFTs can be shown to be fully equivalent to $11D$ supergravity or type IIB supergravity, and the $N=1$ supersymmetric DFT equivalent to $N=1, 10D$ heterotic supergravity, as will be discussed further below.

ExFTs associated with maximal supergravities have been constructed for $8 \ge D \ge 2$ \cite{Hohm:2013vpa,Musaev:2014lna,Hohm:2013uia,Godazgar:2014nqa,Hohm:2014fxa,Baguet:2016jph,Abzalov:2015ega,Musaev:2015ces,Hohm:2015xna,Bossard:2021jix} and a proposal exists for a so-called ``master exceptional field theory'' based on $E_{11}$ \cite{Bossard:2021ebg}. ExFTs  associated with $6D, N=(1,0)$ magical supergravities have been constructed in \cite{Bossard:2023ajq}, and the $N=1$ supersymmetric DFT in \cite{Hohm:2011nu}.  We shall briefly summarize ExFTs based on $E_{n(n)}$ for $n=6,7,8$, and the $N=1$ supersymmetric DFT, since only for these cases the fermionic sectors have also been constructed so far \cite{Musaev:2014lna,Godazgar:2014nqa,Hohm:2014fxa,Hohm:2011nu}. Among many uses of exceptional field theory formalism, it is worth mentioning that it is very efficient in solving the long outstanding problems of consistent truncations around various background geometries \cite{Hohm:2014qga}, and it also provides a remarkably powerful tool for computing the Kaluza-Klein mass spectra around these backgrounds \cite{Malek:2020yue}.

\subsection{Extended geometry}

In the case of ExFTs the extended geometry is formulated in $(D+{\rm dim}\, R_1)$ dimensional generalized space coordinatized by $(x^\m, Y^M)$, where $R_1$ is a suitable representation of $E_{n(n)}$. The  $x$-space is referred to as the external space, and $Y$-space as the internal space. In this geometry the tensors are tensors under $E_{n(n)}\times \mR$, with $\mR$ representing the trombone symmetry. The generalized diffeomorphism with parameter $\Lambda^M$ is supplemented by and extra local $\Sigma$-symmetry with parameter $\Sigma_M$ in the case of $n=8$, and it acts on a generalized vector as \cite{Coimbra:2011ky,Berman:2012vc,Cederwall:2017fjm}
\bea
\delta V^M = \cL_{(\Lambda,\Sigma)}  V^M &=&  \Lambda^N\partial_N V^M -\a_n P_{(adj)}^M{}_{N,}{}^P{}_Q\,\partial_P\Lambda^Q V^N +\beta_n \partial_K \Lambda^K\,V^M
\nn\w2
&& -\delta_{n,8}\, \Sigma_P\, f^{PM}{}_Q V^Q\ ,
\label{DB}
\eea
where  $P_{(adj)}^M{}_{N,}{}^P{}_Q$ is projector onto the adjoint representation of $E_{n(n)}$, and $\a_n= 6,12,60$ for $n=6,7,8$, respectively, and $\beta_n= \frac{1}{9-n}$ which is the trombone weight of a vector field in $D=11-n$ dimensions. The last term in \eq{DB} where $f_{MNP}$ are the structure constants of $E_8$, is needed because when $n=8$ the $\Lambda$-transformations close only up to terms that can be interpreted as $\sigma$-transformations. Gauge connections $(A_\m^M, B_{\m M})$ will be associated with the $(\Lambda,\Sigma)$ transformations. 
In the conventions of \cite{Hohm:2013vpa,Hohm:2013uia,Hohm:2014fxa}, the projectors $P_{adj}$ can be written as\footnote{The Cartan-Killing metrics $\kappa^{\a\b}= (t^\a)^M{}_N (t^\b)_N{}^M$, and $\eta^{MN}= \frac{1}{60} f^{MK}{}_L f^{NL}{}_K$ can be used to raise and lower indices.}
\be
P_{(adj)}^M{}_{N,}{}^P{}_Q = \begin{cases} (t_\a)^M{}_N (t^\a)^P{}_Q & \text{for}\ \ n=6,7 \\  \frac{1}{60} f^M{}_{NL}f^{LP}{}_Q  & \text{for}\ \  n=8
\end{cases}\ ,
\ee
where $t_\a$ are the generators of $E_{6(6)}$and $E_{7(7)}$ in their fundamental representations, and $f_{IJ}{}^K$ are the structure constants of $E_{8(8)}$. The  definitions of covariant field strengths turn out to require the following extra fields:
\begin{align}
E_{6(6)} : & \qquad B_{\m\n M}\ , \quad \ M=1,...,27\ ,
\nn\w2
E_{7(7)}: & \qquad \left( B_{\m\n \a}, B_{\m\n M} \right)\ ,\quad \alpha=1,...,133, \ \ M=1,...,56\ ,
\nn\w2
E_{8(8)}: &  \qquad \left( C_{\m\n}{}^{MN}, C_{\m\n M}{}^N, C_{\m\n}\right)\ ,\quad \ M, N=1,...,248\ ,
\label{fields}
\end{align}
where the two-form $C_{\m\n}{}^{MN}$ is valued in the $3875$ dimensional representation of $E_8$.
It has been established that the transformations \eq{DB} close on all the fields,
\be
[\cL_{(\Lambda_1,\Sigma_1)}, \delta_{(\Lambda_2,\Sigma_2)}] = \delta_{(\Lambda_{12}, \Sigma_{12})}\ , \qquad  (\Lambda_{12}, \Sigma_{12}) \equiv \big[(\Lambda_1,\Sigma_1), (\Lambda_2,\Sigma_2)\big]_E\ ,
\ee    
provided that the following section constraints are imposed \cite{Hohm:2013vpa,Hohm:2013uia,Hohm:2014fxa}
\begin{align}
E_{6(6)}: &\qquad d^{PMN} \ \partial_M \otimes \partial_N =0\ ,
\nn\w2
E_{7(7)}: &\qquad (t_\alpha)^{MN} C_M \otimes C^\prime_N =0\ ,\quad \Omega^{MN} C_M \otimes C^\prime_N =0\ ,
\nn\w2
&\qquad  C_M, C^\prime_M \in \{ \partial_M, B_{\m\n M} \} \ ,
\nn\w2
E_{8(8)}: &\qquad \eta^{MN} C_M \otimes C^\prime_N =0\ ,
\quad f^{PMN} C_M \otimes C^\prime_N =0\ ,\quad \left(P_{(3875)}\right)_{PQ}{}^{MN} C_M \otimes C^\prime_N =0\ ,
\nn\w2 
& \qquad C_M, C^\prime_M \in \{ \partial_M, B_{\m M}, \sigma_M \} \ .
\label{SC}
\end{align}
where $d^{MNP}$ is the totally symmetric invariant tensor of $E_6$, and $\Omega^{MN}$ is the antisymmetric invariant tensor of $E_7$ and $\eta^{MN}$ is the symmetric invariant tensor of $E_8$, and
\be
\left(P_{(3875)}\right)^{MK}{}_{NL} = \frac17 \delta^M_{(N}\delta^K_{L)} -\frac{1}{56}\eta^{MK}\eta_{NL} -\frac{1}{14} f^P{}_N{}^{(M} f_{PL}{}^{K)}\ .
\ee
which projects to the 3875 dimensional representation of $E_8$. The $E$-bracket is defined for $E_6, E_7, E_8$ exceptional field theories in \cite{Hohm:2013vpa,Hohm:2013uia,Hohm:2014fxa}, respectively. Schematically, the only nonvanishing commutators are of the form $[\Lambda,\Lambda] \sim \Lambda$ and $[\Lambda,\Sigma] \sim \Sigma$. The solutions of these constraints will be commented on in the following subsections. The gauge covariant derivatives on any tensors are naturally given by 
\be
D_\m = \partial_\m -\cL_{(A_\m, B_\m)}\ .
\ee
The construction of the covariant field strengths for $(A_\m^M, B_{\m M})$, and the fields listed in \eq{fields}, requires the building of an exceptional field theory analog of the ordinary tensor hierarchy outlined in appendix B. This needs to be done case by case in each spacetime dimension. In the following subsections, we shall note the field strengths for the $(\Lambda, \Sigma)$ gauge fields. Their Bianchi identities provide the definitions of higher form field strengths.

In a theory formulated in the extended geometrical framework, in addition to the internal generalized diffeomorphisms with parameter $\Lambda(x,Y)$, there are also external diffeomorphisms with parameter $\xi^\m(x,Y)$. In the universal sector, they take the form \cite{Hohm:2013vpa,Hohm:2013uia,Hohm:2014fxa}
\bea
\delta e_\m{}^a &=& \xi^\m D_\n e_\m{}^a +D_\m\xi^\n e_\n{}^a\ ,
\nn\w2
\delta \cM_{MN} &=& \xi^\m D_\m \cM_{MN}\ ,
\nn\w2
\delta A_\m^M &=&\begin{cases} \xi^\n \cF_{\n\m}{}^M +\cM^{MN} g_{\m\n} \partial_N \xi^\n & \text{for}\ \ n=6,7 \\
\ve_{\m\n\r} \xi^\n J^{\rh M}+\cM^{MN} g_{\m\n} \partial_N \xi^\n & \text{for}\ \ n=8
\end{cases}
\eea
where $\cM_{MN}$ is the generalized metric in the internal space, the duality equation $\xi^\n \cF_{\n\m}{}^M = \ve_{\m\n\r} \xi^\n J^{\rh M}$ has been used in the case of $n=8$, and 
\be
D_\m e_\n{}^a = \partial_\m e_\n{}^a -A_\m{}^M \partial_M e_\n{}^a -\beta_n\, \left(\partial_M A_\m{}^M\right) e_\n{}^a\ .
\label{cd}
\ee
For the transformations of the remaining fields under the external diffeomorphisms, see \cite{Hohm:2014fxa}. It is a remarkable fact that the external diffeomorphisms fix all the relative coefficients of the bosonic terms in the actions constructed so far, thereby producing the bosonic sectors of maximal supergravities without using local supersymmetry.  

In closing this subsection, it is also useful to give the following universal formulae that provide a universal sector of the actions constructed. Firstly, the generalized Ricci scalar is defined as 
\be
\hR = e e_a{}^\m e_b{}^\n \left( R_{\m\n}{}^{ab}(\omega) +\cF_{\m\n}{}^M  e^{a\rh} \partial_M e_\rho{}^b\right)\ .
\label{F}
\ee
The Riemann tensor $R_{\mu\nu ab}(\omega)$ has the standard expression in terms of the vielbein with all derivatives covariantized as in \eq{cd}, and the second term is needed to ensure covariance not only under the generalized diffeomorphisms but also under the local Lorentz transformations \cite{Hohm:2013nja}. Second, the  potential which arises in the Lagrangians for the exceptional field theories is given by \cite{Hohm:2013vpa,Hohm:2013uia,Hohm:2014fxa}
\bea
V &=& -\frac{1}{4\alpha_n} \cM^{MN} \left( \partial_M \cM^{KL}\right) \left(\partial_N \cM_{KL}\right) +\frac12  \cM^{MN} \left( \partial_M \cM^{KL}\right) \left(\partial_L \cM_{NK}\right)
\nn\w2
&& -\frac12 \left(g^{-1}\partial_M g\right) \partial_N \cM^{MN} -\frac14 \cM^{MN} \left(g^{-1}\partial_M g\right)\left(g^{-1}\partial_N g\right) -\frac14 \cM^{MN} \partial_M g^{\m\n} \partial_N g_{\m\n}
\nn\w2
&& + \frac{1}{7200}  \delta_{n,8}\,  f^{NQ}{}_P f^{MS}{}_R \left( \cM^{PK}\partial_M \cM_{QK} \right) \left( \cM^{RL}\partial_N \cM_{SL} \right)\ ,
\label{UV}
\eea
where the symmetric matrix $\cM_{MN}$ is the `generalized metric' on the coset 
$E_{n(n)}/K_n$. 

Explicit solutions to the section constraints have been found \cite{Hohm:2013vpa} which yield $11D$ or type IIB supergravity compactified on $n$-torus but with dependence on the toroidal coordinates kept. Key to these solutions is the decomposition of the $R_1$ representation under $SL(n) \times GL(1)$ for the $11D$ embedding, and under $SL(n-1)\times SL(2)$ for the type IIB embedding, which are given by \cite{Hohm:2013vpa,Hohm:2013uia,Hohm:2013uia}
\bea
11D: &&  E_{n(n)} \to SL(n)\times GL(1): \ \  R_1 \to n_q + \cdots \ ,
\label{dec}\w2
\mbox{Type IIB:} && E_{n(n)} \to SL(n-1)\times GL(1)\times SL(2): 
\nn\w2
&& R_1 \to (n-1,1)_q + \cdots \ ,
\nn
\eea
where $q$ is the highest $GL(1)$ charge. The section constraints are then solved by (a) restricting the $Y$-dependence of all fields, including gauge parameters, to the $n$ coordinates in the $11D$ embedding, or $n-1$ coordinates in the type IIB embedding, and (b) keeping only $B_{\m\n m}$ and $B_{\m m}$ components of $B_{\m\n M}$ and $B_{\m M}$ nonzero, in the cases of $E_{7(7)}$ and $E_{8(8)}$, respectively. Thus, a generic field  $\Phi(x^\m, Y^M) \to \Phi (x^\m, y^m)$, where $y^m$ corresponds to the $n_q$ or $(n-1,1)_q$ representation. In the case of $E_{6(6)}$, it has been shown that $11D$ supergravity on 5-torus, with the toroidal coordinates kept, is equivalent to the  $E_{6(6)}$ exceptional field theory with the section constraint solved as described above \cite{Hohm:2013vpa}.

In the case of DFTs, focusing on the manifestly duality invariant formulation of $N=1, 10D$ supergravity, the extended geometry is formulated in terms of the coordinates $X^M=({\tilde x}_\mu, x^\mu, y^m)$ which are in the fundamental representation of the duality group $G=O(10,10+n_V)$, with $\mu, \tilde \mu =0,1,...,9$ and $m=1,...,n_V$.  In addition to this global symmetry, the theory has local double Lorentz $H=O(9,1)\times O(1,9+n_V)$ symmetry. The generalized diffeomorphism with parameter $\xi^M$ acts on a generalized vector as
\be
\delta V^M =  \xi^N\partial_N V^M -\left(\partial^M\xi_N-\partial_N\xi^M\right) V^N+ f^M{}_{NP} \xi^N V^P\ ,
\ee
where $M=0,1,...,19+n_V$. The constants $f_{MNP}= f_{MN}{}^L \eta_{LP}$, often referred to as fluxes or gaugings, break the $G$-invariance but the formalism is $G$-covariant, just as in the gauged supergravity theories. In DFT there also exists a dilaton field which transforms as 
\be
\delta d= \xi^M\partial_M d -\frac12 \partial_M \xi^M\ .
\ee
The algebra of these transformations closes provide that the following constraints are imposed 
\begin{align}
& \partial_M \partial^M \cdots =0\ ,\qquad \partial_M\cdots \partial^M \cdots =0\ ,\qquad  f_{MN}{}^P \partial_P =0\ ,
\nn\w2
& f_{MNP}= f_{[MNP]}\ , \qquad f_{[MN}{}^L f_{P]L}{}^Q=0\ .
\label{constraints}
\end{align}
where the ellipses refer to any fields or gauge parameters. 

\subsection{$E_{6(6)}$ exceptional supergravity in $(5+27_c)$ dimensions}

The bosonic sector was constructed in \cite{Hohm:2013vpa}, and the supersymmetric version in \cite{Musaev:2014lna}. We shall summarize the key results below based on these papers. The  field content is given by
\be
\left( g_{\m\n}, \cV_M{}^{ij}, A_\m^M, B_{\m\n\,M}; \psi_\m^i, \chi^{ijk} \right)\ ,
\ee
where $M=1,..,27$ labels the fundamental representation of $E_{6(6)}$, and $\cV_M{}^{ij}$, where $i=1,...,8$ labels the fundamental representation of $USp(8)$, parametrizes the  coset $E_{6(6)}/USp(8)$. The spinors are symplectic Majorana. The two-form potentials have been introduced as duals of the  gauge fields. The formulae summarized in the previous section apply. In the conventions of \cite{Hohm:2013vpa}, the covariant two-form field strength is given by
\bea
\cF_{\m\n}{}^M &=& 2\partial_{[\m} A_{\n]}^M -[A_\m,A_\n]_E^M +10\,d^{MNK} \partial_N B_{\m\n\,K}\ .
\eea
The field strength $\cH_{\m\n\r\,M}$ can be read off from  the Bianchi identity, 
\be
D_{[\m} \cF_{\n\r]}{}^M=\frac{10}{3} d^{MNK} \partial_K \cH_{\m\n\r\,N}\ ,\qquad D_\m = \partial_\m-\cL_{A_\m}\ .
\ee
We also need the scalar current and composite connection defined as
\be
\cP_\m^{ijk\ell} = D_\m \cV_M^{[ij} \cV^{k\ell]M}\ ,\qquad \cQ_{\m i}{}^j = \frac13 \cV_{ik}{}^M D_\m \cV_M{}^{jk} \ ,
\ee
With the above ingredients at hand, the bosonic part of the Lagrangian takes the form \cite{Musaev:2014lna}
\be
e^{-1} \cL =  \hR-\frac14 \cM_{MN} \cF_{\m\n}^M \cF^{\m\n N} -\frac16 \cP_\m{}^{ijk\ell} \cP^\m{}_{ijk\ell} - V(\cM, g) +\frac{\sqrt{10}}{8} e^{-1} \cL_{\rm top}\ ,
\ee
with $(\hR, V)$ from \eq{F} and \eq{UV}, respectively, and $\cM_{MN} = \cV_M{}^{ij} \cV_{N ij}$. The topological Lagrangian is given in \cite{Musaev:2014lna}, where its general variation is also given as 
\be
\delta \cL_{\rm top} = \e^{\m\n\r\s\tau} \Big( \frac34 d_{MNK} \cF_{\m\n}{}^M \cF_{\r\s}{}^N \delta A_\tau{}^K + 5 d^{MNK} \partial_N \cH_{\m\n\r M} \Delta B_{\s\tau K} \Big)\ ,
\ee
with the covariant variation $\Delta B_{\m\n M} \equiv \delta B_{\m\n M} +d_{NKL} A_{[m}{}^K \delta A_{\n]}{}^L$. Using this formula, it is easy to derive the following (projected) duality equation as the field equation of $B_{\m\n M}$:  
\be
d^{MNK} \partial_K \Big( \cM_{MN} \cF^{\m\n N} +\frac{\sqrt{10}}{6}\e^{\m\n\r\s\tau} \cH_{\r\s\tau M} \Big) =0\ .
\ee
This ensures the correct $128_B+128_F$ physical degrees of freedom. Finally, the supertransformations of the fermionic fields are \cite{Musaev:2014lna}
\bea
\delta \psi_\m &=& \cD_\m \e^i -i \sqrt2 \cV^{ij M} \Big(  \Omega_{jk}\nabla^{-}_M (\gamma_\m \e^k) -\frac13 \gamma_\m \nabla^-_M \e_j\Big)\ ,
\nn\w2
\delta \chi^{ijk} &=& \frac{i}{2} \cP_\m{}^{ijk\ell} \Omega_{\ell m} \gamma^\m \e^m +\frac{3}{\sqrt2} \cV^{\llbracket ij M} \nabla^{-}_M \e^{k\rrbracket}\ ,
\eea
where $\llbracket\rrbracket$ means totally antisymmetric and symplectic traceless, $\cD_\m = \cD_\m (A_\n, \omega_\n, \cQ_\n)$ and $\nabla^{-}_M = \nabla_M(\omega_N^{-}, \cQ_n, \Gamma_N)$ with  $\Gamma_N$ the Christoffel connection and 
\be
\omega^{-}_{M ab} \equiv \omega_{M ab} - \frac12 \cM_{MN} \cF_{\m\n}^N e^\m_a  e^\n_b\ .
\ee
For the transformation of all the fields in the theory under the generalized diffeomorphisms and the external diffeomorphisms, see \cite{Musaev:2014lna}. 

\subsection{$E_{7(7)}$ exceptional supergravity in $(4+56)$ dimensions}

The bosonic sector was constructed in \cite{Hohm:2013uia} and the supersymmetric version in \cite{Godazgar:2014nqa}. We shall summarize the key results below based on these papers. The field content is given by
\be
\left( g_{\m\n}, \cV_M{}^A, A_\mu^M, B_{\m\n\,\a}, B_{\m\n\,M} ; \psi_\m^i, \chi^{ijk} \right)\ ,
\label{m}
\ee
where $M=1,...,56$ and $\cV_M{}^A= \big( \cV_M{}^{ij}, \cV_{M ij} \big)$ is a representative of the coset $E_{7(7)}/SU(8)$, satisfying $\cV_{M ij} = \big( \cV_M{}^{ij}\big)^\star$ with $SU(8)$ indices $i,j,...=1,...,8$. The generalized metric is defined as $\cM_{MN}= \cV_{M ij} \cV_N{}^{ij} + \cV_{N ij} \cV_M{}^{ij}$. The dual gauge potentials are introduced so that $A_\m^M = (A_\m^m, A_{\m\,m})$ form the $56$-plet of $E_{7(7)}$. The two-forms $B_{\m\n\,\a}, \alpha=1,...,133$, are introduced as on-shell duals of the scalars. The two form potentials $B_{\m\n\,M}$ is introduced as a new type of field which does not follow from the tensor hierarchy. It is needed for achieving a closed generalized gauge algebra. The fermions are Majorana, and $\chi^{ijk}= \chi^{[ijk]}$. 

The covariant two-form field strength is given by
\bea
\cF_{\m\n}{}^M &=& 2\partial_{[\m} A_{\n]}^M -[A_\m,A_\n]_E^M -12 (t^\alpha)^{MN} \partial_N B_{\m\n\alpha} -\frac12 \Omega^{MN} B_{\m\n N}\ .
\eea
The field strengths $\cH_{\m\n\r\,\a}$ and $H_{\m\n\r M}$ can be read off from the Bianchi identity,
\be
D_{[\m} \cF_{\n\r]}{}^M= -4(t^\a)^{MN} \partial_N H_{\m\n\r \a} -\frac16 \Omega^{MN} H_{\m\n\r N}\ ,\qquad D_\m = \partial_\m-\cL_{A^\m}\ .
\ee
The three forms arising in the tensor hierarchy, namely $C_{\m\n\r}{}^M{}_\a$  and $C_{\m\n\r M}{}^N$, have dropped out by using the section conditions. We also need the scalar current and composite connection defined as \cite{deWit:2007kvg}
\be
\cP_{\m ijk\ell} = i \Omega^{MN} D_\m \cV_{M ij} \cV_{N k\ell}\ ,\qquad \cQ_{\m i}{}^j = \frac{2i}{3} \cV^{N jk}D_\m \cV_{N ki} \ .
\ee
Given these building blocks, the (pseudo) Lagrangian can be written as
\be
e^{-1} \cL =\hR + \frac12 \cP_\m^{ijk\ell} \cP^\m_{ijk\ell} -\frac18 \cM_{MN} \cF^{\m\n M} \cF_{\m\n}{}^N -V(\cM, g) +e^{-1} \cL_{\rm top}\ ,
\ee
with $(\hR, V)$ from \eq{F} and \eq{UV}, respectively. The topological Lagrangian, as well as its general variation
\be
\delta \cL_{\rm top} = -\frac14 \e^{\m\n\r\s} \left( \delta A_\m{}^M D_\n \cF_{\r\s M} +\cF_{\m\n M} \left( 6(t^\a)^{MN} \partial_N \Delta B_{\r\s \a} +\frac14 \Omega^{MN} \Delta B_{\r\s N} \right) \right)\ ,
\ee
with the appropriate covariant variations, are given in \cite{Hohm:2013uia}. It is important to note that this gives a pseudo-action in the sense that it needs to be supplemented by the duality equation
\be
\cF_{\m\n}{}^M = \frac{i}{2} \e_{\m\n\r\s} \Omega^{MN} \cM_{NK} \cF^{\r\s K}\ ,
\label{de4}
\ee
which must be imposed by hand after performing the general variations in getting the equations of motion. The field strengths $\cH_{\m\n\r \a}$ and $\cH_{\m\n\r M}$ obey duality equations which can be derived from the variation of the pseudo-action, and the external curl of the duality equation \eq{de4}. The resulting equations are \cite{Hohm:2013uia}
\bea
\e^{\m\n\r\s} \cH_{\n\r\s \a} &=& -\frac12 (t_\a)_K{}^L \big( D^\m \cM^{KP} \cM_{LP} \big)\ ,
\nn\w2
\frac{1}{12} \e^{\m\n\r\s} \cH_{\n\r\s M} &=& -\frac13 D^\m\cV^{N ij} \partial_M \cV_{N ij} -2i e^\m e^\n_b \Big(\partial_M \omega_\n{}^{ab} -D_\n \Omega_M{}^{ab} \Big)\ ,
\eea
where $\omega_M{}^{ab} := -e^{\n [a} \partial_M e_\n^{b]}$.  The first equation only arises under the projection $(t^\a)^{MN} \partial_N$. 
Finally, the supersymmetry transformations are \cite{Godazgar:2014nqa}
 \bea
 \delta \psi_\m^i &=& 2 \cD_\m\e^i -4i \cV^{M ij} \nabla^{+}_M \left(\gamma_\m \e_j\right)\ ,
 \nn\w2
 \delta \chi^{ijk} &=& -2\sqrt2 \cP_\m{}^{ijkl} \gamma^\m \e_\ell -12\sqrt2 i \cV^{M[ij} \nabla^{+}_M\e^{k]}\ ,
 \eea
where $\cD_\m = \cD_\m (A_\n, \omega_\n, \cQ_\n)$ and $\nabla^+_M = \nabla_M(\omega_N^+, \cQ_n, \Gamma_N)$ with $\Gamma_N$ the Christophel connection and 
\be
\omega^{+}_{M ab} \equiv \omega_{M ab} + \frac14 \cM_{MN} \cF_{\m\n}^N e^\m_a  e^\n_b\ .
\ee
In passing, let us also note that a superspace formulation in which the external spacetime is elevated to a $(4|32)$ dimensional superspace, but with the internal space left intact, was constructed in \cite{Butter:2018bkl}.

\subsection{$E_{8(8)}$ exceptional supergravity in  $(3+248)$ dimensions}

The bosonic sector was constructed in \cite{Hohm:2014fxa}, and the supersymmetric version in \cite{Baguet:2016jph}. We will follow these papers for a brief summary here. The field content of the theory is 
\be
\left( g_{\m\n}, \cV_M{}^X, A_\m^M, B_{\m\,M} ;  \psi_\m^I, \chi^{\dA} \right)\ ,
\label{mm}
\ee
where $M, X=1,...,248$ and $\cV_M{}^X= (\cV_M{}^{IJ}, \cV_M{}^A)$ is the representative of the coset $E_{8(8)}/SO(16)$ in the adjoint representation, satisfying $\cV_M{}^{IJ} =\cV_M{}^{[IJ]}$ with $SO(16)$ vector indices $I,J=1,...,16$, and $SO(16)$ spinor indices $A,\dA=1,...,128$. The spinors are Majorana. The generalized metric is defined as $\cM_{MN}= \cV_M{}^X \cV_N{}^X$. The potential $A_\m^M$ is dual to the scalar fields, while the one-form potential $B_{\m M}$ is introduced is a new type of field which does not follow from  tensor hierarchy. It is needed for achieving the closure of the generalized gauge algebra, as discussed in section 13.1. See \cite{Hohm:2014fxa} for the details.

The Lagrangian is given by \cite{Hohm:2014fxa}
\be
e^{-1}\cL= -\hR +\cP_\m^A \cP^{\m A} +\cL_{\rm top} -V\ ,
\ee
where $\cP_\m{}^A$ is defined by
\be
\cM^{KP} D_\mu \cM_{PL} = 2 f^{MK}{}_L \cV_M{}^A \cP_\mu{}^A\ , \qquad D_\m =\partial_\m -\cL_{(A_\m, B_\m)}\ .
\ee
The topological action is given by \cite{Hohm:2014fxa}
\be
S_{\rm top} = \int d^{248} Y \int_{\cM_4} \left( \cF^M \wedge \cG_M -\frac12 f_{MN}{}^K \cF^M \wedge \partial_K \cF^N \right)\ ,
\ee
with the three-dimensional spacetime residing on  boundary of $\cM_4$. The field strengths are given by
\bea
\cF_{\m\n}{}^M &=& F_{\m\n}{}^M +14 \left(P_{3875}\right)^{MN}{}_{PQ}\partial_N C_{\m\n}{}^{PQ} +\frac14 \partial^M C_{\m\n} +2 f^{MN}{}_P C_{\m\n N}{}^P\ ,
\nn\w2
\cG_{\m\n M} &=& G_{\m\n M} +2\partial_N C_{\m\n M}{}^N +2\partial_M C_{\m\n N}{}^N\ ,
\eea
where $F_{\m\n}{}^M$  and $G_{\m\n M}$ are determined from $[D_\m, D_\n] V^M = -\cL_{(F_{\m\n}, G_{\m\n})} V^M $ and given explicitly in \cite{Hohm:2014fxa}, where it is also shown that all dependence on the two-form fields drop out of the action. The physical degree of freedoms are in accordance with supersymmetry, in view of the fact that the general covariant variation of the action with respect to field $B_{\m M}$ gives the duality equation 
\be
\cF_{\m\n}{}^M =  \frac{1}{60} \e_{\m\n\r} f^{MK}{}_L \left( \cM^{LP} D^\r\cM_{PK} \right) \ ,
\label{de8}
\ee
which holds up to terms of the form $\cO_{\m\n}{}^M$ that vanish when contracted with a field satisfying the section constraints, since the general variation is with respect to $B_{\m M}$ which is subject to the section constraints.  Finally, the supertransformation rules are \cite{Hohm:2014fxa}
\bea
\delta \psi_\m^I &=& \cD_\m \e^i +2i\, \cV^M{}_{IJ} \nabla_M \big(\gamma_\m \e^J\big) +2i\, \cV^M{}_{IJ} \gamma_\m \nabla_M \e^J\ ,
\nn\w2
\delta \chi^{\dA} &=& \frac{i}{2} \gamma^\m \e^i \Gamma^I_{A\dA} \cP_\m^A-2\cV^M{}_A \Gamma^I_{A\dA} \nabla_M \e_I\ .
\eea

\subsection{$N=1$ supersymmetric double field theory}

$N=1, 10D$ supergravity coupled to an arbitrary number of abelian vector multiplets was constructed in \cite{Hohm:2011nu}, and its generalization to non-abelian coupling of $n_V$ vector fields in \cite{Lescano:2021guc}. The main results can be summarized briefly as follows. The extended geometrical framework in this case is based on the coset $G/H= O(10,10+n_V)/(O(9,1)\times O(1,9+n_V)$. The multiplet of fields consists of
\be
( E^M{}_A,\, d,\, \Psi_{\bar a},\, \rho )\ ,
\ee
where $E^M{}_A= \left(E^M{}_a, E^M{}_{\bar a} \right)$, $a=0,1,...,9$, ${\bar a}=0,1,..., 9+n_V$, is the generalized vielbein and $d$ is the dilaton. The gravitino $\Psi_{\bar a}$ and dilatino $\rho$ are Majorana spinors of $O(9,1)$, and the gravitino is a vector of $O(1,9+n_V)$. The bosonic part of the Lagrangian is given by $e^{-2d} R$ where $R$ is the generalized Ricci scalar, thus taking the form \cite{Hohm:2011nu,Lescano:2021guc}
\be
\cL= e^{-2d}\Big[  \tfrac18 \omega_{[ABC]}\omega_{[DEF]} \left( H^{AD} \eta^{BE} \eta^{CF} -\tfrac13 H^{AD} H^{BE} H^{CF}\right) -H^{AB}\left( E_A + \tfrac12 \omega_A \right) \omega_B \Big]\ ,
\ee
where $\omega_A = \omega_{BA}{}^B,\ E_A=-\sqrt{2} E_A{}^M \partial_M$, and the totally antisymmetric part of the spin connection
\be
\omega_{[ABC]} = \Big[\left(E_A E^N{}_B\right) E_{NC} -\tfrac{\sqrt 2}{3} f_{MNP} E^M{}_A E^N{}_B E^P{}_C \Big]_{[ABC]}\ . 
\ee
The symmetric and invertible constant metrics $\eta_{AB}$ and $H_{AB}$ are $H$-invariant metrics, with the latter constrained to satisfy $H_A{}^C H_C{}^B= \delta_A^B$. The constant metric $\eta_{MN}$ is $G$ invariant. The metrics $\eta_{AB}$ and $\eta_{MN}$ are used to raise and lower indices. The supertransformations of the fermionic fields are given by
\be
\delta \Psi_{\bar a} = \nabla_{\bar a} \epsilon\ ,\qquad \delta \rho = -\gamma^a \nabla_a \epsilon\ ,
\ee
where $\epsilon$ is a Majorana spinor of $O(1,9)$, and $\nabla_A\epsilon= \left(E_A -\frac14 \omega_{A bc} \gamma^{bc}\right)\epsilon$. 

\subsection{Consistent Kaluza-Klein reductions}

Generalized Scherk-Schwarz ansatz for the full supersymmetric $E_{6(6)}$ and $E_{7(7)}$ exceptional field theories that uses twist matrices subject to consistency equations, and leading to the field equations of lower dimensional gauged supergravity theories parametrized by an embedding tensor, was achieved in \cite{Hohm:2014qga}. Following \cite{Hohm:2014qga}, let us briefly summarize how this works for $E_{6(6)}$. Under $SL(6)\times SL(2)$ the coordinates in ${\bar \bf 27}$ of $E_{6(6)}$ are decomposed as $Y^M \to (Y^{[AB]}, Y_{A\a})$ with $A=0,1,...,5,\ \a=1,2$, and all fields are taken to depend only on $(x^\m, Y^{0i})$ with $\m=0,1,...,5$ and $i=1,...,5$. Next, in a Scherk-Schwarz reduction scheme, the $E_{6(6)}$ twist matrix is chosen as \cite{Hohm:2014qga}
\be
U_M{}^N = \begin{pmatrix} U_{[AB]}{}^{[CD]} & 0 \\ 0 & \delta^\a_\gamma \left(U^{-1}\right) _C{}^A \end{pmatrix}\ .
\ee
and the following ansatz is made for all the fields
\bea
e_\m{}^a(x,Y) &=& \rho^{-1} (Y) e_\m{}^a (x)\ ,
\nn\w2
\cM_{MN}(x,Y) &=& U_M{}^P(Y) U_N{}^Q(Y) M_{PQ}(x)\ ,
\nn\w2
A_\m{}^M (x,Y) &=& A_\m{}^N(x) \left(U^{-1}\right)_N{}^M (Y) \rho^{-1}(Y)\ ,
\nn\w2
B_{\m\n M}(x,Y) &=& \rho^{-2}(Y) U_M{}^P (Y) B_{\m\n P}(x)\ .
\eea
The consistency of this reduction is shown to impose the conditions
\bea
\Big[ \left(U^{-1}\right)_M{}^P \left(U^{-1}\right)_N{}^Q \partial_P U_Q{}^K \Big]_{(351)} &=& \rho \, \theta_M{}^\a \left(t_\a\right)_N{}^K\ ,
\nn\w2
\Big[(\partial_N-4\left(\rho^{-1}\partial_N\, \rho\right) \Big] \left(U^{-1}\right)_M{}^N &=& 3 \rho\, \vartheta\ ,
\eea
where $(351)$ denotes projection to the $351$-dimensional representation of $E_{6(6)}$, and $\left(\theta_M{}^\a, \vartheta_M\right)$ are the embedding tensor components, the latter associated with the gauging of the trombone symmetry. The solutions of these equations, together with the dictionary that relates exceptional field theories to type $IIB$ supergravity \cite{Hohm:2013vpa}, provides  consistent truncations to $AdS_5\times S^5$ and various hyperboloid compactifications as well. For more details, see \cite{Hohm:2014qga}.

In the case of DFTs, by appropriately choosing the global and local symmetry groups and gaugings, one can describe both the $N=1, 10D$ heterotic supergravity as well as its toroidal compactified version. To this end, following the presentation of \cite{Baron:2017dvb}, the groups $G$ and $H$ discussed in section 12.1 are split as 
\be
G \to \underbrace{O(d,d)}_{G_e}\times \underbrace{O(n,n+n_V)}_{G_i}\ ,\qquad H\to \underbrace{O(d-1,1)\times O(1,d-1)}_{H_e} \times \underbrace{O(n)\times O(n+n_V)}_{H_i}\ ,
\ee
and the matrices $\eta_{AB}, H_{AB}$ and $\eta_{MN}$ are taken to be
\be
\eta_{AB} =\begin{pmatrix} 0 & \delta_b^a & 0\\ \delta_a{}^b &0&0\\ 0&0& \kappa_{mn} \end{pmatrix}\ ,\quad H_{AB} =\begin{pmatrix} g^{ab} & 0 & 0\\ 0 & g_{ab}&0\\ 0&0& M_{\alpha\beta} \end{pmatrix}\ ,\quad \eta_{MN} =\begin{pmatrix} 0 & \delta^\mu{}_\nu & 0\\ \delta_\mu{}^\nu &0&0\\ 0&0& \kappa_{mn} \end{pmatrix}\ ,
\ee
where $\mu,a=0,...,D-1,\ m,\alpha =1,...,2n+n_V$. Here $g_{ab}$ is the flat Minkowski metric $\kappa_{\alpha\beta}$ and $M_{\alpha\beta}$ are two symmetric and constant $H_i$ invariant matrices, and $\kappa_{mn}$ is the $G_i$ invariant metric. A suitable solution to the constraints \eq{constraints} for the reduction to $D=10-n$ dimensions is given by 
\be
f_{MNP}= \delta_M^m \delta_N^n\delta_P^p\,f_{mnp}\ ,\qquad \partial_M=\left({\tilde \partial}^\mu, \partial_\mu, \partial_m\right)= \left(0,\partial_\mu, 0\right)\ ,
\ee
and the generalized vielbein is parametrized as \cite{Baron:2017dvb}
\be
E_M{}^A = \left(
          \begin{array}{ccc}
            e_a{}^\mu & 0 & 0 \\
            -e_a{}^\rho \left(B_{\mu\nu}+\frac12 A_\mu{}^m A_{\nu m}\right)\quad  &\quad e_\mu{}^a \quad &\quad A_\mu{}^p \cV_p{}^\alpha \\
            -e_a{}^\rho A_{\rho m} &0& \cV_m{}^\alpha \\
          \end{array}
        \right)\ ,\qquad e^{-2d}= \sqrt{-g} e^{-2\phi}\ .
\ee
The external double-Lorentz transformations are gauge fixed to the diagonal part corresponding to the single Lorentz group in $D$ dimensions, $\cV_m{}^\alpha$ is the  vielbein on the scalar manifold $G_i/H_i$, and $A_\mu{}^m$ is the Yang-Mills gauge field. After some calculation that also involves field redefinitions, the following action is found \cite{Baron:2017dvb}
\bea
S &=&\int d^D x\,\sqrt{-g}\,e^{-2\phi} \Big[ R+4\nabla_\mu \partial^\mu \phi -4\partial_\mu \partial^\mu \phi -\frac{1}{12} H_{\mu\nu\rho} H^{\mu\nu\rho}
\nn\\
&&  -\frac14 M_{mn} F_{\mu\nu}^m F^{\mu\nu n} +\frac18 \nabla_\m M_{mn} \nabla^\mu M^{mn} -V\Big]\ ,
\eea
where $M_{mn} = M_{\alpha\beta} \cV_m{}^\alpha \cV_n{}^\beta$ and 
\bea
H_{\mu\nu\rho} &=& 3\partial_{[\mu} B_{\nu\rho]} -3 \Big( A_{[\mu}{}^m \partial\nu A_{\rho]m} -\frac13 f_{mnp} A_\mu{}^m A_\nu{}^n A_\rho{}^p\Big)\ ,
\nn\w2
F_{\mu\nu}{}^m &=& 2\partial_{[\mu} A_{\nu]}{}^m -f_{pq}{}^m A_\mu{}^p A_\nu{}^q\ ,
\nn\w2
\nabla_\mu M_{mn} &=& \partial_\mu M_{mn}+f_{pm}{}^q A_\mu{}^p M_{qn} + f_{pn}{}^q A_\mu{}^p M_{mq}\ ,
\eea
and the potential is given by
\be
V =\frac{1}{12} f_{mp}{}^r f_{nq}{}^s M^{mn} M^{pq} M_{rs} +\frac14 f_{mp}{}^q f_{nq}{}^p M^{mn} +\frac16 f_{mnp} f^{mnp}\ ,
\label{pot}
\ee
The constants $f_{mnp}$ break the global duality symmetry $O(n,n+n_V)$, but the action still provides an $O(n,n+n_V)$ covariant description. In fact the action above has the same form as the half-maximal gauged supergravities reviewed in previous sections. For example, taking $n=3$, we have checked that the potential \eq{pot} agrees with the one given in \eq{hm7}, which was obtained directly in $7D$ by means of Noether procedure. 

The reduction of $N=1, 10D$ heterotic supergravity with Yang-Mills symmetry group $K$ was performed in \cite{Lu:2006ah}, leading to a large class of gauged supergravities in $d=10-n$ dimensions, where $n={\rm dim\, G}$. The hidden $O(n,n+ {\rm dim\, K})$ symmetry was uncovered, and local gauge symmetry in  $d$ dimensions as $K\times G\rtimes R^n$ was identified. An alternative route to gauged supergravities in lower dimensions is to perform generalized Scherk-Schwarz reduction in the framework of the DFT. This has been done in \cite{Aldazabal:2011nj,Geissbuhler:2013uka} where $N=4, 4D$ supergravity with its electric gaugings were obtained in agreement with the embedding tensor construction of \cite{Schon:2006kz}. A fuller analysis awaits to be performed to pin down the most general gaugings, and their relationships to higher dimensions.

\subsection*{Acknowledgements}

I especially thank Toine van Proeyen and Axel Kleinschmidt for careful reading of the manuscript, and for their excellent suggestions. I am also grateful to Eric Bergshoeff, Guillaume Bossard, Gabrielle Larios, Yi Pang and Henning Samtleben for many helpful discussions. This work is supported in part by the NSF grant PHYS-2112859.

\begin{appendix} 

\section{ Conventions, and spinors in arbitrary dimensions }

Throughout the paper, it is understood that the supersymmetry transformation rules are given up to terms that are quadratic in the fermionic fields of the supermultiplets involved.  As for the  signature of spacetime, and conventions, we mostly stick to the conventions of the original papers, though some notation changes are employed in certain cases. Nonetheless, it is useful to recall the general properties of  spinors and Dirac gamma-matrices in arbitrary dimensions, which we reproduce here for the reader's convenience. This also gives us an opportunity to correct the typos that appeared in equations (3), (6), (8) and (9) on pages 5 and 6 in \cite{Salam:1989fm}. 

Consider the  Clifford algebra $\{\gamma_\mu, \gamma_\nu\} = 2\eta_{\mu\nu}$ in $(t,s)$ dimensions with the signature 
\be
\eta_{\mu\nu} ={\rm diag} (\underbrace{-,-,...-}_{t\, \rm times}, \underbrace{+,+,...+}_{p\, \rm times})\ .
\ee
The $\gamma$-matrices have the following properties
\be
\gamma_{\m}^{\dg}=(-1)^t A \gamma_{\m} A^{-1}\ ,\qquad \gamma_{\m}^{\star}=\eta B \gamma_{\m} B^{-1}\ ,\qquad
\gamma_{\m}^T=(-1)^t\eta C\gamma_{\m}C^{-1}\ , 
\ee
where 
\be
A =\gamma_0\gamma_1\dots\gamma_{t-1}\ ,\qquad B^T= \e B\ ,\qquad C=BA\ ,\qquad \e=\pm 1\ ,\quad \eta=\pm 1\ .
\ee
It follows that
\be
(C\gamma_{\m_1\m_2\dots\m_n})^T=\e\eta^{t+n}(-1)^{(t-n)(t-n+1)/2} (C\gamma_{\m_1\m_2\dots\m_n})\ .
\ee
Same symmetry property applies to $\gamma_{\m_1\m_2\dots\m_n}C^{-1}$. Note also that $C\gamma_{\m_1\m_2\dots\m_n}$ and $C\gamma_{\m_1\m_2\dots\m_{n+2}}$ have the opposite symmetry property. The Fierz rearrangement formula for two $2^{[d/2]}$ dimensional anti-commuting spinors $\psi$ and $\chi$ in $d$-dimensions, is given by 
\be
\chi {\bar\psi} = -2^{-[d/2]} \sum_{n=0}^{d} (-1)^{[n/2]} \, \gamma^{\n_1...\n_n} \, \left( {\bar\psi} \gamma_{\n_1...\n_n} \chi\right) /n!
\ee
The product of $n$-th rank and $m$-th $\gamma$-matrices can be decomposed as
\bea
\gamma_{\m_1\dots\m_m}\gamma^{\n_1\dots\n_n} & = & \gamma_{\m_1\dots\m_m}{}^{\n_1\dots\n_n}+mn\d_{\m_m}^{\n_1}
\gamma_{\m_1\dots\m_{m-1}}{}^{\n_2\dots\n_n}
\nn\w2
 && +\frac{1}{2!}m(m-1)n(n-1)\d_{\m_m}^{\n_1}\d_{\m_{m-1}}^{\n_2}
\gamma_{\m_1\dots\m_{m-2}}{}^{\n_3\dots\n_n}+\dots\ ,
\eea
where it is understood that the indices $\m_1\dots\m_m$ and $\n_1\dots\nu_n$ are to be totally anti-symmetrized with unit normalization. Finally,  the reality properties of spinors in $(t,s)$ dimensions are tabulated below, where $\Omega^{ij}=-\Omega^{ji}$ is  invariant tensor of the symplectic group. Additional Weyl condition can be imposed for $(t-s)=0\,{\rm mod}\, 4$.

\bigskip

\begin{tabular}{|l|l|l|l|l|l|}
\hline
$(s-t)$ & $\e$ & $ \eta $ & Reality Condition & Spinor \\ \hline
\hline
1,2,8 & +1 & +1 & $\p^*=B\p$ & Majorana \\
6,7,8 & +1 & \ -1 & $\p^*=B\p$ & Pseudo-Majorana \\
4,5,6 & \ -1 & +1 & $(\p_i)^*=\O^{ij}B\p_j$ & Sympl. Majorana \\
2,3,4 & \ -1 & \ -1 & $(\p_i)^*=\O^{ij}B\p_j$ & Pseudo-Sympl-Majorana \\
\hline
\end{tabular}

\section{The embedding tensor formalism}

For a detailed account of the embedding tensor formalism, sometimes referred to as the tensor hierarchy formalism, and for more complete references, see \cite{deWit:2008ta,deWit:2008gc}. Here, we shall primarily follow \cite{deWit:2008gc} to give a brief outline.

Consider a Lie group $G_0$ with  associated Lie algebra $ [t_\a,t_\b]= f_{\a\b}{}^\gamma t_\gamma$. Next, suppose that we wish to gauge a subgroup $G\subset G_0$ with generators $X_M$ given as a linear combination of  generators $t_\a$ as
\be
X_M =\theta_M{}^\a t_\a\ , \qquad M=1,...,\R1\ ,\quad  \a=1,...,{\rm dim}\,G_0\ ,
\ee
where ${\rm R}_1$ is a representation of  group $G$, and  $\theta_M{}^\a$ is a constant matrix, called the embedding tensor. We demand that $X_M$ satisfy a Lie algebra 
\be
[X_M, X_N] = X_{MN}{}^P X_P\ ,
\label{alg}
\ee
with $X_{MN}{}^P$ representing the structure constants. This imposes the condition 
\be
\theta_M{}^\a \theta_N{}^\b f_{\a\b}{}^\gamma + \theta_M{}^\a \left(t_\a\right)_N{}^P \theta_P{}^\gamma =0\ .
\label{qc}
\ee
This, in turn, ensures that the embedding tensor is $G$-invariant. Note that $X_{MN}{}^P$ need not be antisymmetric, and indeed it is useful to define
\be
X_{(MN)}{}^P := Z^P{}_{MN}\ .
\ee
Thus, \eq{alg} implies that 
\be
\theta_P{}^\a Z^P{}_{MN} =0\ .
\ee
Next, one seeks a proper definition of gauge transformation of the gauge field, and a field strength that transforms covariantly under the gauge group $G$. This innocent demand turns out to require the introduction of a hierarchy of $p$-forms, gauge transformations involving shift symmetries, and constant intertwining  tensors that have  following pattern
\bea
p-{\rm forms}:&& \quad A_\m{}^M \qquad  \longrightarrow \quad B_{\m\n}{}^{MN} \quad \longrightarrow \quad C_{\m\n\r}{}^{MNP}  \quad \longrightarrow \quad \cdots
\nn\w2
\mbox{Gauge parameters}: && \quad \Lambda^M  \qquad \longrightarrow \quad \ \Sigma_\m{}^{MN} \quad \longrightarrow \quad \ \Phi_{\m\n}{}^{MNP}  \quad \longrightarrow \quad \cdots
\nn\w2
{\rm Intertwiners}: && Z^M{}_{NP} \quad \longrightarrow \quad Y^{NP}{}_{QRS} \quad \longrightarrow \quad Y^{QRS}{}_{TUVW} \quad \longrightarrow \quad \cdots
\eea
The  gauge transformations are
\bea
\delta A_\m{}^M &=& D_\m \Lambda^M -Z^M{}_{NP} \Sigma_\m{}^{NP}\ ,
\nn\w2
\delta B_{\m\n}{}^{MN}  &=& 2 D_{[\m} \Sigma_{\n]}{}^{MN} -2\Lambda^{\langle M} \cF_{\m\n}^{N\rangle}  -Y^{MN}{}_{P\langle RS \rangle} \Phi_{\m\n}{}^{PRS}  + 2A_{[\m} ^{\langle M} \delta A_{\n]} ^{N\rangle}\ ,
\nn\\
&\vdots &
\eea
where $D_\m\Lambda^M= \partial_M\Lambda^M + X_{KL}{}^M A_\m^K \Lambda^L$, and the notation $\langle MN \rangle$ refers to particular representation obtained by applying the corresponding projector  $\mathbb{P}^{KL}{}_{MN}$ on the symmetric product $R_1 \otimes_{\rm sym} R_1$ as follows 
\be
Z^P{}_{\langle MN \rangle} = Z^P{}_{KL} \mathbb{P}^{KL}{}_{MN}\ .
\ee
The intertwining $Y$-tensors are linear in the embedding tensor and they are the maps
\be
Y^{[p]} : \ R_1^{\otimes (p+1)} \  \longrightarrow\ R_1^{\otimes p}\ ,
\ee
with $\left(Y^{[0]}\right)^\alpha{}_M = \theta_M{}^\alpha$ and $\left(Y^{[1]}\right)^M{}_{PQ} = Z^M{}_{PQ}$, etc., with  the property that
\be
Y^{[p]} \cdot Y^{[p+1]} \approx 0\ ,
\ee
where `$\approx$' means that the expression vanishes a consequence of \eq{qc}. This map has a non-trivial kernel whose complement defines the representation content of  $(p+1)$-forms required for the consistency of  deformed $p$-form gauge algebra.

The $G$-covariant field strengths are constructed as
\bea
\cF_{\m\n}{}^M &=& 2\partial_{[\m} A_{\n]}^M +X_{KL}{}^M A_{[\m}^K A_{\n]}^L +Z^M{}_{KL} B_{\m\n}{}^{KL}\ ,
\nn\w2
\cH_{\m\n\r}{}^{MN} &=& 3 D_{[\m} B_{\n\r]}{}^{MN} + 6A_{[\m}^{\langle M} \partial_\n A_{\r]}^{N\rangle} +2 A_\m^{\langle M} X_{PQ}{}^{N\rangle} A_\n^PA_\r^Q \Big|_{[\m\n\r]} 
\nn\w2
&& +Y^{MN}{}_{P\langle RS\rangle} C_{\m\n\r}^{PRS}\ ,
\nn\\
& \vdots &
\label{FH}
\eea
The  hierarchy can be truncated at a given level if a $p$-form of maximum rank appears in the Lagrangian in such a way that it is projected by a suitable intertwining tensor. For example, to stop  the hierarchy at $B_{\m\nn}^{MN}$, this two-form can only appear in  the combination $Z^P{}_{MN} B_{\m\n}^{MN}$, and the  terms involving $Y^{MN}{}_{PQR}$ will not appear since $Z^P{}_{MN} Y^{MN}{}_{PQR}=0$. For a detailed study of the tensor hierarchy in $D=5,6$ including the construction of all the the field strengths, see \cite{Hartong:2009vc}.

In supergravity theories, the requirement of supersymmetry imposes an additional constraint which is linear in the embedding tensor. This  plays a crucial role in determining the representation content of the embedding tensor. Moreover, in order to have the correct physical degree of freedom count in a given supermultiplet, the representation content of the $p$-forms in the tensor hierarchy must be chosen in such a way that a $p$-form potential in a given representation should come with its dually related $(D-p-2)$-form potential in the conjugate representation. The embedding tensor subjected to the linear and quadratic constraints will determine the form of other constant tensors in the hierarchy such as $Z^P{}_{MN}, Y^{NP}{}_{QRS}$, depending on the dimension of spacetime, the amount for supersymmetry and the supermultiplets involved. In sections 4-10, we have summarized how  this works explicitly.

\newpage

\section{Coset spaces in supergravities}

%
\TableA

\newpage

\thispagestyle{empty}

\newpage

\TableB

\newpage

\TableCC

\end{appendix}

\newpage

{\small
\bibliographystyle{utphys}
\bibliography{references}
}

\end{document}